\documentclass[fleqn,usenatbib,onecolumn]{mnras}

\usepackage{newtxtext}

\usepackage[T1]{fontenc}
\usepackage{ae,aecompl}
\usepackage[normalem]{ulem}

\usepackage[utf8]{inputenc}
\usepackage{graphicx}	
\usepackage{amsmath}	
\usepackage{amssymb}	

\newcommand{\VEC}{\bmath}
\newcommand{\MAT}{\mathbfss}

\newcommand{\hMpc}{\,h^{-1}\, {\rm Mpc}}
\newcommand{\hGpc}{\,h^{-1}\, {\rm Gpc}}

\newcommand{\hk}{\,h\,{\rm Mpc^{-1}}}

\newcommand{\Red}[1]{\textcolor{black}{#1}}

\usepackage{todonotes,ifthen}
\newcounter{mycomment}
\newcommand{\mycomment}[2][]{%
    \refstepcounter{mycomment}%
    \ifthenelse{\equal{#1}{SS}}
    {\todo[color={magenta!40},size=\small, inline]{%
     \textbf{{#1}\themycomment:}~#2}%
    }{\ifthenelse{\equal{#1}{NS}}
    	{\todo[color={red!40},size=\small, inline]{%
     	 \textbf{{#1}\themycomment:}~#2}%
    	}{\ifthenelse{\equal{#1}{FB}}
        {\todo[color={cyan!40},size=\small, inline]{%
     	 \textbf{{#1}\themycomment:}~#2}%
    	}{}}}\noindent}

\title[Covariance matrices of galaxy clustering]{Perturbation theory approach to predict the covariance matrices of the galaxy power spectrum and bispectrum in redshift space}

\author[N. S. Sugiyama et al.]{
Naonori S. Sugiyama$^{1}$\thanks{E-mail: nao.s.sugiyama@gmail.com},
Shun Saito$^{2,3,4}$, Florian Beutler$^{5,6}$, and Hee-Jong Seo$^{7}$
\\
$^{1}$ National Astronomical Observatory of Japan, Mitaka, Tokyo 181-8588, Japan\\
$^{2}$ Department of Physics, Missouri University of Science and Technology, 1315 N. Pine St., Rolla MO 65409, USA\\
$^{3}$ Max-Planck-Institut f\"{u}r Astrophysik, Karl-Schwarzschild-Star{\ss}e 1, D-85740 Garching bei M\"{u}nchen, Germany\\
$^{4}$ Kavli Institute for the Physics and Mathematics of the Universe (WPI), \\
Todai Institutes for Advanced Study, The University of Tokyo, Chiba 277-8582, Japan\\
$^5$ Institute of Cosmology \& Gravitation, University of Portsmouth, Portsmouth, PO1 3FX, UK\\
$^6$ Lawrence Berkeley National Laboratory, 1 Cyclotron Road, Berkeley, CA 94720, USA\\
$^7$ Department of Physics and Astronomy, Ohio University, Clippinger Labs, Athens, OH 45701 
}

\date{}

\pubyear{}

\begin{document}
\label{firstpage}
\pagerange{\pageref{firstpage}--\pageref{lastpage}}
\maketitle

\begin{abstract}

\Red{
In this paper, we predict the covariance matrices of both the power spectrum and the bispectrum, including full non-Gaussian contributions, redshift space distortions, linear bias effects and shot-noise corrections, using perturbation theory (PT). To quantify the redshift-space distortion effect, we focus mainly on the monopole and quadrupole components of both the power and bispectra. We, for the first time, compute the 5- and 6-point spectra to predict the cross-covariance between the power and bispectra, and the auto-covariance of the bispectrum in redshift space. We test the validity of our calculations by comparing them with the covariance matrices measured from the MultiDark-Patchy mock catalogues that are designed to reproduce the galaxy clustering measured from the Baryon Oscillation Spectroscopic Survey Data Release 12. We argue that the simple, leading-order perturbation theory works because the shot-noise corrections for the Patchy mocks are more dominant than other higher-order terms we ignore. In the meantime, we confirm some discrepancies in the comparison, especially of the cross-covariance. We discuss potential sources of such discrepancies. We also show that our PT model reproduces well the cumulative signal-to-noise of the power spectrum and the bispectrum as a function of maximum wavenumber, implying that our PT model captures successfully essential contributions to the covariance matrices. 
}

\end{abstract}


\section{INTRODUCTION}
\label{Sec:Introduction}

It is essential to measure higher order statistics beyond the two-point statistics to extract the full cosmological information, as the galaxy density field at low redshift is strongly non-Gaussian due to non-linear structure formation and galaxy bias. Given the greatly successful analyses of the two-point statistics in galaxy redshift surveys (e.g., see \citealt{Alam:2016hwk} and references therein), the focus has been recently shifting to higher order statistics, i.e., the three-point correlation function or its Fourier space counterpart, the bispectrum. Over the past few years, some applications to use the the three-point statistics to constrain cosmological parameters have been made (as recent works, e.g., ~\citealt{Slepian:2017kfz,Gil-Marin:2017wya,Pearson:2017wtw}). The joint analysis of the two- and three-point statistics will thus continue to be a standard method for analyzing galaxy data in future galaxy redshift surveys: e.g., the Subaru Prime Focus Spectrograph (PFS; ~\citealt{Takada2014PASJ...66R...1T}), the  Dark  Energy  Spectroscopic  Instrument (DESI;~\citealt{Levi2013arXiv1308.0847L}), the \textit{Euclid} mission~\citep{Laureijs:2011gra} and the Wide Field Infrared Survey Telescope (WFIRST;~\citealt{WFIRST}). To correctly interpret the upcoming high quality datasets, accurate modeling not only of the nonlinear power spectrum and the bispecutrm but also of their statistical uncertainties, i.e., covariance matrices, is of crucial importance.

A common approach to estimate the covariance matrix is to utilize hundreds or thousands of synthetic realizations generated by fast approximate schemes to create galaxy catalogues, e.g., the Quick-Particle-Mesh (QPM;~\citealt{White2014MNRAS.437.2594W}) mocks, the MultiDark-Patchy (MD-Patchy;~\citealt{Klypin:2014kpa,Kitaura:2015uqa}) mocks, the Effective Zel'dovich approximation (EZ;~\citealt{Chuang:2014vfa}) mocks, and log-normal mocks~\citep{Agrawal:2017khv}. However, there are two issues for the estimates of the covariance matrix from the mock catalogues. First, such \textit{brute-force} production suffers from the noise due to the finite number of realizations, and uncertainties on the invert covariance matrix estimate propagate directly to increased uncertainties on cosmological parameters~\citep{Hartlap:2006kj,Taylor:2012kz,Dodelson:2013uaa,Percival:2013sga,Taylor:2014ota}. The situation becomes worse in the joint analysis of the power spectrum and the bispectrum, because it requires about 10 or 20 times larger number of data bins than the power spectrum only analysis~\citep{Sugiyama2018}, substantially increasing the required number of independent realizations. To reduce the computational cost of mock generation, various other approaches have been proposed for the power spectrum analysis~\citep{Hamilton:2005dx,Pope:2007vz,Schneider:2011wf,Paz:2015kwa,Pearson:2015gca,Padmanabhan:2015vlf,OConnell:2015src,Howlett:2017vwp,Escoffier:2016qnf,Takahashi:2018pze}. Second, aforementioned fast mock generation schemes are typically designed to only reproduce the observed 2-point function for a target galaxy sample, and hence it is not entirely clear if they can reproduce the 3-point functions at the same time. These difficulties associated with the mock-based approach motivate an analytical approach as a complementary way to estimate the covariance matrix. An analytical approach does not suffer from the issues discussed above, albeit its accuracy should be confirmed by \textit{ideal} mock galaxy simulations. Therefore it is highly desirable to develop an analytic model to predict the covariance matrices.

From a theoretical point of view, the covariance matrix can be expressed with unconnected and connected parts, the so-called Gaussian and non-Gaussian terms, respectively. To evaluate the non-Gaussian terms, one needs to take into account statistics beyond the power spectrum and bispectrum: namely, the trispectrum, the 5-point and the 6-point spectrum in Fourier space, as we will explicitly show later. In the case of the two-point statistics,  the analytical expression in the Gaussian limit was first shown in e.g., \citet{Feldman:1993ky}; subsequently, perturbation theory was often adopted to assess the impact of the non-Gaussian term (the trispectrum) on the matter (halo) power spectrum covariances ~\citep{Meiksin:1998mu,Eisenstein:1999jg,Smith:2008ut,Carron:2014hja,Bertolini:2015fya,Mohammed:2016sre,Barreira:2017sqa,Barreira:2017kxd,Howlett:2017vwp} and the weak lensing power spectrum covariances~\citep{Scoccimarro:1999kp,Reischke:2016ana,Barreira:2017fjz}. In addition, the halo model approach~(for a review, see~\citealt{Cooray:2002dia}) has been used to estimate the covariance matrix for the matter (halo)~\citep{Neyrinck:2006xd,Neyrinck:2006zi,Wu:2013pia,Mohammed:2014lja,Ginzburg:2017mgf,Takada:2013bfn} and weak lensing power spectra~\citep{Cooray:2000ry,Takada:2007fq,Takada:2008fn,Takahashi:2018pze}. Since these analytical models are unable to capture full non-linear gravitational effects on small scales, one also often uses $N$-body simulations for a better understanding of the non-Gaussian effect and for testing the validity of these models~\citep{Rimes:2005dz,Neyrinck:2007bp,Takahashi:2009bq,Takahashi:2009ty,Sato:2009ct,Ngan:2011cc,dePutter:2011ah,Blot:2014pga,Blot:2015cvj,Blot:2018oxk}. 

The bispectrum covariance is far less well understood compared to the power spectrum case. \cite{Kayo:2012nm,Kayo:2013aha} addressed the amount of information included in the lensing power and bispectra using the halo model approach as well as ray-tracing simulations, but took into account the contribution only from the 1-halo term to the 6-point function. \cite{Chan:2016ehg,Chan:2017fiv} measured the 3-dimensional (3D) matter and halo bispectrum covariances from $N$-body simulations in real space and computed a part of non-Gaussian terms of the bispectrum covariance using perturbation theory, ignoring the 6-point function. \Red{\cite{Gualdi:2017iey,Gualdi:2018pyw} have computed a part of non-Gaussian terms of the bispectrum covariance, as well as the cross-covariance between the power and bispectra,
using perturbation theory in redshift-space.} \cite{Colavincenzo:2018cgf} has compared the bispectrum ``monopole'' measurements and their covariances in both real and redshift spaces from a set of different methods for the efficient generation of approximate dark matter catalogues.

Despite all the work discussed above, there is still a limited understanding of the analytical covariance in \textit{redshift space}. The observed 3D galaxy clustering is distorted along the line of sight (LOS) because of the peculiar velocities of galaxies~\citep{Kaiser1987MNRAS.227....1K}, known as redshift space distortion (RSD; see~\citealt{Hamilton:1997zq} for a review). An additional anisotropic signal arises due to the conversion from the observed redshifts into radial distances with incorrect cosmological parameters, which is known as the Alcock-Paczy\'{n}ski (AP) effect~\citep{Alcock:1979mp}. To single out only the anisotropic signal, we commonly decompose the power spectrum (e.g.,~\citealt{Hamilton:1997zq}) and the bispectrum~\citep{Scoccimarro:1999ed,Sugiyama2018,Slepian:2017lpm} into multipole components regarding the angle with respect to the LOS direction. In particular, we have recently proposed a new decomposition formalism for the redshift-space bispectrum, based on the tri-polar spherical harmonic (TriPoSH) formalism~\citep{Sugiyama2018}, and showed the advantages of our novel formalism including the survey window function; throughout this paper, we adopt our decomposed multipole components of the bispectrum to explore their covariances. We note that, while the Gaussian term of the covariance of the power spectrum multipoles has been well studied in~\cite{Grieb:2015bia,Li:2018scc}, the theoretical modeling of non-Gaussian contributions to the covariances of the power spectrum and especially of the bispectrum has not been investigated in detail.

The aim of this paper is therefore to develop a simple model for the galaxy power spectrum and bispectrum covariances, including the full non-Gaussian parts up to the 6-point spectrum, the RSD effect, the linear bias and the shot-noise corrections, which will enable us to compare our calculations with the covariances estimated from realistic mock simulations. For this purpose, we adopt the standard perturbation theory (PT; for a review, see~\citealt{Bernardeau:2001qr}) and compute up to the leading order, i.e., the tree level. This requires the fifth order of perturbation solutions of density fluctuations, because the trispectrum (4-point), the 5-point and the 6-point spectra calculations require the third, fourth and fifth order of the perturbation theory, respectively. We test the validity of our PT calculations by comparing them with the Multidark Patchy mock catalogues~\citep{Klypin:2014kpa,Kitaura:2015uqa} that are designed to reproduce the galaxy data set measured from the Baryon Oscillation Spectroscopic Survey Data Release 12 (BOSS DR12;~\citealt{Alam:2015mbd}). We remark that, in our previous work~\citep{Sugiyama2018}, we showed that the Patchy mock catalogs can reproduce the bispectrum multipoles of the observed BOSS data. We will show good agreement of the PT calculations with the mock results of the covariance matrices for the power spectrum and bispectrum up to the scales of $\sim 0.2\hk$.

To predict the covariances of the power and bispectrum measured from discrete samples, i.e., galaxy samples, the Poisson shot-noise remains an issue, which is highly relevant to the off-diagonal component in the covariances. As pointed out in~\cite{Smith:2008ut} using $N$-body simulations, there is a difference between the covariance matrices estimated from the halo power spectrum measurements \textit{without} and \textit{with} the shot-noise subtraction: namely, the off-diagonal correlation that is present in the shot-noise \textit{uncorrected} covariance are suppressed in the shot-noise \textit{corrected} covariance. This result implies that, since the total number of halos (galaxies) varies between realizations, there are additional sources of covariance that originate from the cross-correlation between the halo (galaxy) power spectrum and the number density as well as the variance of the number density, but such correlations relevant to the shot-noise term are mostly removed by the shot-noise subtraction. \cite{Chan:2016ehg} discussed a similar behavior for the halo bispectrum covariance. \Red{In the case of the two-point correlation function, \citet{OConnell:2015src} has shown the covariance expression accounting for removing the self-counting of galaxies corresponding to the shot-noise term in Fourier space.} In this paper, we work in Fourier space and will show the analytic expressions of the shot-noise \textit{corrected} covariances in both the power spectrum and bispectrum, and address its impact on the non-Gaussian contribution of the covariances.

The plan of this paper is as follows. Section~\ref{Sec:CovarianceModel} presents the analytical expressions of the covariance matrices of the power spectrum and the bispectrum. Section~\ref{Sec:Multipoles} reviews the decomposition formalism of the redshift-space power spectrum and bispectrum. We test the validation of our analytical calculations by comparing them with the Patchy mock results in Section~\ref{Sec:ComparisonWithMocks}. To estimate the impact of the non-Gaussian errors, we compute the cumulative signal-to-noise ratios in \ref{Sec:SN}. The conclusions and discussion are summarized in Section~\ref{Sec:Conclusions}. Throughout this paper, we adopt a flat $\Lambda$CDM cosmology that is the same as used in the Patchy mocks: $(\Omega_{\Lambda},\Omega_{\rm m}, \Omega_{\rm b},\sigma_8,h) = (0.693, 0.307, 0.0480, 0.829, 0.678)$. 

\section{Covariance model}
\label{Sec:CovarianceModel}

The goal of this paper is to provide an analytic model of the covariance matrices for both the power spectrum and the bispectrum on the basis of PT, and to compare them with the covariance matrices measured from the Patchy mock catalogs to test the validity of the perturbation theory calculations. In this section, we present {\it exact} formulae to describe the covariance matrices. More concretely, we show the representations of the power, bi, tri, 5-point and 6-point spectra in a discretized picture in Section~\ref{Sec:DescretizedDensityFields}, and derive the analytic expressions of the auto-covariances of both the power spectrum and the bispectrum, as well as the cross-covariance between the power spectrum and the bispectrum, with the shot-noise term subtracted from the power- and bi-spectrum measurements, in Sections~\ref{Sec:PP}, \ref{Sec:PB} and \ref{Sec:BB}. \Red{Throughout this paper, we ignore non-linear corrections on the covariance matrix such as higher-order loop corrections, including the Finger-of-God effect, non-linear bias effects, and the super-sample covariance (SSC) effect. We summarize the limitation of our analytical calculations in Section~2.6.}

For notational simplicity, we omit to denote the redshift- and LOS-dependence on all statistics that we compute: e.g., the redshift-space power spectrum $P(\VEC{k},\hat{n},z)$, with $\VEC{k}$, $\hat{n}$ and $z$ being respectively wavevector, the unit vector orienting to the LOS direction and redshift, is represented just as $P(\VEC{k})$. In Appendix~\ref{Ap:PT}, we present our PT approach to evaluate the expressions shown here.

\subsection{$N$-point Statistics in a discretized picture}
\label{Sec:DescretizedDensityFields}

Following~\citet{Peebles1980lssu.book}, the Fourier transform of the galaxy number density, $n(\VEC{x})$, is discretized as follows:
\begin{eqnarray}
	n(\VEC{k}) = \int d^3x\, e^{-i\VEC{k}\cdot\VEC{x}}\, n(\VEC{x}) 
	           \to \sum_i\, n_i\, e^{-i\VEC{k}\cdot\VEC{x}_i},
\end{eqnarray}
where we divide the space into the infinitesimal grid cells of volume $\delta V$. Each cell is allowed to contain at most one galaxy, which can be described by the occupation number $n_i = n(\VEC{x}_i)\delta V$ such that $n_i=1$ if the $i$-th cell has a galaxy, otherwise $n_i=0$, and therefore, it satisfies $n_i = n_i^2 = \dots = n_i^n$. The total number of galaxies, $N$, is then given by $N = \sum_i \langle n_i \rangle_{\rm c}$, where $\langle \cdots \rangle_{\rm c}$ denotes a cumulant of the ensemble average. The background (unperturbed) number density, which is defined as $\bar{n} = N/V$ with $V$ being the survey volume, is measured by the $\VEC{k}=0$ mode of the ensemble average of the galaxy number density: namely, $\langle n(\VEC{k})\rangle_{\rm c} = (2\pi)^3\delta_{\rm D}\left( \VEC{k} \right) \bar{n}$, where $\delta_{\rm D}$ represents the Dirac delta function. This fact indicates that $n(\VEC{k})$ at $\VEC{k}\neq \VEC{0}$ is a perturbed quantity.

The power spectrum is then represented in the discretized description as
\begin{eqnarray}
	\frac{V}{N^2} \left\langle n(\VEC{k})n(\VEC{k}') \right\rangle_{\rm c}
	&=& \frac{(2\pi)^3\delta_{\rm D}\left( \VEC{k}+\VEC{k}' \right)}{V}\,
	\left[  \frac{V}{N^2}\sum_{i\neq j}\left\langle n_in_j \right\rangle_{\rm c} e^{-i\VEC{k}\cdot(\VEC{x}_{i}-\VEC{x}_j)}
	+ \frac{V}{N^2}\sum_i \left\langle n_i^2\right\rangle_{\rm c}   \right]\nonumber\\
	&=&  \frac{(2\pi)^3\delta_{\rm D}\left( \VEC{k}+\VEC{k}' \right)}{V}\, \left[ P(\VEC{k}) + \frac{1}{\bar{n}} \right],
	\label{Eq:nnP}
\end{eqnarray}
where $\VEC{k}\neq \VEC{0}$, $\VEC{k}'\neq \VEC{0}$, and in the second line of the above expression, we used $\sum_i \langle n_i^2 \rangle_{\rm c} = \sum_i \langle n_i \rangle_{\rm c}=N$. From Eq.~(\ref{Eq:nnP}), the power spectrum estimator is derived as
\begin{eqnarray}
	\widehat{P}(\VEC{k},\VEC{k}') 
	&=& \frac{V}{N^2} \sum_{i,j} n_i n_j e^{-i\VEC{k}\cdot\VEC{x}_{i}}e^{-i\VEC{k}'\cdot\VEC{x}_j}
	- \frac{V}{N^2} \sum_{i=j} n_i^2 e^{-i(\VEC{k}+\VEC{k}')\cdot\VEC{x}_{i}} \nonumber \\
	&=& \frac{V}{N^2}\sum_{i\neq j}n_in_j  e^{-i\VEC{k}\cdot\VEC{x}_{i}}e^{-i\VEC{k}'\cdot\VEC{x}_j},
	\label{Eq:P_estimator}
\end{eqnarray}
which satisfies
\begin{eqnarray}
	\langle \widehat{P}(\VEC{k},\VEC{k}') \rangle_{\rm c} =  \frac{(2\pi)^3\delta_{\rm D}\left( \VEC{k}+\VEC{k}' \right)}{V} P(\VEC{k}).
\end{eqnarray}
We stress that the condition in this estimator, $i\neq j$, ensures that the Poisson shot noise, i.e., the second term in the first line of Eq.~(\ref{Eq:P_estimator}), is subtracted out.

Similarly, the estimators of higher order statistics, the bispectrum, the trispectrum, the 5-point spectrum and the 6-point spectrum, are given by
\begin{eqnarray}
	\widehat{B}(\VEC{k}_1,\VEC{k}_2,\VEC{k}_3)&=& 
	(V^2/N^3)\sum_{i\neq j \neq k} 
	 n_in_j n_k\,  e^{-i\VEC{k}_1\cdot\VEC{x}_{i}}e^{-i\VEC{k}_2\cdot\VEC{x}_{j}}e^{-i\VEC{k}_3\cdot\VEC{x}_{k}}	\nonumber\\
	\widehat{T}(\VEC{k}_1,\VEC{k}_2,\VEC{k}_3,\VEC{k}_4)&=&  
	(V^3/N^4)\sum_{i\neq j \neq k \neq l} 
	 n_i n_j n_k n_l\,  e^{-i\VEC{k}_1\cdot\VEC{x}_{i}}e^{-i\VEC{k}_2\cdot\VEC{x}_{j}}e^{-i\VEC{k}_3\cdot\VEC{x}_{k}} e^{-i\VEC{k}_4\cdot\VEC{x}_{l}} 
	\nonumber\\
	\widehat{P}_5(\VEC{k}_1,\VEC{k}_2,\VEC{k}_3,\VEC{k}_4,\VEC{k}_5) &=&
	(V^4/N^5) \hspace{-0.2cm} \sum_{i \neq j \neq k \neq l \neq m} \hspace{-0.2cm} 
	 n_i n_j n_k n_l n_m \, 
	e^{-i\VEC{k}_1\cdot\VEC{x}_{i}}e^{-i\VEC{k}_2\cdot\VEC{x}_{j}}e^{-i\VEC{k}_3\cdot\VEC{x}_{k}} e^{-i\VEC{k}_4\cdot\VEC{x}_{l}} e^{-i\VEC{k}_5\cdot\VEC{x}_{m}} 
	\nonumber\\
	\widehat{P}_6(\VEC{k}_1,\VEC{k}_2,\VEC{k}_3,\VEC{k}_4,\VEC{k}_5,\VEC{k}_6)&=& 
	(V^5/N^6) \hspace{-0.3cm}  \sum_{i \neq j \neq k \neq l \neq m \neq n} \hspace{-0.3cm}       
	 n_i n_j n_k n_l n_m n_n\, 
	e^{-i\VEC{k}_1\cdot\VEC{x}_{i}}e^{-i\VEC{k}_2\cdot\VEC{x}_{j}}e^{-i\VEC{k}_3\cdot\VEC{x}_{k}} e^{-i\VEC{k}_4\cdot\VEC{x}_{l}} e^{-i\VEC{k}_5\cdot\VEC{x}_{m}} e^{-i\VEC{k}_6\cdot\VEC{x}_{n}}.
	\label{Eq:estimators}
\end{eqnarray}
Note that all these estimators are contributed only by density correlations among different positions. Namely, the self-counting of galaxies is all removed: e.g., $i\neq j \neq k$ includes the condition $i \neq k$, and $i\neq j \neq k\neq l$ means $i \neq k$, $i \neq l$ and $j \neq l$. Similar results hold also for $i\neq j \neq k \neq l \neq m$ and $i \neq j \neq k \neq l \neq m \neq n$. Therefore, these estimators do not include the shot noise effect. They satisfy
\begin{eqnarray}
	\langle \widehat{B}(\VEC{k}_1,\VEC{k}_2,\VEC{k}_3) \rangle_{\rm c}
	&=& \frac{(2\pi)^3\delta_{\rm D}\left( \VEC{k}_{123} \right)}{V} B(\VEC{k}_1,\VEC{k}_2,\VEC{k}_3)\nonumber\\
	\langle \widehat{T}(\VEC{k}_1,\VEC{k}_2,\VEC{k}_3,\VEC{k}_4) \rangle_{\rm c}
	&=&  \frac{(2\pi)^3\delta_{\rm D}\left( \VEC{k}_{1234} \right)}{V} T(\VEC{k}_1,\VEC{k}_2,\VEC{k}_3,\VEC{k}_4)\nonumber\\
	\langle \widehat{P}_5(\VEC{k}_1,\VEC{k}_2,\VEC{k}_3,\VEC{k}_4,\VEC{k}_5) \rangle_{\rm c}
	&=&  \frac{(2\pi)^3\delta_{\rm D}\left( \VEC{k}_{12345} \right)}{V} P_5(\VEC{k}_1,\VEC{k}_2,\VEC{k}_3,\VEC{k}_4,\VEC{k}_5)\nonumber\\
	\langle \widehat{P}_6(\VEC{k}_1,\VEC{k}_2,\VEC{k}_3,\VEC{k}_4,\VEC{k}_5,\VEC{k}_6) \rangle_{\rm c}
	&=&  \frac{(2\pi)^3\delta_{\rm D}\left( \VEC{k}_{123456} \right)}{V} P_6(\VEC{k}_1,\VEC{k}_2,\VEC{k}_3,\VEC{k}_4,\VEC{k}_5,\VEC{k}_6),
\end{eqnarray}
where $\VEC{k}_1\neq \VEC{0}$, $\VEC{k}_2\neq \VEC{0}$, $\VEC{k}_3\neq \VEC{0}$, $\VEC{k}_4\neq \VEC{0}$, $\VEC{k}_5\neq \VEC{0}$, $\VEC{k}_6\neq \VEC{0}$, and $\VEC{k}_{1\dots n} = \VEC{k}_1+\dots+\VEC{k}_n$. We stress again that these higher order statistics defined above are the quantities with the self-counting of galaxies, i.e. the shot-noise, removed. To correctly estimate the signal of the measured statistics, the shot-noise subtraction is crucial, because the amplitude of the bispectrum monopole becomes comparable to or even smaller than that of the corresponding shot-noise term at scales of $k\gtrsim 0.1\hk$ for the BOSS sample~\citep{Sugiyama2018}. We expect similar results for higher order statistics. As we will show in Section \ref{subsec:shot-noise}, the shot-noise subtraction is also crucial for suppressing the off-diagonal component of the covariance.

\subsection{Power spectrum covariance}
\label{Sec:PP}

The power spectrum covariance can be decomposed into two contributions:
\begin{eqnarray}
	{\rm Cov}\big[  \widehat{P}(\VEC{k}),\widehat{P}(\VEC{k}')  \big]
	= 
	{\rm Cov}\big[  \widehat{P}(\VEC{k}),\widehat{P}(\VEC{k}')  \big]_{PP}
	+
	{\rm Cov}\big[  \widehat{P}(\VEC{k}),\widehat{P}(\VEC{k}')  \big]_{T},
\end{eqnarray}
where the first and second terms on the right hand side (RHS) denote the Gaussian and non-Gaussian parts, respectively.
The subscripts ``$PP$'' and ``$T$'' indicate that the Gaussian part consists of the product of two power spectra ($P$), and that the non-Gaussian part arises from the trispectrum ($T$), respectively. 

To correctly derive the auto-covariance of the galaxy power spectrum with shot-noise corrections, 
we use the discretized representation of the power spectrum estimator (\ref{Eq:P_estimator});
then, the power spectrum covariance can be represented as
\begin{eqnarray}
	{\rm Cov}\big[  \widehat{P}(\VEC{k}),\widehat{P}(\VEC{k}')  \big]
	 &=&  \left( \frac{V}{N^2} \right)^2 \sum_{i\neq j}\sum_{k\neq l} e^{-i\VEC{k}\cdot(\VEC{x}_{i}-\VEC{x}_j)}e^{-i\VEC{k}'\cdot(\VEC{x}_{k}-\VEC{x}_l)} 
	\left[ \langle n_in_k\rangle_{\rm c} \langle n_jn_l\rangle_{\rm c} +  \langle n_in_l\rangle_{\rm c} \langle n_jn_k\rangle_{\rm c} + \langle n_in_jn_kn_l\rangle_{\rm c} \right].
	\label{Eq:CovPP}
\end{eqnarray}
The first two terms on the RHS correspond to the Gaussian part, given by
\begin{eqnarray}
	{\rm Cov}\big[ \widehat{P}(\VEC{k}),\widehat{P}(\VEC{k}') \big]_{PP}
	&=&  \left( \frac{V}{N^2} \right)^2 \bigg[  \sum_{i,k}\langle n_in_k \rangle_{\rm c} e^{-i\VEC{k}\cdot\VEC{x}_{i}}e^{-i\VEC{k}'\cdot\VEC{x}_{k}}  \bigg] 
	\bigg[ \sum_{j,l}\langle n_jn_l \rangle_{\rm c} e^{i\VEC{k}\cdot\VEC{x}_{j}} e^{i\VEC{k}'\cdot\VEC{x}_{l}}  \bigg] + \mbox{(1 perm.)}.
\end{eqnarray}
Using Eq.~(\ref{Eq:nnP}), the Gaussian term becomes the well-known form
\begin{eqnarray}
	{\rm Cov}\big[ \widehat{P}(\VEC{k}),\widehat{P}(\VEC{k}') \big]_{PP}
	=  \frac{(2\pi)^3\delta_{\rm D}(\VEC{k}+\VEC{k}')+(2\pi)^3\delta_{\rm D}(\VEC{k}-\VEC{k}')}{V} \left[  P^{(\rm N)}(\VEC{k})\right]^2,
	\label{Eq:cov_PP_Gaussian}
\end{eqnarray}
where we define the power spectrum with the shot-noise term as
\begin{eqnarray}
	P^{(\rm N)}(\VEC{k}) \equiv P(\VEC{k}) + \frac{1}{\bar{n}}.
	\label{Eq:P_N}
\end{eqnarray}

As well known, here the shot noise effect on the power spectrum covariance remains even after shot noise correction due to the contribution from e.g., $i = k$ or $j=l$ terms. 
The third term on the RHS of Eq.~(\ref{Eq:CovPP}), i.e., the non-Gaussian part, is given by
\begin{eqnarray}
	{\rm Cov}\big[ \widehat{P}(\VEC{k}),\widehat{P}(\VEC{k}') \big]_{T} 
	&=& (V/N^2)^2\sum_{(i\neq j), (k\neq l)}
	\langle n_in_jn_k n_l\ \rangle_{\rm c}\, e^{-i\VEC{k}\cdot\VEC{x}_{i}}e^{i\VEC{k}\cdot\VEC{x}_{j}}  e^{-i\VEC{k}'\cdot\VEC{x}_{k}} e^{i\VEC{k}'\cdot\VEC{x}_{l}} \nonumber \\
	&=&(V/N^2)^2
	\sum_{i\neq j \neq k \neq l}
	\langle n_in_jn_k n_l\ \rangle_{\rm c}\,  e^{-i\VEC{k}\cdot\VEC{x}_{i}} e^{i\VEC{k}\cdot\VEC{x}_{j}} e^{-i\VEC{k}'\cdot\VEC{x}_{k}} e^{i\VEC{k}'\cdot\VEC{x}_{l}} \nonumber \\
	&+& (V/N^2)^2\Big[
	\sum_{(i\neq j \neq l), (i=k)}
	\langle n_i^2 n_j n_l\ \rangle_{\rm c}\, e^{-i(\VEC{k}+\VEC{k}')\cdot\VEC{x}_{i}} e^{i\VEC{k}\cdot\VEC{x}_{j}} e^{i\VEC{k}'\cdot\VEC{x}_{l}} 
	+ \mbox{(3 perms.)} \Big]\nonumber \\
	&+& (V/N^2)^2\Big[
	\hspace{-0.3cm}\sum_{\substack{(i\neq j),(i=k),(j=l)}}\hspace{-0.3cm} \langle n_i^2 n_j^2\rangle_{\rm c}\, e^{-i(\VEC{k}+\VEC{k}')\cdot\VEC{x}_{i}} e^{i(\VEC{k}+\VEC{k}')\cdot\VEC{x}_{j}}
	+ \mbox{(1 perm.)} \Big].
	\label{Eq:TT}
\end{eqnarray}
From~Eq.~(\ref{Eq:estimators}), we finally derive
\begin{eqnarray}
	{\rm Cov}\big[ \widehat{P}(\VEC{k}),\widehat{P}(\VEC{k}') \big]_{T}
	= \frac{1}{V}\, T^{(\rm N)}(\VEC{k},-\VEC{k},\VEC{k}',-\VEC{k}'),
	\label{Eq:CovPP_T}
\end{eqnarray}
where the trispectrum term including shot-noise is defined as
\begin{eqnarray}
	 T^{(\rm N)}(\VEC{k}_1,\VEC{k}_2,\VEC{k}'_1,\VEC{k}'_2) &\equiv& T(\VEC{k}_1,\VEC{k}_2,\VEC{k}'_1,\VEC{k}'_2) \nonumber \\
	&+& \frac{1}{\bar{n}} \left[  B(-\VEC{k}_1-\VEC{k}'_1, \VEC{k}_1,\VEC{k}'_1)
               	                    + B(-\VEC{k}_1-\VEC{k}'_2, \VEC{k}_1,\VEC{k}'_2)
			            + B(-\VEC{k}_2-\VEC{k}'_1, \VEC{k}_2,\VEC{k}'_1)
               	                    + B(-\VEC{k}_2-\VEC{k}'_2, \VEC{k}_2,\VEC{k}'_2)
			    \right] \nonumber \\
	&+& \frac{1}{\bar{n}^2}\left[ P(\VEC{k}_1+\VEC{k}'_1) + P(\VEC{k}_1+\VEC{k}'_2) \right].
	\label{Eq:T_N}
\end{eqnarray}
\Red{The expression of the power spectrum covariance matrix shown in this subsection, Eqs.~(\ref{Eq:cov_PP_Gaussian}) and (\ref{Eq:T_N}), corresponds to Eqs.~(2.19), (2.20) and (2.21) in~\cite{OConnell:2015src}, which are derived for the two-point correlation function in configuration space. We shall extend the calculations of the power spectrum covariance to the covariances associated with the bispectrum in Sections~\ref{Sec:PB} and \ref{Sec:BB}.
}

\subsection{The shot-noise correction and comparison with previous works}
\label{subsec:shot-noise}
Our result of the non-Gaussian term of the power spectrum covariance, Eq.~(\ref{Eq:CovPP_T}), is different from the expression derived in some of previous works~(e.g., \citealt{Meiksin:1998mu,Smith:2008ut,Chan:2016ehg}). In this subsection, we show that the difference originates from the Poisson shot-noise subtraction in the estimator.

Since the non-Gaussian part of the power spectrum covariance results from the four-point function, let us first compute the trispectrum including the shot-noise terms \citep[see e.g.,][]{Matarrese:1997sk}:
\begin{eqnarray}
	\frac{V^3}{N^4}
	\left \langle n(\VEC{k}_1) n(\VEC{k}_2) n(\VEC{k}_3) n(\VEC{k}_4) \right\rangle_{\rm c}
	&=& 
	\frac{(2\pi)^3\delta_{\rm D}\left( \VEC{k}_{1234} \right) }{V}
	\Big\{
	T(\VEC{k}_1,\VEC{k}_2,\VEC{k}_3,\VEC{k}_4) \nonumber \\
	&+& (1/\bar{n})[ 
		B(-\VEC{k}_1-\VEC{k}_2,\VEC{k}_1,\VEC{k}_2) + B(-\VEC{k}_1-\VEC{k}_3,\VEC{k}_1,\VEC{k}_3) + B(-\VEC{k}_1-\VEC{k}_4,\VEC{k}_1,\VEC{k}_4)\nonumber \\
	&&
		\hspace{0.49cm}	+ \  B(-\VEC{k}_2-\VEC{k}_3,\VEC{k}_2,\VEC{k}_3) + B(-\VEC{k}_2-\VEC{k}_4,\VEC{k}_2,\VEC{k}_4) + B(-\VEC{k}_3-\VEC{k}_4,\VEC{k}_3,\VEC{k}_4) ]\nonumber \\
	 &+& (1/\bar{n}^2)[
		 P(\VEC{k}_1) + P(\VEC{k}_2) +  P(\VEC{k}_3) + P(\VEC{k}_4) + P(\VEC{k}_1+\VEC{k}_2) + P(\VEC{k}_1+\VEC{k}_3) +  P(\VEC{k}_1+\VEC{k}_4)  ] \nonumber \\
	 &+& (1/\bar{n}^3)  \Big\}.
	\label{Eq:T_shotnoise}
\end{eqnarray}
Hence, we find the non-Gaussian term in the covariance of the power spectrum with the shot-noise contribution as \citep[see e.g., Eqs~(7) and (8) in][]{Meiksin:1998mu} 
\begin{eqnarray}
	\left( \frac{V}{N^2} \right)^2	\left\langle |n(\VEC{k})|^2 |n(\VEC{k}')|^2\right\rangle_{\rm c}
	&=& 
	\frac{1}{V}
	\Big\{ T(\VEC{k},-\VEC{k},\VEC{k}',-\VEC{k}') \nonumber \\
	&+& (1/\bar{n})[ B(\VEC{0},\VEC{k},-\VEC{k})+ B(\VEC{0},\VEC{k}',-\VEC{k}')  \nonumber\\
	&& \hspace{0.49cm}	+ \ 
	B(-\VEC{k}-\VEC{k}',\VEC{k},\VEC{k}') + B(-\VEC{k}+\VEC{k}',\VEC{k},-\VEC{k}) + B(\VEC{k}-\VEC{k}',-\VEC{k},\VEC{k}') 
	 + B(\VEC{k}+\VEC{k}',-\VEC{k},-\VEC{k}') ]\nonumber \\
	&+& (1/\bar{n}^2)[ 2 P(\VEC{k}) +  2P(\VEC{k}') + P(\VEC{0}) + P(\VEC{k}+\VEC{k}') +  P(\VEC{k}-\VEC{k}') ] \nonumber \\
	&+& (1/\bar{n}^3) \Big\}.
	\label{Eq:CovPP_T_sub}
\end{eqnarray}
Meanwhile, the covariance of the power spectrum with the shot-noise subtracted can be expressed as \citep[see e.g., Eqs~(83) in][]{Smith:2008ut}
\begin{eqnarray}
	{\rm Cov}\big[ \widehat{P}(\VEC{k}),\widehat{P}(\VEC{k}') \big]_{T}
	&=& \frac{V^2}{N^4} \left\langle |n(\VEC{k})|^2 |n(\VEC{k}')|^2 \right\rangle_{\rm c}
	 -  \frac{V^2}{N^4} \left\langle |n(\VEC{k})|^2 \bigg(  \sum_i n_i \bigg) \right\rangle_{\rm c} \nonumber\\
	&-&  \frac{V^2}{N^4} \left\langle \bigg(\sum_i n_i \bigg)|n(\VEC{k}')|^2  \right\rangle_{\rm c}
	 +  \frac{V^2}{N^4} \left\langle \bigg(  \sum_i n_i \bigg)\bigg(  \sum_j n_j \bigg)  \right\rangle_{\rm c}.
	 \label{Eq:CovPP_T_2}
\end{eqnarray}
Now, we derive the analytical expressions of the second term (likewise the third term) as 
\begin{eqnarray}
	\frac{V^2}{N^4}\left\langle \bigg(\sum_i n_i\bigg)  |n(\VEC{k})|^2\right\rangle_{\rm c}
	&=& 
	\frac{V^2}{N^4}\sum_{i, j, k} \langle n_i n_j n_k\rangle_{\rm c}e^{-i\VEC{k}\cdot(\VEC{x}_j-\VEC{x}_k)} \nonumber\\
	&=& 
	\frac{V^2}{N^4}
	\left( 
	\sum_{(i\neq j), (j\neq k), (i\neq k)} 
	+ 
	\sum_{(i= j), (i\neq k)} 
	+ 
	\sum_{ (i= k),(i\neq j)} 
	+ 
	\sum_{(j =k),(i\neq j)} 
	+ 
	\sum_{i=j=k} 
	\right)
	\langle n_i n_j n_k\rangle_{\rm c}e^{-i\VEC{k}\cdot(\VEC{x}_j-\VEC{x}_k)} \nonumber\\
	&=& \frac{1}{V} \left( \frac{1}{\bar{n}} \right)
	\left[   B(\VEC{0}, \VEC{k}, -\VEC{k}) + \frac{2}{\bar{n}}P(\VEC{k})+ \frac{1}{\bar{n}}P(\VEC{0}) + \frac{1}{\bar{n}^2} \right]
	\label{Eq:CovPP_T_shotnoise_1}
\end{eqnarray}
and the last term becomes 
\begin{eqnarray}
	\frac{V^2}{N^4}\left\langle \bigg(\sum_i n_i\bigg) \bigg( \sum_j n_j\bigg)\right\rangle_{\rm c}
	&=& 
	\frac{V^2}{N^4}\left(  \sum_{i \neq j} +  \sum_{i=j} \right)\langle n_i n_j \rangle_{\rm c} \nonumber\\
	&=& \frac{1}{V}\left( \frac{1}{\bar{n}^2} \right) \left[ P(\VEC{0})  + \frac{1}{\bar{n}} \right].
	\label{Eq:CovPP_T_shotnoise_2}
\end{eqnarray}
Substituting Eqs.~(\ref{Eq:CovPP_T_sub}), (\ref{Eq:CovPP_T_shotnoise_1}) and (\ref{Eq:CovPP_T_shotnoise_2}) into Eq.~(\ref{Eq:CovPP_T_2}), we recover the formula, Eq.~(\ref{Eq:CovPP_T}). In other words, the shot-noise subtraction in the power spectrum measurements cancels with some of the non-Gaussian terms. This suggests that the non-Gaussian error of the power spectrum estimation is suppressed if the shot noise is subtracted out, which is contrary to the Gaussian error of the power spectrum where shot noise subtraction does not alter the error. Since we are not interested in the shot-noise component as a desired signal, we argue that the shot noise should be subtracted to ensure the higher signal to noise. 

It is worth noting that the correlations relevant to the shot-noise terms in Eq.~(\ref{Eq:CovPP_T_sub}) are caused by the same reason as the so-called local mean effect ~\citep{dePutter:2011ah}, which comes from long-wavelength modes of density fluctuations beyond survey area. These effects arise from the fact that one does not know the true mean number density of galaxies, but instead has to rely on an estimate of the mean density within a finite survey volume. Consequently, the observed mean density is modulated by the $\VEC{k}=\VEC{0}$ mode, which is referred as the \textit{beat} or \textit{zero} mode. In fact, Eq.~(\ref{Eq:CovPP_T_sub}) includes some terms estimated at $\VEC{k}=\VEC{0}$. To correctly estimate these beat mode terms, we have to account for the survey window effect; for instance, $B(\VEC{0},\VEC{k},-\VEC{k})$ in Eq.~(\ref{Eq:CovPP_T_sub}) can be replaced by
\begin{eqnarray}
	B(\VEC{0},\VEC{k},-\VEC{k}) \to \frac{1}{V}\int \frac{d^3\varepsilon}{(2\pi)^3} |W(\VEC{\varepsilon})|^2 B(\VEC{\varepsilon},\VEC{k},-\VEC{k}-\VEC{\varepsilon}),
\end{eqnarray}
where $V$ denotes the survey volume, $W(\VEC{\varepsilon})$ represents the Fourier transform of a given survey selection function, and $\VEC{\varepsilon}$ corresponds to the beat mode. If we assume that the survey window is effectively a delta function in Fourier space, $W(\VEC{\varepsilon})=(2\pi)^3\delta_{\rm D}(\VEC{\varepsilon})$, the RHS recovers the left hand side (LHS) that has no window effect. In this paper, we compare our models with the mock data for which we know the true mean density, and therefore, we ignore the impact of the wrong shot-noise subtraction with the wrong assumption of mean number density. We nevertheless stress here that the contributions from the beat mode and some of the other terms that appear in Eq.~(\ref{Eq:CovPP_T_sub}), e.g., $(1/\bar{n}) B(\VEC{0},\VEC{k},-\VEC{k})$ and a constant $1/\bar{n}^3$, are not necessary to be computed as long as we measure the shot-noise subtracted power spectrum and estimate the corresponding covariance, because these terms cancel with the shot-noise subtraction effect (\ref{Eq:CovPP_T_2}), yielding our result, Eq.~(\ref{Eq:CovPP_T}). In the following sections we will apply the same techniques discussed here for the power spectrum covariance to the bispectrum covariance: namely, we estimate the covariance of the bispectrum after the shot-noise subtraction and with no beat mode term appearing in the shot-noise terms.

\subsection{Cross-covariance between the power spectrum and the bispectrum}
\label{Sec:PB}

The cross-covariance between the power spectrum and the bispectrum has two sources:
\begin{eqnarray}
	{\rm Cov}\big[ \widehat{P}(\VEC{k}),\widehat{B}(\VEC{k}_1,\VEC{k}_2,\VEC{k}_3) \big]
	&=& 
	{\rm Cov}\big[ \widehat{P}(\VEC{k}),\widehat{B}(\VEC{k}_1,\VEC{k}_2,\VEC{k}_3) \big]_{PB}
	+
	{\rm Cov}\big[ \widehat{P}(\VEC{k}),\widehat{B}(\VEC{k}_1,\VEC{k}_2,\VEC{k}_3) \big]_{P_5},
\end{eqnarray}
where $\VEC{k}_1+\VEC{k}_2+\VEC{k}_3=\VEC{0}$, and the first term on the RHS with the subscript ``$PB$'' consists of a product of the power spectrum and the bispectrum, and the second term with ``$P_5$'' arises from the 5-point power spectrum. Unlike the power spectrum covariance, these two terms both come from non-Gaussian effects.

The ``$PB$'' term is given by
\begin{eqnarray}
	{\rm Cov}\big[ \widehat{P}(\VEC{k}),\widehat{B}(\VEC{k}_1,\VEC{k}_2,\VEC{k}_3) \big]_{PB}
	\hspace{-0.25cm}&=& \hspace{-0.25cm} \left( \frac{V^3}{N^5} \right) \sum_{i\neq j}\sum_{k\neq l \neq m}
	e^{-i\VEC{k}\cdot( \VEC{x}_i-\VEC{x}_j )}e^{-i\VEC{k}_1\cdot\VEC{x}_k}e^{-i\VEC{k}_2\cdot\VEC{x}_l}e^{-i\VEC{k}_3\cdot\VEC{x}_m}
	\left[    \langle n_i n_k \rangle_{\rm c}\langle n_jn_l n_m\rangle_{\rm c} + \mbox{(5 perms.)}   \right] \nonumber \\
	\hspace{-0.25cm}&=&\hspace{-0.25cm} \left( \frac{V^3}{N^5} \right) \left[  \sum_{i, k}\, \langle n_i n_k \rangle_{\rm c}\,
		e^{-i\VEC{k}\cdot\VEC{x}_i}e^{-i\VEC{k}_1\cdot\VEC{x}_k}  \right]
	\left[ \sum_{j, l\neq m} \, \langle n_jn_l n_m\rangle_{\rm c}\, e^{i\VEC{k}\VEC{x}_j}e^{-i\VEC{k}_2\cdot\VEC{x}_l}e^{-i\VEC{k}_3\cdot\VEC{x}_m}  \right]+ \mbox{(5 perms.)},  \nonumber\\
\end{eqnarray}
and from Eq.~(\ref{Eq:estimators}), we obtain
\begin{eqnarray}
	{\rm Cov}\big[ \widehat{P}(\VEC{k}),\widehat{B}(\VEC{k}_1,\VEC{k}_2,\VEC{k}_3) \big]_{PB}
	\hspace{-0.25cm}&=&\hspace{-0.25cm} \frac{(2\pi)^3\delta_{\rm D}\left( \VEC{k}+\VEC{k}_1 \right)}{V}\, 2\, P^{(\rm N)}(\VEC{k}_1) B^{(\rm N)}(\VEC{k}_1,\VEC{k}_2,\VEC{k}_3) \nonumber\\
	\hspace{-0.25cm}&+&\hspace{-0.25cm} \frac{(2\pi)^3\delta_{\rm D}\left( \VEC{k}+\VEC{k}_2 \right)}{V}\, 2\, P^{(\rm N)}(\VEC{k}_2) B^{(\rm N)}(\VEC{k}_2,\VEC{k}_1,\VEC{k}_3) \nonumber\\
	\hspace{-0.25cm}&+&\hspace{-0.25cm} \frac{(2\pi)^3\delta_{\rm D}\left( \VEC{k}+\VEC{k}_3 \right)}{V}\, 2\, P^{(\rm N)}(\VEC{k}_3) B^{(\rm N)}(\VEC{k}_3,\VEC{k}_1,\VEC{k}_2),
	\label{Eq:covPB}
\end{eqnarray}
where the bispectrum with shot-noise terms is given by
\begin{eqnarray}
	B^{(\rm N)}(\VEC{k}_1,\VEC{k}_2,\VEC{k}_3) = B(\VEC{k}_1,\VEC{k}_2,\VEC{k}_3) + \frac{1}{\bar{n}}\left[P(\VEC{k}_2) + P(\VEC{k}_3) \right].
	\label{Eq:B_N}
\end{eqnarray}
Note that the above shot-noise terms in the bispecrtrum are not the same as the normal bispectrum shot-noise terms, i.e., $(1/\bar{n})[P(\VEC{k}_1)+P(\VEC{k}_2)+P(\VEC{k}_3) ] + (1/\bar{n}^2)$~\citep{Matarrese:1997sk}. This discrepancy comes from the same reason as the case of the power spectrum covariance, as discussed in Section~\ref{subsec:shot-noise}. Namely, since we measure the power and bispectra with their shot-noise terms subtracted, some of the shot-noise corrections in the corresponding covariance are offset by correlation terms associated with the mean number density just like Eqs.~(\ref{Eq:CovPP_T_shotnoise_1}) and (\ref{Eq:CovPP_T_shotnoise_2}) in the power spectrum case. Furthermore, it is worth noting that the shot-noise terms in ${\rm Cov}[P,B]_{PB}$, e.g., $ [P(\VEC{k}_1)+1/\bar{n}]\times(1/\bar{n}) [P(\VEC{k}_2) + P(\VEC{k}_3)]$, never vanish even in the Gaussian limit. Thus, taking account of the shot-noise terms in non-Gaussian covariances is important for the joint analysis of the power spectrum and the bispectrum.

Following the calculation of the trispectrum contribution to the power spectrum covariance (\ref{Eq:TT}), the ``$P_5$'' term can be calculated as
\begin{eqnarray}
	{\rm Cov}\big[ \widehat{P}(\VEC{k}),\widehat{B}(\VEC{k}_1,\VEC{k}_2,\VEC{k}_3) \big]_{P_5} 
	&=& \left( \frac{V^3}{N^5} \right) \sum_{i\neq j}\sum_{k \neq l \neq m}  \langle n_i n_j n_kn_l n_m\rangle_{\rm c}
	e^{-i\VEC{k}\cdot( \VEC{x}_i-\VEC{x}_j )}e^{-i\VEC{k}_1\cdot\VEC{x}_k}e^{-i\VEC{k}_2\cdot\VEC{x}_l}e^{-i\VEC{k}_3\cdot\VEC{x}_m} 
	\nonumber\\
	&=& 
	(V^3/N^5)\sum_{i \neq j \neq k \neq l \neq m} 
	\langle  n_i n_j n_k n_l n_m \rangle_{\rm c}\,
	e^{-i\VEC{k}\cdot\VEC{x}_{i}}e^{i\VEC{k}\cdot\VEC{x}_{j}}e^{-i\VEC{k}_1\cdot\VEC{x}_{k}} e^{-i\VEC{k}_2\cdot\VEC{x}_{l}} e^{-i\VEC{k}_3\cdot\VEC{x}_{m}} \nonumber\\
	&+&(V^3/N^5) \Big[\hspace{-0.5cm}\sum_{(i \neq j \neq l \neq m),\, (i=k)}\hspace{-0.5cm} 
	\langle n_i^2 n_j n_l n_m \rangle_{\rm c}\,
	e^{-i(\VEC{k}+\VEC{k}_1)\cdot\VEC{x}_{i}}e^{i\VEC{k}\cdot\VEC{x}_{j}} e^{-i\VEC{k}_2\cdot\VEC{x}_{l}} e^{-i\VEC{k}_3\cdot\VEC{x}_{m}} + \mbox{(5 perms.)}\Big]\nonumber\\
	&+&(V^3/N^5) \Big[\hspace{-0.5cm}\sum_{(i\neq j \neq m),\, (i=k),\, (j=l)} \hspace{-0.5cm} 
	\langle n_i^2 n_j^2 n_m \rangle_{\rm c}\,
	e^{-i(\VEC{k}+\VEC{k}_1)\cdot\VEC{x}_{i}}e^{-i(\VEC{k}_2-\VEC{k})\cdot\VEC{x}_{j}} e^{-i\VEC{k}_3\cdot\VEC{x}_{m}} + \mbox{(5 perms.)}\Big].
\end{eqnarray}
The above expression can be simplified to
\begin{eqnarray}
	{\rm Cov}\big[\widehat{P}(\VEC{k}),\widehat{B}(\VEC{k}_1,\VEC{k}_2,\VEC{k}_3) \big]_{P_5} 
	= \frac{1}{V} P_5^{(\rm N)}(\VEC{k},-\VEC{k},\VEC{k}_1,\VEC{k}_2,\VEC{k}_3),
\end{eqnarray}
where
\begin{eqnarray}
	P^{(\rm N)}_5(\VEC{k},-\VEC{k},\VEC{k}_1,\VEC{k}_2,\VEC{k}_3) &=& P_5(\VEC{k},-\VEC{k},\VEC{k}_1,\VEC{k}_2,\VEC{k}_3) \nonumber \\
	&+& \frac{1}{\bar{n}}\Big[ T(\VEC{k}+\VEC{k}_1, -\VEC{k}, \VEC{k}_2,\VEC{k}_3) + T(\VEC{k}+\VEC{k}_2, -\VEC{k}, \VEC{k}_1,\VEC{k}_3) + T(\VEC{k}+\VEC{k}_3, -\VEC{k}, \VEC{k}_1,\VEC{k}_2) 
		\nonumber \\
	&+&  \hspace{0.4cm} T(-\VEC{k}+\VEC{k}_1, \VEC{k}, \VEC{k}_2,\VEC{k}_3) + T(-\VEC{k}+\VEC{k}_2, \VEC{k}, \VEC{k}_1,\VEC{k}_3) + T(-\VEC{k}+\VEC{k}_3, \VEC{k}, \VEC{k}_1,\VEC{k}_2) \Big] 
	\nonumber \\
	&+& \frac{1}{\bar{n}^2}\Big[ B(\VEC{k}+\VEC{k}_1, \VEC{k}_2-\VEC{k}, \VEC{k}_3)+ B(\VEC{k}+\VEC{k}_1, \VEC{k}_3-\VEC{k}, \VEC{k}_2)+B(\VEC{k}+\VEC{k}_2, \VEC{k}_3-\VEC{k}, \VEC{k}_1)
	\nonumber \\
	&+&  \hspace{0.5cm} 	
	B(-\VEC{k}+\VEC{k}_1, \VEC{k}_2+\VEC{k}, \VEC{k}_3)+ B(-\VEC{k}+\VEC{k}_1, \VEC{k}_3+\VEC{k}, \VEC{k}_2)+B(-\VEC{k}+\VEC{k}_2, \VEC{k}_3+\VEC{k}, \VEC{k}_1)
	\Big].
	\label{Eq:P5_N}
\end{eqnarray}

\subsection{Bispectrum covariance}
\label{Sec:BB}

The auto-covariance of the bispectrum has four sources:
\begin{eqnarray}
	{\rm Cov}\big[ \widehat{B},\widehat{B} \big]
	&=& 
	 {\rm Cov}\big[ \widehat{B},\widehat{B} \big]_{PPP}
	+{\rm Cov}\big[ \widehat{B},\widehat{B} \big]_{BB}
	+{\rm Cov}\big[ \widehat{B},\widehat{B} \big]_{PT}
	+{\rm Cov}\big[ \widehat{B},\widehat{B} \big]_{P_6},
\end{eqnarray}
where we omitted to denote the dependence of wavevector on the bispectrum for notational simplicity, and the term with the subscript ``$PPP$'' consists of a product of three power spectra, the ``$BB$'' term a product of two bispectra, the ``$PT$'' term a product of power spectrum and trispectrum, and the ``$P_6$'' term the 6-point spectrum.

The expression of the ``$PPP$'' term, which is the Gaussian part, is well known (e.g., see \citealt{Sefusatti:2006pa}):
\begin{eqnarray}
	{\rm Cov}\big[ \widehat{B}(\VEC{k}_1,\VEC{k}_2,\VEC{k}_3),\widehat{B}(\VEC{k}'_1,\VEC{k}'_2,\VEC{k}'_3) \big]_{PPP}
	&=& \Bigg[\frac{(2\pi)^3\delta_{\rm D}\left( \VEC{k}_1+\VEC{k}'_1 \right)(2\pi)^3\delta_{\rm D}\left( \VEC{k}_2+\VEC{k}'_2 \right)}{V} + \mbox{(5 perms.)} \Bigg]\nonumber \\
	&\times&
	P^{(\rm N)}(\VEC{k}_1) P^{(\rm N)}(\VEC{k}_2) P^{(\rm N)}(\VEC{k}_3)
	\label{Eq:covPPP_perm}
\end{eqnarray}
with $\VEC{k}_1+\VEC{k}_2+\VEC{k}_3=\VEC{0}$ and $\VEC{k}'_1+\VEC{k}'_2+\VEC{k}'_3=\VEC{0}$.
The ``$BB$'' and ``$PT$'' terms are given by
\begin{eqnarray}
	{\rm Cov}\big[ \widehat{B}(\VEC{k}_1,\VEC{k}_2,\VEC{k}_3),\widehat{B}(\VEC{k}'_1,\VEC{k}'_2,\VEC{k}'_3) \big]_{BB}
	&=& \left( \frac{V^2}{N^3} \right) 
	\Big[\sum_{(i\neq j), l}\langle n_in_jn_l\rangle_{\rm c}e^{-i\VEC{k}_1\cdot\VEC{x}_i}e^{-i\VEC{k}_2\cdot\VEC{x}_j}e^{-i\VEC{k}'_1\cdot\VEC{x}_l} \Big]\nonumber \\
	&\times&\left( \frac{V^2}{N^3} \right)\Big[\sum_{k,\, (m\neq n)} \langle n_kn_mn_n\rangle_{\rm c}e^{-i\VEC{k}_3\cdot\VEC{x}_k}e^{-i\VEC{k}'_2\cdot\VEC{x}_m}e^{-i\VEC{k}'_3\cdot\VEC{x}_n}+ \mbox{(8 perms.)} \Big] 
	\nonumber\\
	{\rm Cov}\big[ \widehat{B}(\VEC{k}_1,\VEC{k}_2,\VEC{k}_3),\widehat{B}(\VEC{k}'_1,\VEC{k}'_2,\VEC{k}'_3) \big]_{PT} 
	&=&\left( \frac{V}{N^2} \right) 
	\Big[\sum_{i,\,l}\langle n_in_l\rangle_{\rm c}e^{-i\VEC{k}_1\cdot\VEC{x}_i}e^{-i\VEC{k}'_1\cdot\VEC{x}_l} \Big] \nonumber \\
	&\times&\left( \frac{V^3}{N^4} \right)\Big[\sum_{(j\neq k),\, (m\neq n)} \langle n_jn_kn_mn_n\rangle_{\rm c}
	e^{-i\VEC{k}_2\cdot\VEC{x}_j}e^{-i\VEC{k}_3\cdot\VEC{x}_k}e^{-i\VEC{k}'_2\cdot\VEC{x}_m}e^{-i\VEC{k}'_3\cdot\VEC{x}_n}+(\mbox{8 perms.)}\Big]. \nonumber\\
\end{eqnarray}
Using Eq.~(\ref{Eq:estimators}), these above expressions become
\begin{eqnarray}
	{\rm Cov}\big[ \widehat{B}(\VEC{k}_1,\VEC{k}_2,\VEC{k}_3),\widehat{B}(\VEC{k}'_1,\VEC{k}'_2,\VEC{k}'_3) \big]_{BB}
	=  \frac{(2\pi)^3\delta_{\rm D}\left( \VEC{k}_1-\VEC{k}'_1 \right)}{V} 
	B^{(\rm N)}(\VEC{k}_1,\VEC{k}_2,\VEC{k}_3) B^{(\rm N)}(\VEC{k}'_1,\VEC{k}'_2,\VEC{k}'_3) + \mbox{(8 perms.)}, 
	\label{Eq:covBB_term}
\end{eqnarray}
\begin{eqnarray}
	{\rm Cov}\big[ \widehat{B}(\VEC{k}_1,\VEC{k}_2,\VEC{k}_3),\widehat{B}(\VEC{k}'_1,\VEC{k}'_2,\VEC{k}'_3) \big]_{PT}
	= \frac{(2\pi)^3\delta_{\rm D}\left( \VEC{k}_1+\VEC{k}'_1 \right)}{V}  P^{(\rm N)}(\VEC{k}_1) T^{(\rm N)}(\VEC{k}_2,\VEC{k}_3,\VEC{k}'_2,\VEC{k}'_3) + \mbox{(8 perms.)}.
	\label{Eq:covPT_term}
\end{eqnarray}
Finally, just like the calculation of ${\rm Cov}[ P,P ]_T$ and ${\rm Cov}[ P,B ]_{P_5}$, the ``$P_6$'' term is given by
\begin{eqnarray}
	{\rm Cov}\big[ \widehat{B}(\VEC{k}_1,\VEC{k}_2,\VEC{k}_3),\widehat{B}(\VEC{k}'_1,\VEC{k}'_2,\VEC{k}'_3) \big]_{P_6} 
	&=& \frac{1}{V} \Big\{ P_6(\VEC{k}_1,\VEC{k}_2,\VEC{k}_3,\VEC{k}'_1,\VEC{k}'_2,\VEC{k}'_3)  \nonumber \\
	&+& \frac{1}{\bar{n}} \Big[ P_5(\VEC{k}_1+\VEC{k}'_1, \VEC{k}_2,\VEC{k}_3,\VEC{k}'_2,\VEC{k}'_3) + \mbox{(8 perms.)}  \Big]\nonumber\\
	&+& \frac{1}{\bar{n}^2} \Big[  T(\VEC{k}_1+\VEC{k}'_1, \VEC{k}_2+\VEC{k}'_2,\VEC{k}_3,\VEC{k}'_3) + \mbox{(17 perms.)}\Big]\nonumber\\
	&+& \frac{1}{\bar{n}^3} \Big[ B(\VEC{k}_1+\VEC{k}'_1, \VEC{k}_2+\VEC{k}'_2,\VEC{k}_3+\VEC{k}'_3) + \mbox{(5 perms.)}  \Big]\Big\}.
	\label{Eq:covP6_term}
\end{eqnarray}
The full expressions of ${\rm Cov}[B,B]_{PPP}$, ${\rm Cov}[B,B]_{BB}$, ${\rm Cov}[B,B]_{PT}$ and ${\rm Cov}[B,B]_{P_6}$ without abbreviating expressions using ``perms.'' 
are summarized in Appendix~\ref{Ap:equations}.

\subsection{Computation of analytical PT expressions and its limitations}
\label{Sec:Prescription}

We compute the covariances up to the leading order of the standard PT, and summarize the necessary expressions in Appendix \ref{Ap:PT}. In practice, the computation requires the linear power spectrum, $P_{\rm lin}(k)$, which we generate with {\scriptsize CLASS}~\citep{Lesgourgues:2011re}. In numerically computing the covariance matrices, we use {\scriptsize CUBA}, a publicly available library for multidimensional numerical integration~\footnote{\url{http://www.feynarts.de/cuba/}}. The cosmological parameters used in our calculations are shown at the end of Section~\ref{Sec:Introduction}.

Let us here clarify the assumptions and limitations in our approach. First of all, we focus on the leading-order (i.e., the tree-level) PT terms and ignore any loop corrections. Similarly, we only include the linear bias, and ignore any higher order bias parameters such as $b_2$, $b_3$ and $b_{\rm K}$, where $b_2$, $b_3$, $b_{\rm K}$ are the second- and third-order local biases, and the tidal bias, respectively \citep[see e.g.,][]{Saito:2014aa,Desjacques:2018xx}. In terms of RSD, we do take into account some of the nonlinear terms through the higher-order kernels such as $Z_{2}$ (see Appendix \ref{Ap:PT}), but do not include fully nonlinear term such as the Finger-of-God (FOG) suppression \citep[e.g.,][]{Taruya:2010aa}. Studying the impact of these additional non-linear effects on the power and bispectrum covariances are left to future work.

In addition, we do not take account of any survey window effects in predicting the covariance matrices. The survey window effect suppresses the amplitude of the covariance on large scales~\citep{Li:2018scc}, and also generate additional sources to the covariance arising from long-wavelength fluctuations beyond the survey area, the so-called super-sample covariance (SSC;~\citealt{Hamilton:2005dx,Rimes:2005dz,dePutter:2011ah,Takada:2013bfn,Chan:2017fiv}). Because of the suppression effect, our analytical calculations of the covariance are somewhat overestimated on large scales compared to the mock results. However, as shown in~\citep{Li:2018scc}, such large scale contributions to the covariance does not significantly affect cumulative signal-to noise ratios (see Section~\ref{Sec:SN}), since the signal to noise for the modes close to the survey size is expected to be extremely small. For the SSC effect, we completely ignore it, because the Patchy mocks with which we compare our results do not also include the SSC effect correctly. Besides, at least in the case of the galaxy power spectrum, it is considered that the two main contributions to the SSC effect on galaxy clustering, the beat-coupling effect~\citep{Hamilton:2005dx} and the local mean effect~\citep{dePutter:2011ah}, tend to cancel each other out~\citep{dePutter:2011ah}, so we assume that the total contributions of the SSC effect to the covariance is small enough to be ignored. \Red{More recently, \citet{Wadekar:2019rdu} provided detailed studies of the SSC effect on the power spectrum covariance. Our results on the signal-to-noise ratios of the power spectrum monopole and quadrupole are quantitatively consistent with Fig.~10 in \cite{Wadekar:2019rdu}, which takes the SSC terms into account. For instance, our $({\rm S/N})_{P_{0}}(k_{\rm max}=0.2)\sim 160$ and $({\rm S/N})_{P_{2}}(k_{\rm max}=0.2)\sim 25$  (see Fig. 6 in Section~\ref{Sec:SN}). This fact would support the finding in the previous works. In the bispectrum case, \citet{Chan:2017fiv} showed for dark matter in real space that relative to the bispectrum covariance without the SSC effect, the magnitude of SSC of the bispectrum is roughly an order of magnitude smaller than the power spectrum case. Thus, these works imply that the SSC effect may be sub-dominant in the covariance of the galaxy clustering. The remaining work on the super-survey effect is to account for the RSD effect and to compute the quadrupole (and hexadecapole) component of the bispectrum covariance, which is left for future works.} 

\Red{
Finally, the survey window effect introduces two effects on the Gaussian terms of ${\rm Cov}\left[ P,P \right]$ and ${\rm Cov}\left[ B,B \right]$. The first one is to suppress the amplitude of the power spectrum at the $\sim 20\%$ level at large scales such as $k\sim0.01\hk$ for a BOSS-like survey region, indicating that the amplitude of the covariance is also suppressed at the large scales. The second one is to generate correlations between different $k$-modes in the covariance; e.g., even the Gaussian term of ${\rm Cov}\left[ P,P \right]$ can have contributions to the off-diagonal elements. In the case of the power spectrum, these effects have been studied in details by \cite{Li:2018scc}. We ignore the effect in this paper just for simplicity and leave the work on extending to the bispectrum covariance case for future topics.
}

\section{Decomposition formalism of the covariance matrix}
\label{Sec:Multipoles}

\subsection{Decomposition formalism of the power spectrum and the bispectrum}
\label{Sec:DecompositionFormalism}

So far we have presented the analytic expressions to describe the covariance matrices for the full 3D power spectrum and the bispectrum. In the following, we decompose the full 3D power spectrum and the bispectrum into the form that is practically more convenient in analyzing actual galaxy surveys. The galaxy power and bispectra have the angular-dependence along the LOS direction $\hat{n}$ due to the RSD and the AP effects. To quantify such anisotropic signals, it is useful to expand the power and bispectra in orthogonal functions, i.e., Legendre polynomials and tri-polar spherical harmonics, respectively, and to spherically average them around wavevectors and the LOS with the orthogonal functions weighted to give multipole components~\citep{Hamilton:1997zq,Sugiyama2018}:
\begin{eqnarray}
	P_{\ell}(k) &=& (2\ell+1) \int \frac{d^2\hat{k}}{4\pi}\int \frac{d^2\hat{n}}{4\pi}\,
	{\cal L}_{\ell}(\hat{k}\cdot\hat{n})\, P(\VEC{k}) \nonumber \\
	B_{\ell_1\ell_2L}(k_1,k_2) &=& N_{\ell_1\ell_2L}H_{\ell_1\ell_2L}^2
	\int \frac{d^2\hat{k}_1}{4\pi}\int \frac{d^2\hat{k}_2}{4\pi}\int \frac{d^2\hat{n}}{4\pi}\, 
	{\cal S}_{\ell_1\ell_2L}(\hat{k}_1,\hat{k}_2,\hat{n})\, B(\VEC{k}_1,\VEC{k}_2,-\VEC{k}_{12}),
	\label{Eq:PB_multipole}
\end{eqnarray}
where ${\cal L}_{\ell}$ denotes the Legendre polynomials at $\ell$-th order, $N_{\ell_1\ell_2L}=(2\ell_1+1)(2\ell_2+1)(2L+1)$, and $H_{\ell_1\ell_2L} = \left( \begin{smallmatrix} \ell_1 & \ell_2 & L \\ 0 & 0 & 0 \end{smallmatrix}  \right)$ filters $\ell_1+\ell_2+L={\rm even}$ components with the bracket with $3$ multipole indices, $(\dots)$, being the Wigner-$3j$ symbol. The base function ${\cal S}_{\ell_1\ell_2L}$ is defined as
\begin{eqnarray}
	{\cal S}_{\ell_1\ell_2L}(\hat{k}_1,\hat{k}_2,\hat{n}) 
    &=& 
   \frac{1}{H_{\ell_1\ell_2L}} \sum_{m_1m_2M}  \left( \begin{smallmatrix} \ell_1 & \ell_2 & L \\ m_1 & m_2 & M \end{smallmatrix}  \right) 
   y_{\ell_1}^{m_1}(\hat{k}_1) y_{\ell_2}^{m_2}(\hat{k}_2) y_L^M(\hat{n}),
    \label{Eq:Slll}
\end{eqnarray} 
where $y_{\ell}^m=\sqrt{4\pi/(2\ell+1)}\,Y_{\ell}^m$ is a normalized spherical harmonic function, which reduces to the Legendre polynomial if $L=0$: ${\cal S}_{\ell \ell L=0}(\hat{k}_1,\hat{k}_2,\hat{n}) = {\cal L}_{\ell}(\hat{k}_1\cdot\hat{k}_2)$. One of the most remarkable features of this decomposition formalism, especially for the bispectrum, is that the final result is independent of the choice of the coordinate system, so one can choose any convenient coordinate system for numerical computation: e.g., the coorinate system taking the $\hat{k}_1$ or the LOS $\hat{n}$ as the $z$-axis~\citep{Scoccimarro:1999ed,Slepian:2017lpm}. The multipole indexes, $\ell$ and $L$, in the power spectrum and the bispectrum respectively mean the expansion with respect to the LOS direction; they should be even, because the anisotropies arising from the RSD or AP effect are axially symmetric around the LOS in the framework of Newtonian gravity. If we consider General Relativistic corrections, the odd mode, i.e., $\ell={\rm odd}$ and $L={\rm odd}$, may appear (e.g., for the bispectrum, see~\citealt{Clarkson:2018dwn}). We can then single out only the anisotropic signal by computing the $\ell>0$ mode for the power spectrum and $L>0$ for the bispectrum. The power and bispectrum multipoles defined above satisfy three fundamental statistical properties of the Universe: homogeneity, isotropy and parity-symmetry. From these conditions, the bispectrum multipoles become non-zero if $H_{\ell_1\ell_2L}\neq 0$. (For more details of the bispectrum multipoles, see \citealt{Sugiyama2018}.) Throughout this paper, we refer to the $\ell=0$ ($L=0$) mode for the power spectrum (bispectrum) as ``monopole'', and to the $\ell=2$ ($L=2$) mode as ``quadrupole''.

Using Eq.~(\ref{Eq:PB_multipole}) the covariances of the power and bispectrum multipoles can be written as
\begin{eqnarray}
	{\rm Cov}\left[ P_{\ell}(k),\, P_{\ell'}(k') \right]
	&=&  (2\ell+1)(2\ell'+1)\int \frac{d\hat{k}}{4\pi}\int \frac{d\hat{k}'}{4\pi}\int \frac{d\hat{n}}{4\pi}\,
	{\cal L}_{\ell}(\hat{k}\cdot\hat{n})\, {\cal L}_{\ell'}(\hat{k}'\cdot\hat{n})\, {\rm Cov}\left[ P(\VEC{k}),\, P(\VEC{k}') \right] \nonumber \\
	{\rm Cov}\left[ P_{\ell}(k),\, B_{\ell_1\ell_2L}(k_1,k_2) \right]
	&=& (2\ell+1) N_{\ell_1\ell_2L}H_{\ell_1\ell_2L}^2\int \frac{d\hat{k}}{4\pi}\int \frac{d\hat{k}_1}{4\pi}\int \frac{d\hat{k}_2}{4\pi}\int \frac{d\hat{n}}{4\pi}\,
	{\cal L}_{\ell}(\hat{k}\cdot\hat{n})\, {\cal S}_{\ell_1\ell_2L}(\hat{k}_1,\hat{k}_2,\hat{n}) \nonumber \\
	&\times& {\rm Cov}\left[ P(\VEC{k}),\, B(\VEC{k}_1,\VEC{k}_2,-\VEC{k}_{12}) \right] \nonumber \\
	{\rm Cov}\left[ B_{\ell_1\ell_2L}(k_1,k_2), \, B_{\ell'_1\ell'_2L'}(k'_1,k'_2) \right]
	&=& N_{\ell_1\ell_2L} N_{\ell'_1\ell'_2L'}H_{\ell_1\ell_2L}^2H_{\ell'_1\ell'_2L'}^2
	\int \frac{d\hat{k}_1}{4\pi}\int \frac{d\hat{k}_2}{4\pi}\int \frac{d\hat{k}'_1}{4\pi}\int \frac{d\hat{k}'_2}{4\pi}\int \frac{d\hat{n}}{4\pi}
	 \nonumber \\
	&\times& 
	{\cal S}_{\ell_1\ell_2L}(\hat{k}_1,\hat{k}_2,\hat{n})\,  {\cal S}_{\ell'_1\ell'_2L'}(\hat{k}'_1,\hat{k}'_2,\hat{n})\,
	{\rm Cov}\left[ B(\VEC{k}_1,\VEC{k}_2,-\VEC{k}_{12}), B(\VEC{k}'_1,\VEC{k}'_2,-\VEC{k}'_{12}) \right],
	\label{Eq:cov_multipole}
\end{eqnarray}
where $\VEC{k}_{12}=\VEC{k}_1+\VEC{k}_2$. Thus, the covariance of the multipoles is directly related to that of the 3D power spectrum and the bispectrum. The calculations of these expressions require three, four and five two-dimensional angular integrals for ${\rm Cov}[P,P]$, ${\rm Cov}[P,B]$ and ${\rm Cov}[B,B]$, respectively. Because of the rotational invariance, we can reduce the number of integral dimensions by three by choosing a specific coordinate system; therefore, the actually required numbers of integral dimensions are $3\, (=6-3)$, $5\, (=8-3)$ and $7\, (=10-3)$ for ${\rm Cov}[P,P]$, ${\rm Cov}[P,B]$ and ${\rm Cov}[B,B]$, respectively. In this paper, we adopt the following coordinates: for ${\rm Cov}[P,P]$,
\begin{eqnarray}
	\hat{k}  &=& \{\sin\theta_k, 0, \cos\theta_k\} \nonumber \\
	\hat{k}' &=& \{\sin\theta_{k'} \cos\varphi_{k'}, \sin\theta_{k'} \sin\varphi_{k'}, \cos\theta_{k'}\} \nonumber \\
	\hat{n}  &=& \{0,0,1\},
	\label{Eq:PP_coordinate}
\end{eqnarray}
for ${\rm Cov}[P,B]$,
\begin{eqnarray}
	\hat{k}   &=& \{\sin\theta_k, 0, \cos\theta_k\}\nonumber \\
	\hat{k}_1 &=& \{\sin\theta_{k_1} \cos\varphi_{k_1}, \sin\theta_{k_1} \sin\varphi_{k_1}, \cos\theta_{k_1}\}\nonumber \\
	\hat{k}_2 &=& \{\sin\theta_{k_2} \cos\varphi_{k_2}, \sin\theta_{k_2} \sin\varphi_{k_2}, \cos\theta_{k_2}\}\nonumber \\
	\hat{n}   &=& \{0,0,1\},
	\label{Eq:PB_coordinate}
\end{eqnarray}
and for ${\rm Cov}[B,B]$,
\begin{eqnarray}
	\hat{k}_1  &=& \{\sin\theta_{k_1}, 0, \cos\theta_{k_1}\}\nonumber \\
	\hat{k}_2  &=& \{\sin\theta_{k_2} \cos\varphi_{k_2}, \sin\theta_{k_2} \sin\varphi_{k_2}, \cos\theta_{k_2}\}\nonumber \\
	\hat{k}'_1 &=& \{\sin\theta_{k'_1} \cos\varphi_{k'_1}, \sin\theta_{k'_1} \sin\varphi_{k'_1}, \cos\theta_{k'_1}\}\nonumber \\
	\hat{k}'_2 &=& \{\sin\theta_{k'_2} \cos\varphi_{k'_2}, \sin\theta_{k'_2} \sin\varphi_{k'_2}, \cos\theta_{k'_2}\}\nonumber \\
	\hat{n}  &=&   \{0,0,1\},
\end{eqnarray}
where we fixed the LOS to the $z$-axis. The above discussion on the number of integral dimensions is the case only for the connected parts, which arise from the trispectrum, the 5-point spectrum and the 6-point spectrum for ${\rm Cov}[P,P]$, ${\rm Cov}[P,B]$ and ${\rm Cov}[B,B]$, respectively. As we will explicitly show in the next subsection, the required number of integrals for the unconnected parts becomes even smaller, because we can analytically calculate the angular integrals relevant to the Dirac delta function.

\subsection{Further simplification of the unconnected parts}

In this subsection, we provide analytical calculations of the unconnected parts of the covariance. We begin with the simplest case, the power spectrum covariance, in Section~\ref{Sec:CovPP}. We will then extend it to the cross-covariance between $P$ and $B$, and the auto bispectrum covariance in Sections~\ref{Sec:CovPB} and \ref{Sec:CovBB}.

\subsubsection{Power spectrum covariance}
\label{Sec:CovPP}

Inserting Eq.(\ref{Eq:cov_PP_Gaussian}) in Eq.~(\ref{Eq:cov_multipole}), one trivially obtains
\begin{eqnarray}
	{\rm Cov}\left[ P_{\ell}(k),\, P_{\ell'}(k') \right]_{PP}
	&=& 2 (2\ell+1)(2\ell'+1)\frac{(2\pi)^3\delta_{\rm D}(k-k')}{ 4\pi k^2V }
	\int \frac{d\mu}{2}\,
	{\cal L}_{\ell}(\mu)\, {\cal L}_{\ell'}(\mu)\, 
	 \left[  P(\VEC{k}) + \frac{1}{\bar{n}} \right]^2,
	 \label{Eq:cov_PP_multipole_Gaussian_base}
\end{eqnarray}
where $\mu = \hat{k}\cdot\hat{n}$, and we used the relation
\begin{eqnarray}
	\delta_{\rm D}\left( \VEC{k} - \VEC{k}' \right) = \frac{1}{k^2}\delta_{\rm D}\left( k - k' \right) \delta_{\rm D}\big( \hat{k}- \hat{k}' \big).
\end{eqnarray}
When discretizing the delta function, it is common to use the following relation
\begin{eqnarray}
	\delta_{\rm D}\left( k-k' \right) \to \frac{1}{\Delta k} \delta^{(\rm K)}_{k k'},
	\label{Eq:DeltaToKronecker}
\end{eqnarray}
where $\delta_{\rm K}$ represents the Kronecker delta defined such that $\delta^{(\rm K)}_{kk'}=1$ if $k=k'$, otherwise zero, and $\Delta k $ denotes the width of $k$-bins. Equation~(\ref{Eq:cov_PP_multipole_Gaussian_base}) then becomes the well-known form \citep[e.g.,][]{Taruya:2010aa,Taruya:2011aa}
\begin{eqnarray}
	{\rm Cov}\left[ P_{\ell}(k),\, P_{\ell'}(k') \right]_{PP}
	&=& 2 (2\ell+1)(2\ell'+1)\frac{\delta^{(\rm K)}_{kk'}}{N_{\rm mode}(k)}
	\int \frac{d\mu}{2}\,
	{\cal L}_{\ell}(\mu)\, {\cal L}_{\ell'}(\mu)\, 
	 \left[  P(\VEC{k}) + \frac{1}{\bar{n}} \right]^2,
	 \label{Eq:CovPP_G}
\end{eqnarray}
where $N_{\rm mode}(k) = 4\pi k^2 \Delta k V/ (2\pi)^3$ corresponds to the number of independent Fourier modes in each $k$-bin. 
Thus, the unconnected part of the covariance is reduced to a one-dimensional integral, and has the dependence of the bin width.

\subsubsection{The cross-covariance}
\label{Sec:CovPB}

For the cross-covariance, we substitute Eq.~(\ref{Eq:covPB}) into Eq.~(\ref{Eq:cov_multipole});
then, we can analytically calculate the integral over the angle $\hat{k}$:
\begin{eqnarray}
	{\rm Cov}\left[ P_{\ell}(k),\, B_{\ell_1\ell_2L}(k_1,k_2) \right]_{PB} 
	&=& 2\, (2\ell+1)\, N_{\ell_1\ell_2L}H_{\ell_1\ell_2L}^2
	\int \frac{d \cos\theta_{k_1}}{2}\int \frac{d \cos \theta_{k_2} d\varphi_{k_2}}{4\pi}{\cal S}_{\ell_1\ell_2L}(\hat{k}_1,\hat{k}_2,\hat{n}) \nonumber \\
	&\times& \hspace{-0.25cm}
	\Bigg\{
	\frac{\delta^{(\rm K)}_{k k_1}}{N_{\rm mode}(k_1)}{\cal L}_{\ell} (\hat{k}_1\cdot\hat{n}) P^{(\rm N)}(\VEC{k}_1) B^{(\rm N)}(\VEC{k}_1,\VEC{k}_2,\VEC{k}_3)\nonumber \\
	&+&  \frac{\delta^{(\rm K)}_{k k_2}}{N_{\rm mode}(k_2)}
	{\cal L}_{\ell} (\hat{k}_2\cdot\hat{n})  P^{(\rm N)}(\VEC{k}_2) B^{(\rm N)}(\VEC{k}_2,\VEC{k}_1,\VEC{k}_3) \nonumber \\
	&+& 
	\frac{\delta^{(\rm K)}_{k k_3}}{N_{\rm mode}(k_3)}
	{\cal L}_{\ell} (\hat{k}_3\cdot\hat{n}) P^{(\rm N)}(\VEC{k}_3) B^{(\rm N)}(\VEC{k}_3,\VEC{k}_1,\VEC{k}_2)
	 \Bigg\},
	\label{Eq:covPB_multipole_base}
\end{eqnarray}
where the bispectrum satisfies the triangle condition $\VEC{k}_3 = -\VEC{k}_1-\VEC{k}_2$. Here, we take $\hat{n}$ as the $z$-axis and adopt the same coordinates as Eq.~(\ref{Eq:PP_coordinate}).

Note that it is impractical to numerically compute $\delta^{\rm (K)}_{k k_3}$ in the last line of Eq.~(\ref{Eq:covPB_multipole_base}), because $k_3$ is continuous due to its dependence of an angle between $\hat{k}_1$ and $\hat{k}_2$. To approximately estimate the last term on the RHS of Eq.~(\ref{Eq:covPB_multipole_base}) including the binning effect, we adopt a top-hat function instead of the Kronecker delta and make the following replacement
\begin{eqnarray}
	\frac{\delta^{(\rm K)}_{kk'}}{N_{\rm mode}(k)} \to \frac{W(k,k')}{\widetilde{N}_{\rm mode}(k,k')}
	\label{Eq:new_Nmode}
\end{eqnarray}
where
\begin{eqnarray}
	W(k,k') = \begin{cases} 1 & |k-k'| < \Delta k/2 \\ 0 & \mbox{otherwise} \end{cases},
	\label{Eq:new_delta}
\end{eqnarray}
and 
\begin{eqnarray}
	\widetilde{N}_{\rm mode}(k,k') =  \frac{4\pi k k' \Delta k V}{ (2\pi)^3}.
\end{eqnarray}
Then, Eq.~(\ref{Eq:covPB_multipole_base}) becomes
\begin{eqnarray}
	{\rm Cov}\left[ P_{\ell}(k),\, B_{\ell_1\ell_2L}(k_1,k_2) \right]_{PB} 
	&=& 2\, (2\ell+1)\, N_{\ell_1\ell_2L}H_{\ell_1\ell_2L}^2
	\int \frac{d \cos\theta_{k_1}}{2}\int \frac{d \cos \theta_{k_2} d\varphi_{k_2}}{4\pi}{\cal S}_{\ell_1\ell_2L}(\hat{k}_1,\hat{k}_2,\hat{n}) \nonumber \\
	&\times& \hspace{-0.25cm}
	\Bigg\{
	\frac{W(k,k_1)}{\widetilde{N}_{\rm mode}(k,k_1)}{\cal L}_{\ell} (\hat{k}_1\cdot\hat{n}) P^{(\rm N)}(\VEC{k}_1) B^{(\rm N)}(\VEC{k}_1,\VEC{k}_2,\VEC{k}_3)\nonumber \\
	&+&  \frac{W(k,k_2)}{\widetilde{N}_{\rm mode}(k,k_2)}
	{\cal L}_{\ell} (\hat{k}_2\cdot\hat{n}) P^{(\rm N)}(\VEC{k}_2) B^{(\rm N)}(\VEC{k}_2,\VEC{k}_1,\VEC{k}_3) \nonumber \\
	&+&  \frac{W(k,k_3)}{\widetilde{N}_{\rm mode}(k,k_3)}
		{\cal L}_{\ell} (\hat{k}_3\cdot\hat{n}) P^{(\rm N)}(\VEC{k}_3) B^{(\rm N)}(\VEC{k}_3,\VEC{k}_1,\VEC{k}_2)
	 \Bigg\}.
	\label{Eq:covPB_multipole}
\end{eqnarray}

\subsubsection{Bispectrum covariance}
\label{Sec:CovBB}

As mentioned in Section~\ref{Sec:BB}, the unconnected parts of the bispectrum covariance have three sources: one Gaussian term, ${\rm Cov}[B,B]_{PPP}$, and two non-Gaussian parts, ${\rm Cov}[B,B]_{BB}$ and ${\rm Cov}[B,B]_{PT}$. Since their analytical expressions are quite lengthy, we restrict ourselves here to only one term appearing in the $PPP$, $BB$ and $PT$ terms and present their full expressions without using "perms." in Appendix~\ref{Ap:equations}.

Using the same approach as in the case of the cross-covariance, the Gaussian part is given by
\begin{eqnarray}
	{\rm Cov}\left[ B_{\ell_1\ell_2L}(k_1,k_2), \, B_{\ell'_1\ell'_2L'}(k'_1,k'_2) \right]_{PPP}
	&=& N_{\ell_1\ell_2L} N_{\ell'_1\ell'_2L'}H_{\ell_1\ell_2L}^2H_{\ell'_1\ell'_2L'}^2\, V
	\int \frac{d \cos\theta_{k_1}}{2}\int \frac{d \cos \theta_{k_2} d\varphi_{k_2}}{4\pi}
	 \nonumber \\
	&\times& 
	{\cal S}_{\ell_1\ell_2L}(\hat{k}_1,\hat{k}_2,\hat{n})\,  {\cal S}_{\ell'_1\ell'_2L'}(\hat{k}_1,\hat{k}_2,\hat{n})\,
	\frac{W(k_1,k_1')}{\widetilde{N}_{\rm mode}(k_1,k_1')}\frac{W(k_2,k_2')}{\widetilde{N}_{\rm mode}(k_2,k_2')} \nonumber\\
	&\times&  P^{(\rm N)}(\VEC{k}_1) P^{(\rm N)}(\VEC{k}_2) P^{(\rm N)}(\VEC{k}_3) + \mbox{(5 perms.)}.
\end{eqnarray}
where we used the same coordinates as Eq.~(\ref{Eq:PP_coordinate}).
The two non-Gaussian unconnected terms are calculated as follows:
\begin{eqnarray}
	&&{\rm Cov}\left[ B_{\ell_1\ell_2L}(k_1,k_2), \, B_{\ell'_1\ell'_2L'}(k'_1,k'_2) \right]_{BB} \nonumber \\
	&=& N_{\ell_1\ell_2L} N_{\ell'_1\ell'_2L'}H_{\ell_1\ell_2L}^2H_{\ell'_1\ell'_2L'}^2
	\int \frac{d \cos \theta_{k_1}}{2}\int \frac{d\hat{k}_2}{4\pi}\int \frac{d\hat{k}'_2}{4\pi}
	\nonumber \\
	&\times& 
	{\cal S}_{\ell_1\ell_2L}(\hat{k}_1,\hat{k}_2,\hat{n})\,  {\cal S}_{\ell'_1\ell'_2L'}(\hat{k}_1',\hat{k}'_2,\hat{n})\,
	\frac{W\left( k_1,k'_1 \right)}{\widetilde{N}_{\rm mode}(k_1,k'_1)} 
	B^{(\rm N)}(\VEC{k}_1,\VEC{k}_2, \VEC{k}_3) B^{(\rm N)}(\VEC{k}'_1, \VEC{k}'_2, \VEC{k}'_3) + \mbox{(8 perms.)}
	\label{Eq:covBB_BB2}
\end{eqnarray}
with $\VEC{k}_1=\VEC{k}'_1$,
and
\begin{eqnarray}
	&&{\rm Cov}\left[ B_{\ell_1\ell_2L}(k_1,k_2), \, B_{\ell'_1\ell'_2L'}(k'_1,k'_2) \right]_{PT} \nonumber \\
	&=& N_{\ell_1\ell_2L} N_{\ell'_1\ell'_2L'}H_{\ell_1\ell_2L}^2H_{\ell'_1\ell'_2L'}^2
	\int \frac{d \cos \theta_{k_1}}{2}\int \frac{d\hat{k}_2}{4\pi}\int \frac{d\hat{k}'_2}{4\pi}
	\nonumber \\
	&\times& 
	{\cal S}_{\ell_1\ell_2L}(\hat{k}_1,\hat{k}_2,\hat{n})\,  {\cal S}_{\ell'_1\ell'_2L'}(\hat{k}'_1,\hat{k}'_2,\hat{n})\,
	\frac{W\left( k_1,k'_1 \right)}{\widetilde{N}_{\rm mode}(k_1,k'_1)} 
	P^{(\rm N)}(\VEC{k}_1) T^{(\rm N)}(\VEC{k}_2,\VEC{k}_3,\VEC{k}'_2, \VEC{k}'_3) + \mbox{(8 perms.)}
	\label{Eq:covBB_PT2}
\end{eqnarray}
with $\VEC{k}_1=-\VEC{k}'_1$, 
where we adopt the same coordinates as Eq.~(\ref{Eq:PB_coordinate}).

\section{Comparison with the Patchy mocks}
\label{Sec:ComparisonWithMocks}

\begin{table}
\centering
\begin{tabular}{cccccccc}
\hline\hline
sample & $z$ & $V$ $[\hGpc]^{3}$ & $\bar{n}/10^{4}$ $[\hMpc]^{-3}$ & $f\, \sigma_8 (z)$ & $b\, \sigma_8 (z)$  & $\sigma_8 (z)$ & $\Delta k$ $[\hk]$ \\
\hline  
NGC ($0.4<z<0.6$) & $0.51$ &  $1.76$ & $3.26$ & $0.48$ & $1.27$ & $0.64$ & $0.02$ \\
\hline  
\end{tabular}
\caption{
Parameters required for calculations of the covariance matrix from analytical expressions.
The assumed galaxy sample is the BOSS NGC sample at the redshift range of $0.4<z<0.6$.
This table shows the corresponding parameters of the sample:
from left to right, the mean redshift $z$, survey volume $V$, mean number density $\bar{n}$, growth rate function $f\sigma_8$
, linear bias parameter $b \sigma_8$ and rms matter density fluctuation on scales of $8\hMpc$.
The right end column shows the width of $k$-bins $\Delta k$, because the covariance depends on the bin width.
}
\label{Table:parameters}
\end{table}

To test the validity of our analytical calculations in perturbation theory (PT), we compare them with the covariance matrices measured from the Patchy mock catalogs. The outline of this section is as follows: First, we present the prescription of how to measure the power and bispectra from the mocks in Section~\ref{Sec:MeasurementsFromMocks}. The parameters required for theoretical predictions are summarized in Table~\ref{Table:parameters}. Since the primary goal of this paper is to investigate the properties of the covariance of the galaxy clustering in redshift space, we focus especially on the auto- and cross-covariances relevant to the monopole and quadrupole components. Namely, for the power spectrum covariance, we compute ${\rm Cov}\left[ P_0,P_0 \right]$, ${\rm Cov}\left[ P_0,P_2 \right]$ and ${\rm Cov}\left[ P_2,P_2 \right]$ in Figure~\ref{fig:covPP}. Since the decomposed bispectra (\ref{Eq:PB_multipole}) have an infinite number of multipole terms, $B_{000}$, $B_{110}$, $B_{220}$, etc., for the monopole component ($L=0$), and $B_{202}$, $B_{112}$, $B_{222}$, etc., for the quadrupole component ($L=2$), we first restrict our attention to the lowest order of each of the monopole and quadrupole components, i.e., $B_{000}$ and $B_{202}$. Then, we compute the following four cross-covariances between the power and bispectra, ${\rm Cov}\left[ P_0,B_{000} \right]$, ${\rm Cov}\left[ P_0,B_{202} \right]$, ${\rm Cov}\left[ B_{202}, P_0 \right]$ and ${\rm Cov}\left[ P_2,B_{202} \right]$ in Figure~\ref{fig:covPB}. For the auto-covariance of the bispectrum, Figure~\ref{fig:covBB} shows ${\rm Cov}\left[ B_{000},B_{000} \right]$, ${\rm Cov}\left[ B_{000},B_{202} \right]$ and ${\rm Cov}\left[ B_{202}, B_{202} \right]$. After that, we study the covariances relevant to higher order multipoles of the monopole bispectrum and show in Figure~\ref{fig:covBB_higher} ${\rm Cov}\left[ B_{110},B_{110} \right]$, ${\rm Cov}\left[ B_{220},B_{220} \right]$ and ${\rm Cov}\left[ B_{000},B_{110} \right]$. Finally, we reveal the scales where the shot-noise term becomes dominant on the covariances for both the power and bispectrum cases through Figure~\ref{fig:cov_Shotnoise}.

\subsection{Measurements from the Patchy mocks}
\label{Sec:MeasurementsFromMocks}

The Patchy mocks~\citep{Klypin:2014kpa,Kitaura:2015uqa} have been calibrated to an $N$-body simulation based reference sample using approximate galaxy solvers and analytical-statistical biasing models, and incorporate observational effects including the survey geometry, veto mask and fiber collisions. According to the cosmological analysis of the BOSS DR12 galaxies~\citep{Alam:2016hwk}, we divide the range of observed redshift in the BOSS survey into three bins, $0.3<z<0.5$, $0.4<z<0.6$ and $0.5<z<0.75$ 
for two distinct sky regions (North and South Galactic Caps); \Red{in this paper, as a demonstration, we decide to use the combined (i.e., CMASS plus LOWZ) sample in the middle redshift bin of North Galactic Cap (NGC) only, which corresponds to the mean redshift $z=0.51$.} To estimate the sample covariance matrix, we measure the power spectrum and bispectrum multipoles from all available $2048$ realizations\footnote{\url{https://www.sdss.org/dr12/}}, using estimators based on the Fast Fourier Transform (FFT) schemes (see \citealt{Bianchi2015MNRAS.453L..11B,Scoccimarro2015PhRvD..92h3532S,Hand:2017irw,Sugiyama:2017ggb} for the power spectrum and \citealt{Scoccimarro2015PhRvD..92h3532S,Slepian:2016qwa,Sugiyama2018} for the bispectrum. In particular, for details of how to measure the bispectrum multipoles~(\ref{Eq:PB_multipole}), see Section $4$ in ~\citealt{Sugiyama2018}). The $k$-range that we measure is $0.02\hk < k < 0.2\hk$ with $10$ bins; thus, the width between $k$-bins is $\Delta k =0.02\hk$.

Just for simplicity, we focus on the $k_1=k_2$ elements of the bispectrum multipoles $B_{\ell_1\ell_2L}(k_1,k_2)$. Then, the bispectrum multipoles are characterized by only one wavenumber $k$ like the power spectrum multipoles, which helps reducing computational time. As shown in \cite{Sugiyama2018}, the $k_1=k_2$ elements of the bispectrum multipoles dominate the signal-to-noise ratios for the monopole components, because the corresponding covariance matrix is nearly diagonal, like in the case of the power spectrum (see Figure~$4$ in~\citealt{Sugiyama2018}). Therefore, we suppose that studying the $k_1=k_2$ elements is a good approximation to investigate the impact of the bispectrum covariance on the signal-to-noise ratio.

The sample covariance matrix from the mocks are estimated as follows. Let $\VEC{X}$ and $\VEC{Y}$ be data vectors of measured quantities. The cross-covariance matrix of $\VEC{X}$ and $\VEC{Y}$ is then given by
\begin{eqnarray}
	\MAT{C}_{XY} = \frac{1}{N_{\rm mock}-1} \sum_r^{\rm N_{\rm mock}}
	\left( \VEC{X}^{(r)}- \overline{\VEC{X}} \right)^{\rm T}
	\left( \VEC{Y}^{(r)}- \overline{\VEC{Y}} \right),
\end{eqnarray}
where $N_{\rm mock}=2048$ is the number of the Patchy mocks, $\VEC{X}^{(r)}$ ($\VEC{Y}^{(r)}$) is the data vector obtained from the $r$-th mock, and the mean expectation value over the mocks $\overline{\VEC{X}}$ ($\overline{\VEC{Y}}$ ) is given by $\overline{\VEC{X}} = (1/N_{\rm mock})\sum_r^{N_{\rm mock}} \VEC{X}^{(r)}$. In the case of $\VEC{X}=\VEC{Y}$, we can estimate the auto-covariance of $\VEC{X}$. In this paper, we set the data vector $\VEC{X}$ ($\VEC{Y}$) to $\{P_{\ell}(k_i)\}$ ($\{P_{\ell'}(k_j)\}$) or $\{B_{\ell_1\ell_2L}(k_i,k_i)\}$ ($\{B_{\ell'_1\ell'_2L'}(k_j,k_j)\}$), where the indexes $i$ and $j$ run over the number of $k$-bins, i.e., $i, j=1,2,\dots,10$. Since the bispectrum multipoles that we compute in this paper depends only on one wavenumber, all covariance matrices we focus, i.e., ${\rm Cov}[P,P]$, ${\rm Cov}[P,B]$ and ${\rm Cov}[B,B]$, are functions of two wavenumbers $k_i$ and $k_j$, and the $ij$ elements of $\MAT{C}_{XY}$ are given by $[\MAT{C}_{XY}]_{ij} = C_{XY}(k_i, k_j)$. Given the covariance matrix, the $ij$ elements of the correlation coefficient matrix are defined as
\begin{eqnarray}
	[\MAT{r}_{XY}]_{ij} = \frac{ C_{XY}(k_i,k_j) }{ \sqrt{C_{XX}(k_i,k_i) C_{YY}(k_j,k_j) } }.
	\label{Eq:CorrelationMatrix}
\end{eqnarray}

The theoretical prediction of the covariance requires the values of the survey volume, $V$, and the mean number density, $\bar{n}=N/V$,
with the total number of galaxies within the survey area $N$.
We compute the mean number of galaxies in the mocks: $N = (1/N_{\rm mock})\sum_{r=1}^{N_{\rm mock}} N^{(r)} = 573\,012$,
where $N^{(r)}$ is the number of galaxies in the $r$-th mock catalogue.
One way to estimate the volume of the survey with a complicated geometry is to use the number density measured from a synthetic random catalogue with the same survey geometry as the galaxy sample,
$n_{\rm ran}(\VEC{x})$. Then, the survey volume can be computed by
\begin{eqnarray}
	 V = \frac{N_{\rm ran}^2}{\int d^3x\, [n_{\rm ran}(\VEC{x})]^2}
\end{eqnarray}
with the number of particles included in the random catalogue, $N_{\rm ran}$. In the case of the sample we use, the survey volume is $V=1.76\, [\hGpc]^3$; therefore, the number density is $\bar{n}=3.26\times 10^{-4}\, [\hMpc]^{-3}$.

The two parameters relevant to the amplitude of the power and bispectra, $b\sigma_8$ and $f\sigma_8$, are set to the expectation values for the Patchy mocks,  $b\sigma_8 =1.27$ and $f\sigma_8 =0.48$, where $b$ denotes the linear bias parameter, $\sigma_8$ the rms matter fluctuation on scales of $8\hMpc$ at a given redshift, and $f \sigma_8 = d \ln \sigma_8/ \ln a$ is the logarithmic growth rate. All of the parameters we use in this paper are summarized in Table~\ref{Table:parameters}.

\subsection{Results}

\begin{figure}
	\includegraphics[width=\columnwidth]{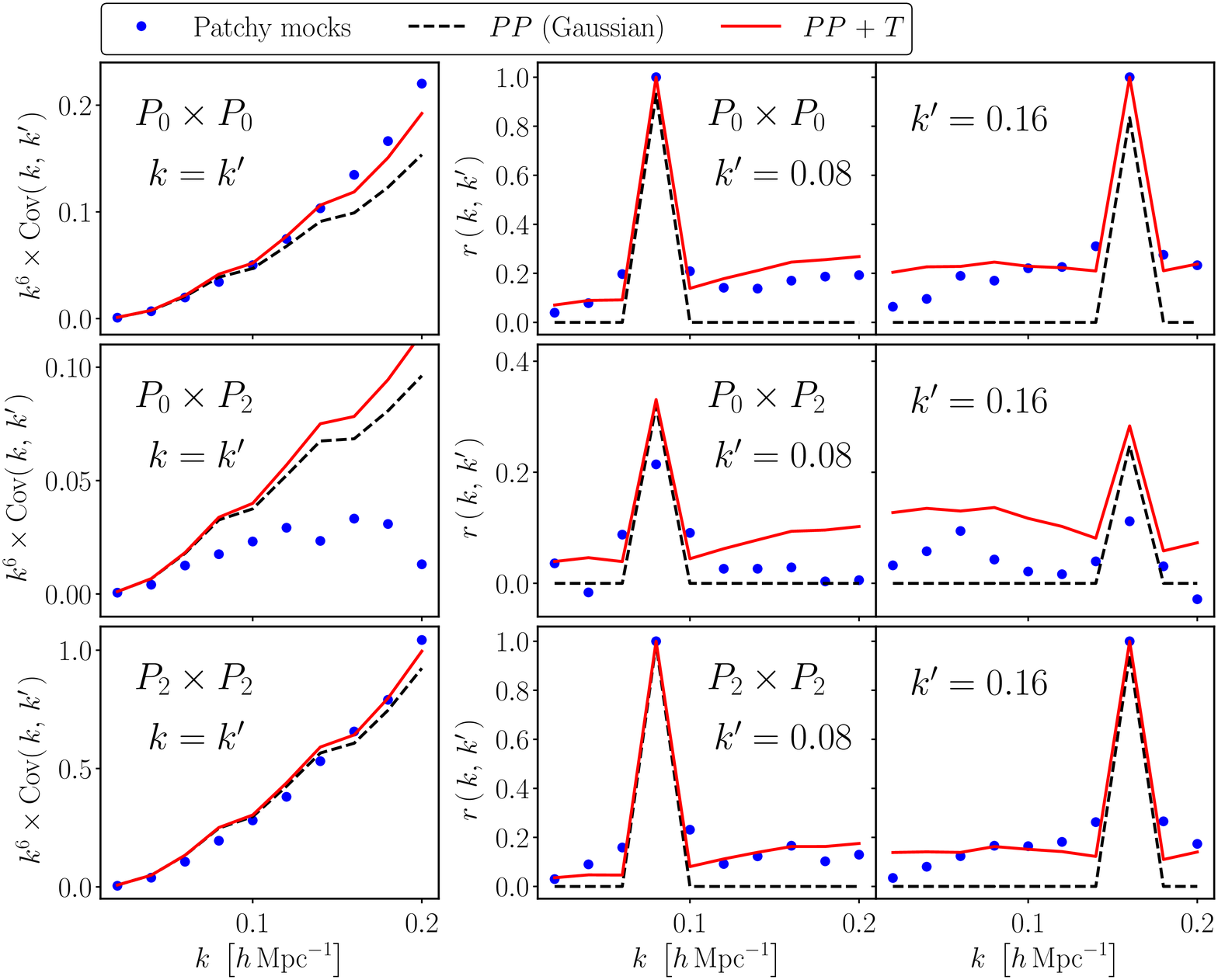}
	\caption{
	Comparison of the power spectrum multipole covariance matrices and the correlation coefficient matrices measured from the Patch mocks (blue symbols)
	with those computed by the perturbation theory (PT).
	Two PT terms, the Gaussian term (black dashed) and a full model with the trispectrum contribution (red solid), are shown.
	\textit{Left panels}: diagonal elements of three power spectrum covariances,
	${\rm Cov}\left[ P_0,P_0 \right]$, ${\rm Cov}\left[ P_0,P_2 \right]$
	and ${\rm Cov}\left[ P_2,P_2 \right]$ from top to bottom, are shown.
	The diagonal elements shown in the figure are multiplied by $k^6$ for display purposes.
	\textit{Middle and right panels}: the corresponding correlation coefficients between different wave vectors, $r(k,k')$, are shown as a function of $k$ 
	for a list of fixed $k' = 0.08\hk$ (middle) and $0.16\hk$ (right).
	The correlation coefficient from the Gaussian part is defined by dividing the Gaussian covariance by the full covariance model
	as given in Eq.~(\ref{Eq:r_PP_PP}).
	}
	\label{fig:covPP}
\end{figure}

\begin{figure}
	\includegraphics[width=\columnwidth]{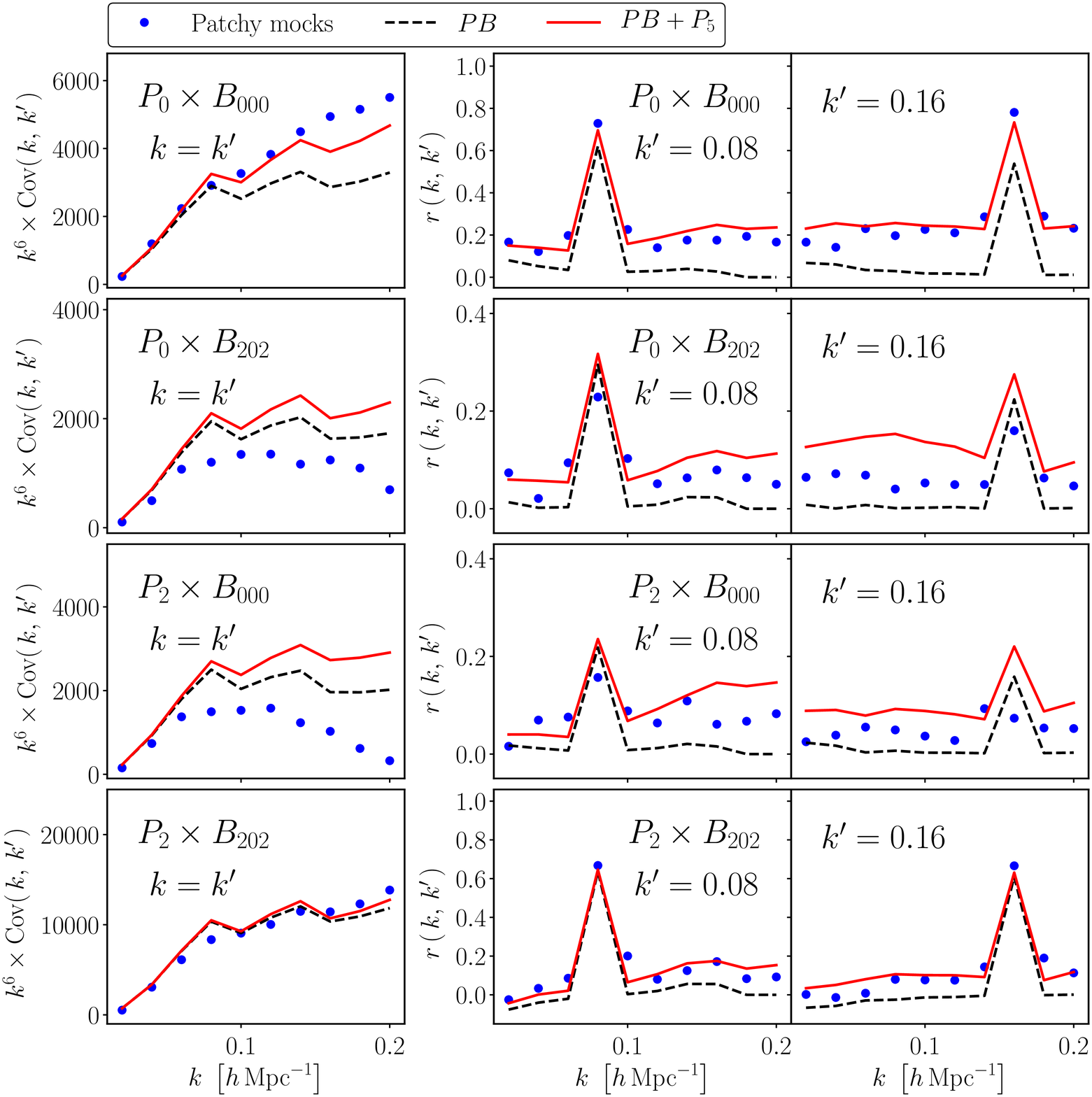}
	\caption{
		Similar plots to Figure~\ref{fig:covPP}.
		This figure shows the cross-covariance between the power spectrum and the bispectrum
		for both the monopole ($P_0$ and $B_{000}$) and the quadrupole ($P_2$ and $B_{202}$) components.
		From top to bottom, the four covariances 
		${\rm Cov}\left[ P_0, B_{000} \right]$, ${\rm Cov}\left[ P_0, B_{202} \right]$,
		${\rm Cov}\left[ P_2, B_{000} \right]$ and ${\rm Cov}\left[ P_2, B_{202} \right]$,
		and the corresponding correlation coefficients are shown.
		The diagonal elements shown in the left panels are multiplied by $k^6$ for display purposes.
		The $PB$ and $P_5$ contributions to the covariance matrix are plotted by black dashed lines and red solid lines, respectively,
		while the mock measurements are shown by blue points.
	}
	\label{fig:covPB}
\end{figure}

\begin{figure}
	\includegraphics[width=\columnwidth]{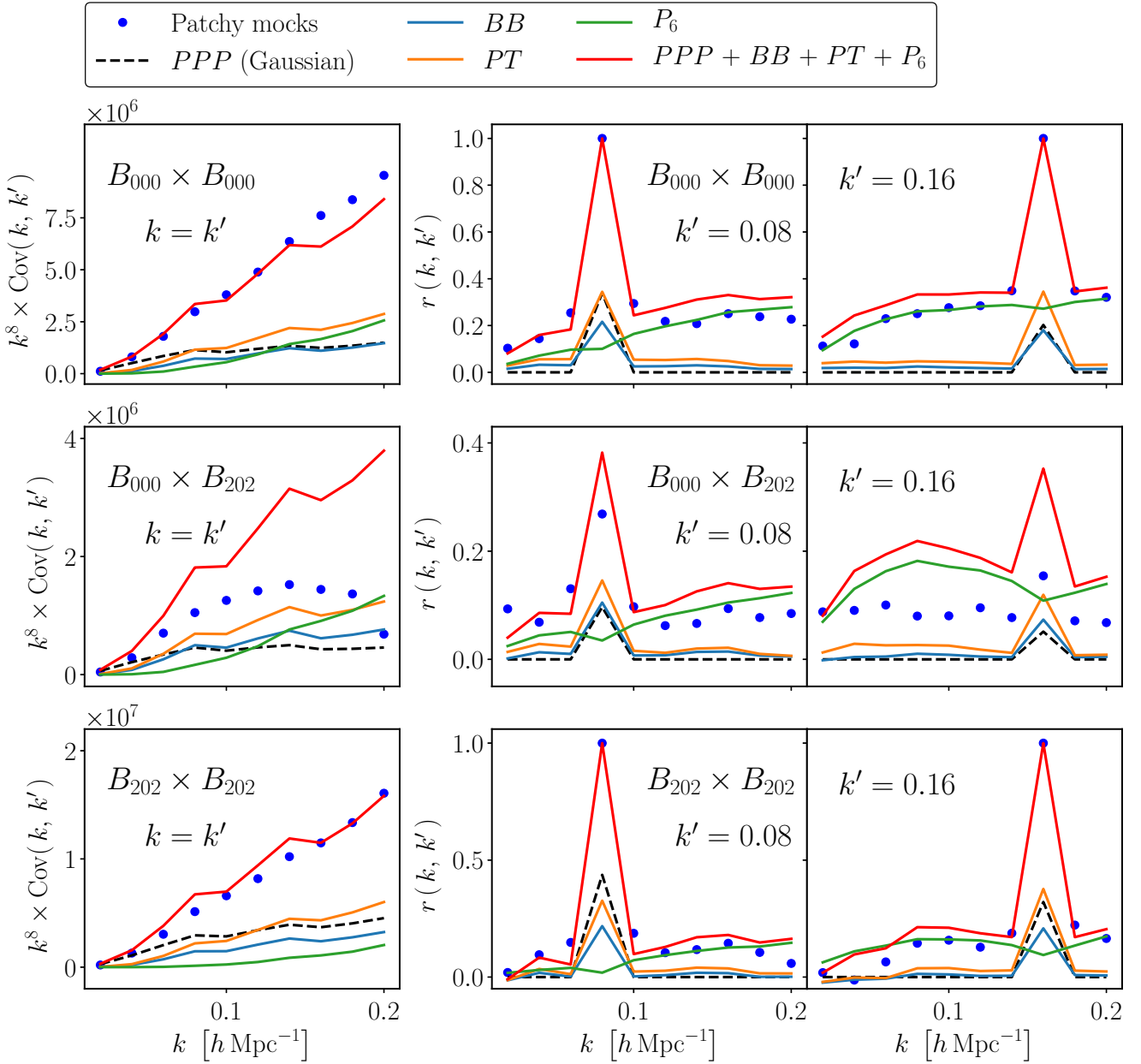}
	\caption{
		Similar plots to Figure~\ref{fig:covPP}.
		This figure shows the three cross-covariances,
		${\rm Cov}\left[ B_{000}, B_{000} \right]$, ${\rm Cov}\left[ B_{000}, B_{202} \right]$ and 
		${\rm Cov}\left[ B_{202}, B_{202} \right]$, and the corresponding correlation coefficients.
		The diagonal elements shown in the left panels are multiplied by $k^8$ for display purposes.
		The four sources to the bispectrum covariances,
		the Gaussian limit (black dashed), the $PT$ term (blue solid), the $BB$ term (orange solid) and
		the $6$th spectrum (green solid), and a full model summing up all four contributions (red solid) are shown.
	}
	\label{fig:covBB}
\end{figure}

\begin{figure}
	\includegraphics[width=\columnwidth]{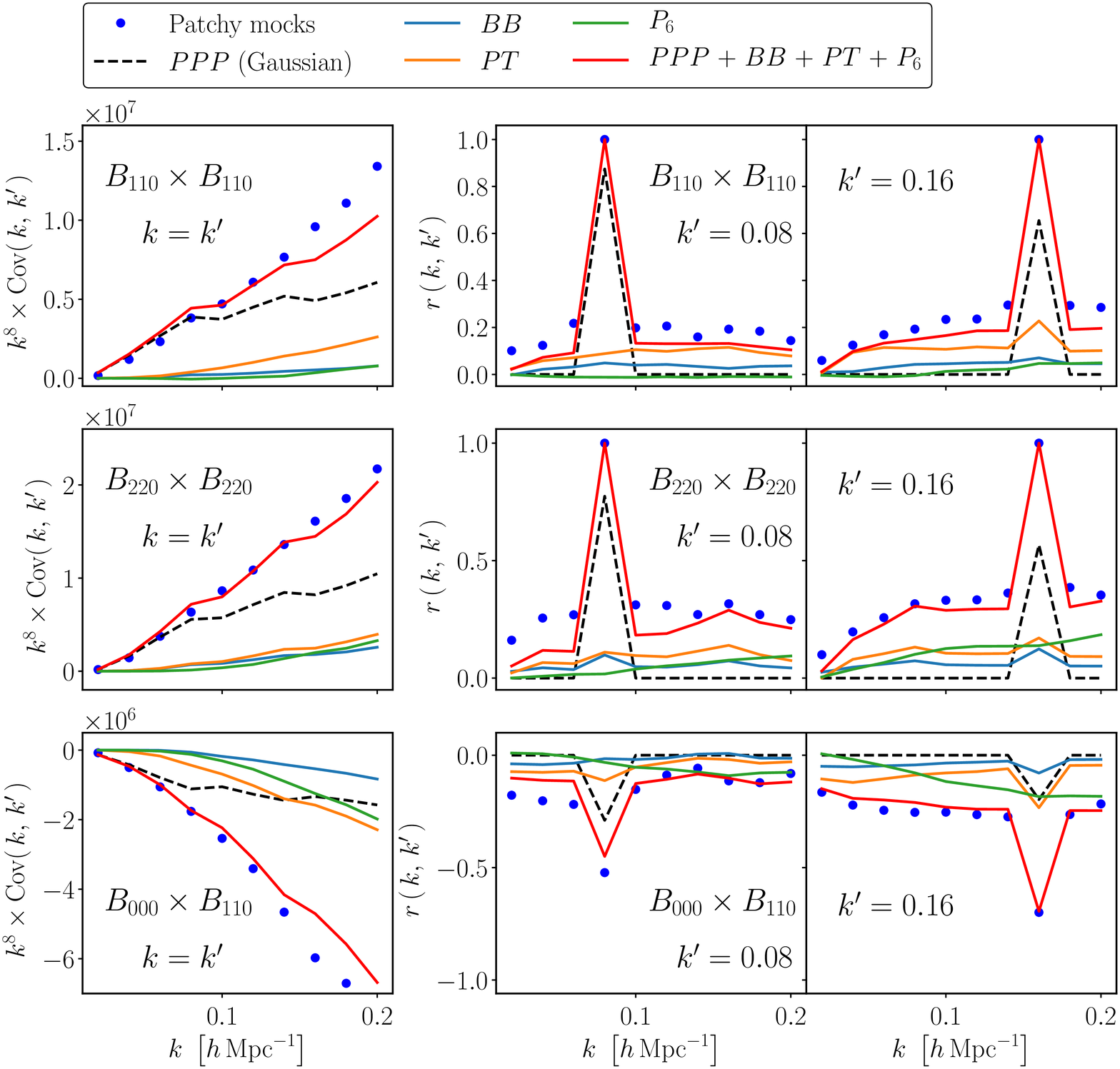}
	\caption{
		Similar plots to Figure~\ref{fig:covBB}.
		This figure focuses on higher multipole terms of the monopole bispectrum: namely, $B_{110}$, $B_{220}$.
		This figure shows 
		${\rm Cov}\left[ B_{110}, B_{110} \right]$, ${\rm Cov}\left[ B_{220}, B_{220} \right]$ and 
		${\rm Cov}\left[ B_{000}, B_{110} \right]$, and the corresponding correlation coefficients.
	}
	\label{fig:covBB_higher}
\end{figure}

\begin{figure}
	\includegraphics[width=\columnwidth]{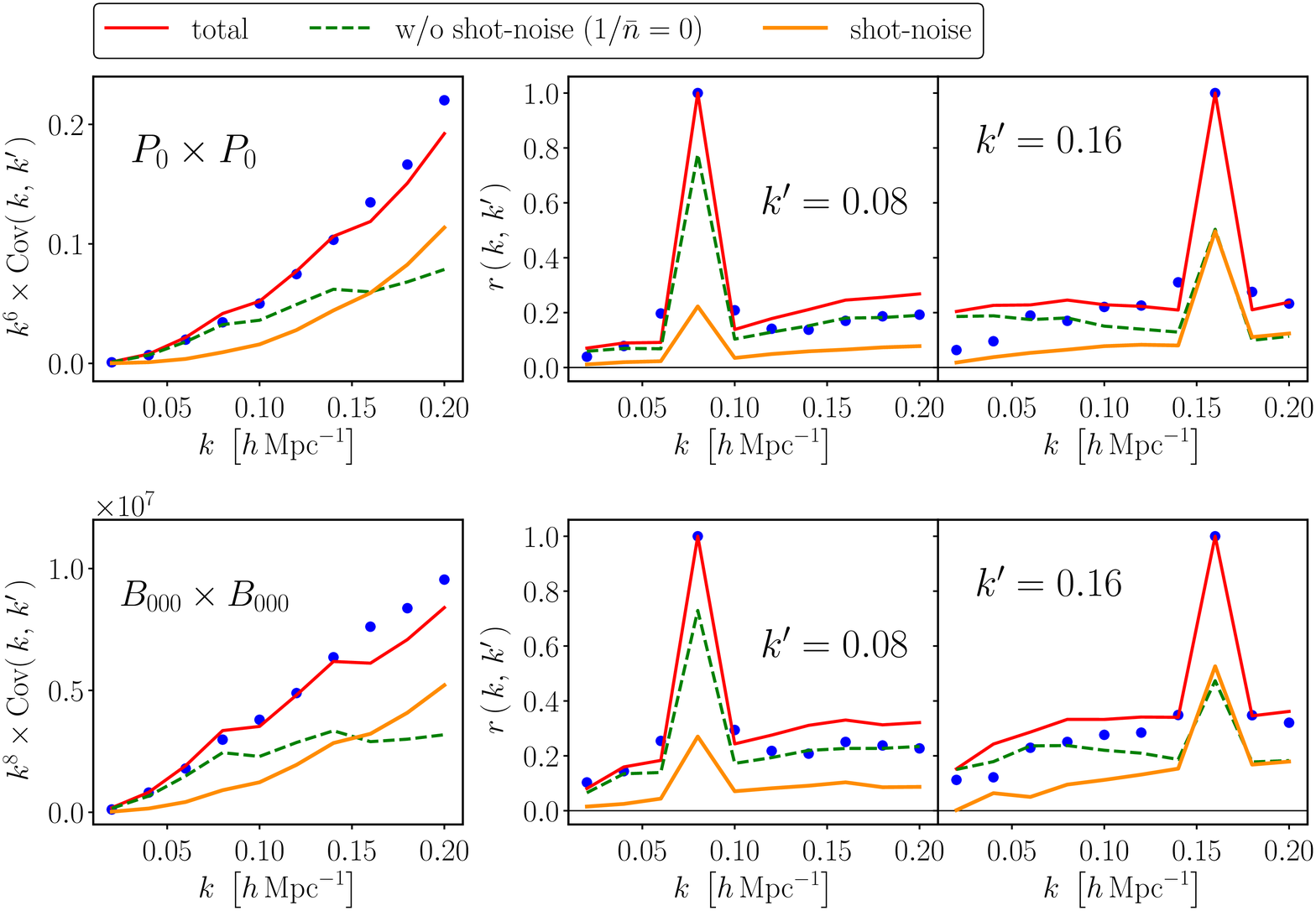}
	    \caption{The effect of shot noise.
		Upper and bottom panels are the same as the upper panels of Figure~\ref{fig:covPP} and Figure~\ref{fig:covBB},
		respectively,
		and the red solid lines and blue points are the same as those plotted in Figure~\ref{fig:covPP} and Figure~\ref{fig:covBB}.
		As additional lines,
		"non-shot noise contributions" (green dashed) versus "shot noise-only contributions" (orange solid) are shown.
		The shot-noise contributions dominate the covariance matrix on smaller scales than $k\sim 0.15\hk$ in the Patchy mocks corresponding to the BOSS survey.
		}
	\label{fig:cov_Shotnoise}
\end{figure}

We show the covariance matrices of the power spectrum multipoles in Figure~\ref{fig:covPP}. Each panel shows the Gaussian part (black dashed line, referred to as ``$PP$''), a full model adding the trispectrum contribution (red solid line, ``PP+T'') and the measurement from the Patchy mocks (blue points). The left panels of Figure~\ref{fig:covPP} show, from top to bottom, the diagonal elements of three covariances, ${\rm Cov}[P_0,P_0]$, ${\rm Cov}[P_0,P_2]$ and ${\rm Cov}[P_2,P_2]$, multiplied by $k^6$ for display purposes. The middle and right panels show the corresponding correlation coefficients. Since the correlation coefficients are characterized by two wavenumbers, $k$ and $k'$, we fix $k'$ to, e.g., $k'=0.08\hk$ and $k'=0.16\hk$ to plot the off-diagonal elements as a function of $k$; therefore, the peak positions at $k'=0.08\hk$ and $k'=0.16\hk$ correspond to the diagonal elements.

When computing the correlation coefficient matrix in perturbation theory, we want to know how each term in the covariance matrix, e.g., the Gaussian part ${\rm Cov}\left[ P,P \right]_{PP}$, contributes to the off-diagonal elements of the correlation coefficient matrix. To see that, we define the correlation coefficient matrix estimated only from the Gaussian part as the Gaussian part of the covariance matrix divided by the full covariance matrix including both the Gaussian and non-Gaussian parts. Namely, let $\MAT{C}\left[ PP \right]$ and $\MAT{C}\left[ PP+T \right]$ be the covariance matrices computed from the $PP$ (Gaussian) term and from the summation of the $PP$ and $T$ terms, respectively; then, the correlation coefficient matrix from the Gaussian part is given by
\begin{eqnarray}
	r_{ij}\left[ PP\right] = \frac{C_{ij}\left[ PP\right]}{\sqrt{C_{ii}\left[ PP+T \right]C_{jj}\left[ PP+T \right]}},
	\label{Eq:r_PP_PP}
\end{eqnarray}
which is plotted by the black dashed lines in the middle and right panels of Figure~\ref{fig:covPP}. Replacing $\MAT{C}\left[ PP \right]$ by $\MAT{C}\left[ PP+T \right]$ leads to the standard definition of the correlation matrix (\ref{Eq:CorrelationMatrix}), which is shown by red solid lines. Note that $\MAT{r}\left[ PP \right]$ defined this way no longer becomes unity at the diagonal element $k=k'$ even for the auto-covariances, ${\rm Cov}[P_0,P_0]$ and ${\rm Cov}[P_2,P_2]$. For the cross-covariance between the power and bispectra and the auto-covariance of the bispectrum, we also adopt a similar definition of the correlation coefficient and investigate the behavior of each part of the correlation coefficient using perturbation theory.

For the auto-covariances, ${\rm Cov}[P_0,P_0]$ and ${\rm Cov}[P_2,P_2]$, we find that the PT calculations are in remarkable agreement with the measurements in the Patchy mocks at about $20\%$ accuracy in the quasi-linear regime (up to $k \sim 0.2\hk$ at $z=0.51$). Surprisingly, the PT calculations can explain even the off-diagonal elements of the covariance matrix of ${\rm Cov}[P_0,P_0]$ (top right panels) and ${\rm Cov}[P_2,P_2]$ (bottom right panels), which arise from the trispectrum. On the other hand, for the cross-covariance ${\rm Cov}[P_0,P_2]$ (middle panels), we find that the PT calculations of the diagonal terms ($k=k'$) are larger than the Patchy mock results by a factor of about $1.5$ around $k\sim0.1\hk$ and get worse on smaller scales. We believe that this failure is caused by the fact that the cross-covariance between the monopole and the quadrupole arises mainly from quadrupole components, while the auto-covariances of both the monopole and the quadrupole are from the monopole components at the leading order. For instance, in the Gaussian part (Eq.~\ref{Eq:CovPP_G}), a product of two Legendre polynomials ${\cal L}_{\ell}(\mu) {\cal L}_{\ell'}(\mu)$ leads to ${\cal L}_0 {\cal L}_0 = {\cal L}_0$ for $(\ell,\ell')=(0,0)$, ${\cal L}_0 {\cal L}_2 = {\cal L}_2$ for $(\ell,\ell')=(0,2)$, and ${\cal L}_2 {\cal L}_2 = (1/5){\cal L}_0 + (2/7){\cal L}_2 + (18/35){\cal L}_4$ for $(\ell,\ell')=(2,2)$. Namely, the cross-covariance has the quadrupole component as the leading contribution, and should be sensitive to uncertainties on the velocity field, i.e., the RSD effect, while the auto-covariances is dominated mainly by the monopole term. \Red{The discrepancies of the cross-covariance could be explained by higher-order corrections relevant to the RSD effect, e.g., the Fingers-of-God effect, to some extent. The quadrupole power spectrum itself also becomes smaller than the corresponding mock measurement on small scales, as shown in Fig.~\ref{fig:pkbk} likely due to the FOG effect.}

Next we turn to the cross-covariance of the power spectrum and the bispectrum in Figure~\ref{fig:covPB}, where it shows the contributions from the $PB$ term (black dashed) and a full model adding the $5$-point spectrum, $P_5$ (red solid). From top to bottom, this figure shows ${\rm Cov}[P_0,B_{000}]$, ${\rm Cov}[P_0,B_{202}]$, ${\rm Cov}[P_2,B_{000}]$ and ${\rm Cov}[P_2,B_{202}]$, and the corresponding correlation coefficients between $k$ and $k'$. Similarly to the case of the power spectrum covariance, the PT calculations can well reproduce the Patchy mock results for both ${\rm Cov}[P_0,B_{000}]$ and ${\rm Cov}[P_2,B_{202}]$, and show that the monopole (quadrupole) power spectrum is strongly correlated with the monopole (quadrupole) bispectrum: their correlation coefficients at $k=k'$ are as large as $0.7$, that is, $70\%$ of the perfect correlation between $P_0(k)$ and $B_{000}(k)$. While the $PB$ term provides small contributions to the off-diagonal elements of the covariance matrix, the $P_5$ term dominates the off-diagonal elements. Because of similar reasons to the case of the power spectrum covariance, for ${\rm Cov}[P_0,B_{202}]$ and ${\rm Cov}[P_2,B_{000}]$, we find a significant departure of the PT calculations from the Patchy mock results on small scales.

We move on to the bispectrum covariance in Figure~\ref{fig:covBB}. As mentioned in Section~\ref{Sec:BB}, the bispectrum covariance has four sources:
the Gaussian term (black dashed, referred to as ``$PPP$''), the $PT$ term (blue solid, ``$PT$''), 
the $BB$ term (orange solid, ``$BB$'') and the $6$-point spectrum (green solid, ``$P_6$'').
From top to bottom, this figure shows 
${\rm Cov}[B_{000},B_{000}]$, ${\rm Cov}[B_{000},B_{202}]$ and ${\rm Cov}[B_{202},B_{202}]$,
and the corresponding correlation coefficients.
For diagonal elements of these covariances,
the non-Gaussian terms, $PT$, $BB$ and $P_6$, become comparable to or larger than the Gaussian term on quasi-nonlinear scales around $k\sim 0.1\hk$.
On the other hand, for the off-diagonal elements, the $P_6$ term becomes dominant,
and the $PP$, $PT$ and $BB$ terms are small so that they can be ignored.

The behavior of each contribution to the bispectrum covariance (the $PPP$, $BB$, $PT$ and $P_6$ contributions) 
found from Figure~\ref{fig:covBB} dramatically change for other multipole components 
of the bispectrum. For instance, we compute ${\rm Cov}\left[ B_{110}, B_{110} \right]$, ${\rm Cov}\left[ B_{220}, B_{220} \right]$ and ${\rm Cov}\left[ B_{000}, B_{110} \right]$ from top to bottom in Figure~\ref{fig:covBB_higher}.
For ${\rm Cov}\left[ B_{110}, B_{110} \right]$ and ${\rm Cov}\left[ B_{220}, B_{220} \right]$,
the Gaussian term dominates the diagonal elements, and 
the contributions from the $PT$ and $BB$ terms to the off-diagonal elements are comparable to or larger than the $P_6$ term.
The cross-covariance ${\rm Cov}\left[ B_{000}, B_{110} \right]$ yields negative correlations because of the negative signal of $B_{110}$ as shown in Figure~\ref{fig:pkbk},
and all contributions to the covariance, the $PPP$, $PB$, $BB$ and $P_6$ terms, 
are required to reproduce the diagonal and off-diagonal elements of the covariance matrix measured from the Patchy mocks.

Finally, to see how important the shot-noise effect is in the covariance estimates, we show, in Figure~\ref{fig:cov_Shotnoise}, ${\rm Cov}\left[ P_0,P_0 \right]$ and ${\rm Cov}\left[ B_{000},B_{000} \right]$ coming from all terms relevant to the shot-noise (orange solid line), e.g., the second term on the RHS of Eq.~(\ref{Eq:P_N}), and the second and third rows of Eq.~(\ref{Eq:T_N}) for $Cov[P,P]$, where we refer to them as the shot-noise terms. We compare them with the shot-noise independent terms, which are derived by setting $1/\bar{n}$ to zero in our analytical expressions shown in Section~\ref{Sec:CovarianceModel} (green dashed line). The shot-noise contribution dominates the covariances of both the power and bispectra on smaller scales than $k\sim0.15\hk$, as expected; at $k=0.2\hk$ they are about 1.5 times larger than the non-shotnoise contribution. For ${\rm Cov}[P_0,P_0]$, assuming the Gaussian limit, the scale where the shot-noise becomes dominant is $\sim0.18\hk$, which is smaller than $k=0.15\hk$ shown in Figure~\ref{fig:cov_Shotnoise}, indicating that the non-Gaussian shot-noise terms (the second and third rows of Eq.~(\ref{Eq:T_N})) play an important role even in diagonal elements of the power spectrum covaraince. For ${\rm Cov}[B_{000},B_{000}]$, the non-Gaussian shot-noise effect is more significant than the power spectrum case, because its diagonal elements are dominated by the non-Gaussian terms as shown in Fig.~\ref{fig:covBB}.

\section{Signal-to-noise ratio}
\label{Sec:SN}

\begin{figure}
	\includegraphics[width=\columnwidth]{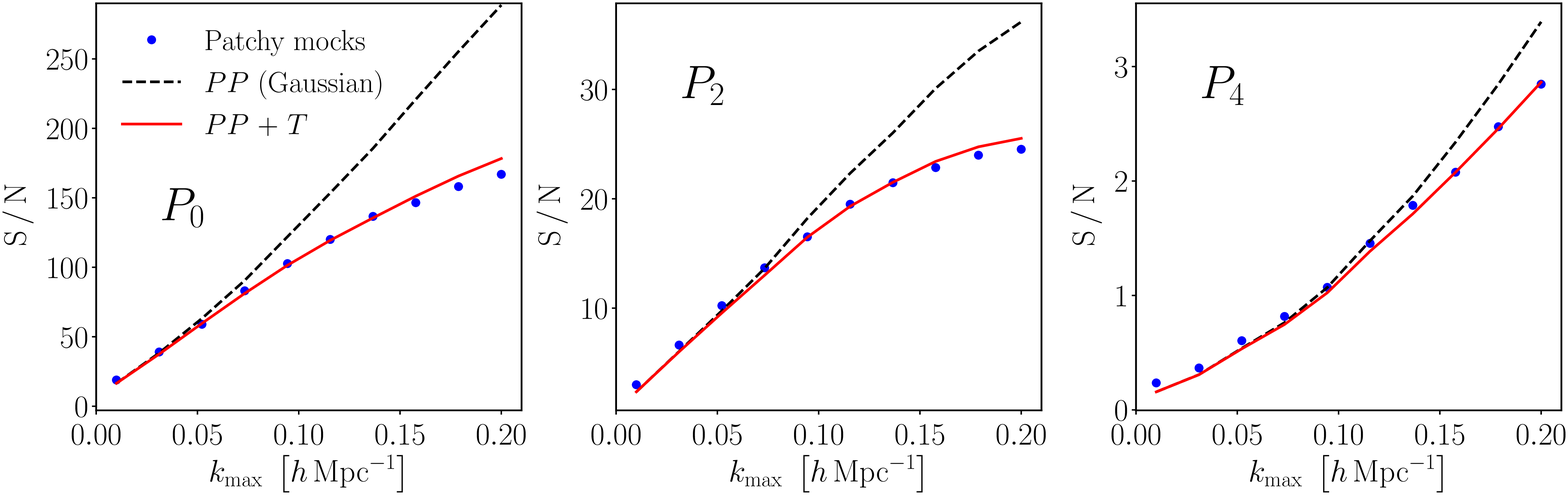}
	\includegraphics[width=\columnwidth]{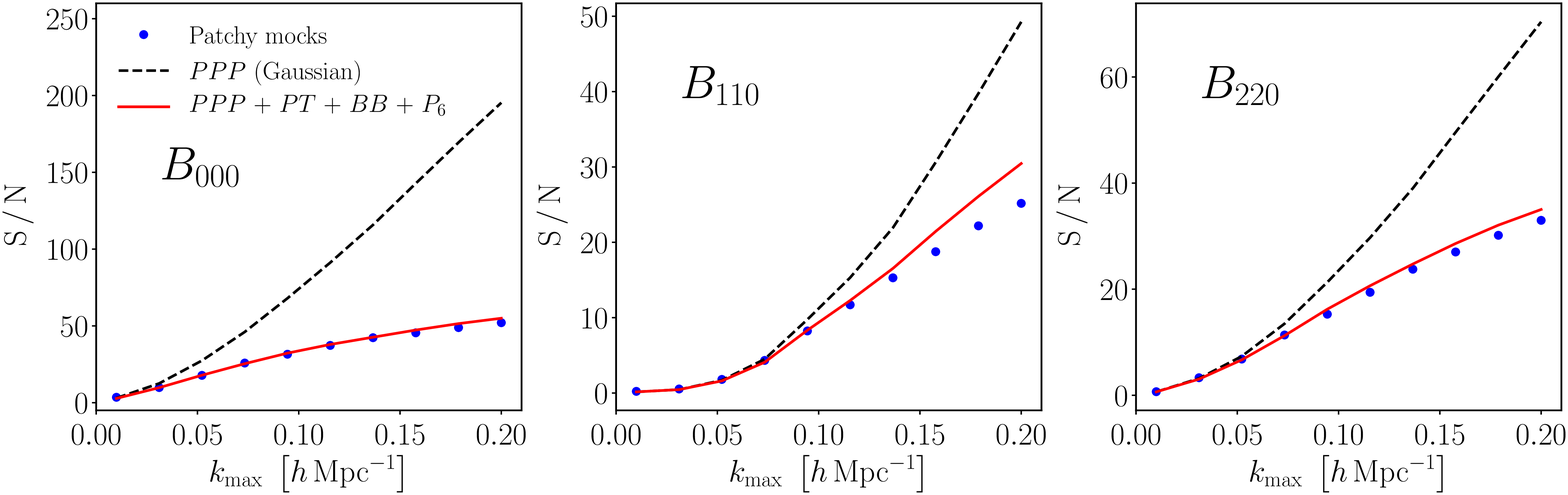}
	\includegraphics[width=\columnwidth]{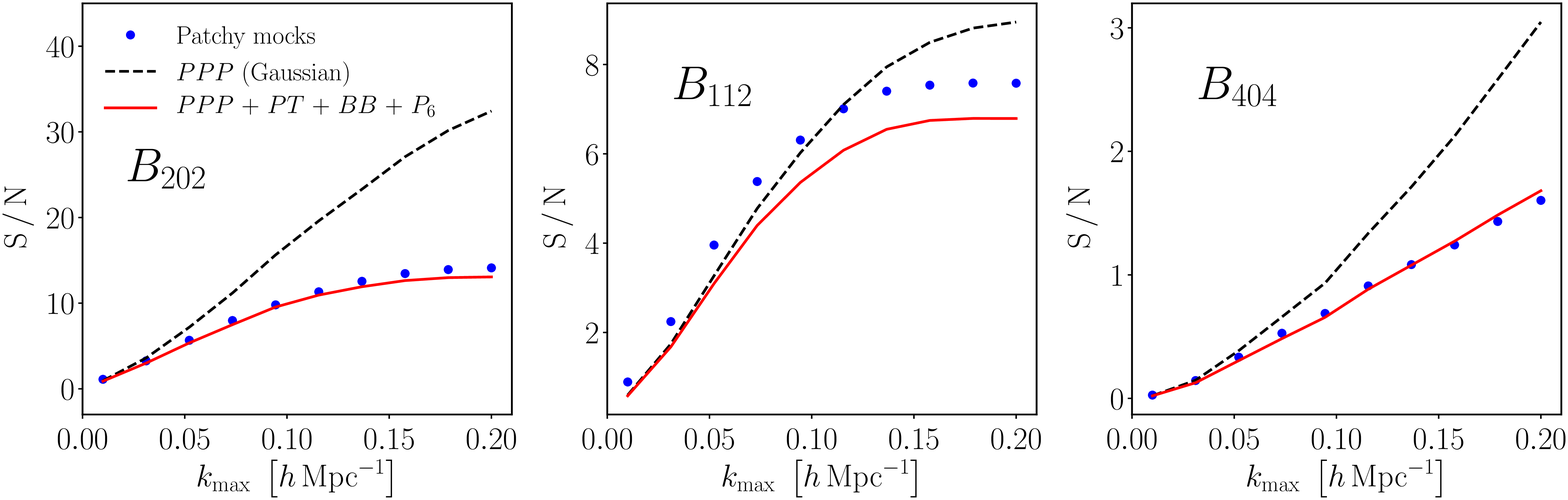}
	\caption{
	Cumulative signal-to-noise ratios as a function of $k_{\rm max}$ in redshift space for 
	the power spectrum and bispectrum multipoles, where the information over $0.01\hk \leq k \leq k_{\rm max}$ is included.
	The blue symbols show the ${\rm S/N}$s measured from the Patchy mocks,
	and the black dashed lines and the red solid lines show the PT calculations from the Gaussian part and the full model adding the non-Gaussian part, respectively.
	Note that for a fair comparison, we used the mean power and bispectrum multipoles measured from the Patchy mocks 
	as the signals to compute the ${\rm S/N}$ in both cases of the mock measurements and the PT calculations;
	therefore, the difference among the three predictions in each panel arises only from the difference of the covariance estimates.
	}
	\label{fig:SN}
\end{figure}

\begin{figure}
	\hspace{3.3cm}
	\includegraphics[width=0.55\columnwidth]{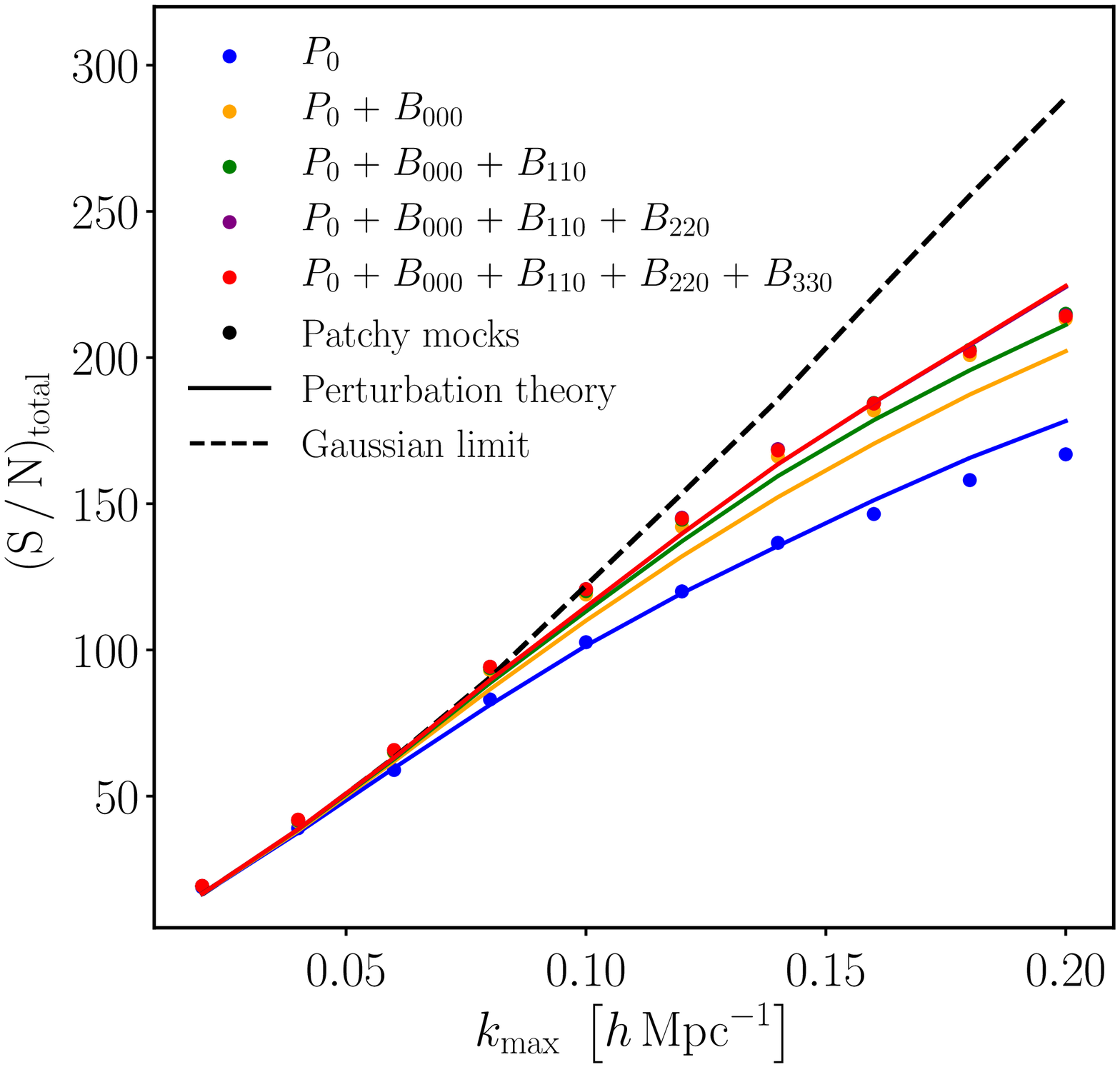}
	\caption{
		Total signal-to-noise ratio of the amplitude of the galaxy density flucuation as a function of $k_{\rm max}$,
		which jointly uses both the power spectrum and the bispectrum.
		The blue symbols, the blue line and the black dashed line are the same as 
		the blue symbols, the red line and the black dashed line plotted in upper left panel of Figure~\ref{fig:SN}, respectively.
		The orange, green, purple and red symbols progressively add one bispectrum multipole term.
		The colored solid lines show the corresponding PT predictions.
		In the Patchy mocks corresponding to the BOSS NGC survey, the total {\rm S/N} reproduces the Gaussian prediction up to $\sim 80\%$,
		while the power spectrum ${\rm S/N}$ reaches up to $\sim 60\%$.
		The total ${\rm S/N}$ well converses when compugin up to $B_{220}$.
	}
	\label{fig:SN_total}
\end{figure}

A useful way to compress and quantify the multi-dimensional offset between our analytical covariance model in Section~\ref{Sec:ComparisonWithMocks} and the covariance of the mocks is to estimate the cumulative signal-to-noise (${\rm S/N}$) ratios of the power spectrum and the bispectrum amplitudes. Such difference in the signal to noise of the amplitudes will propagate to the errors of the final cosmological parameters. Let $\MAT{C}_{XX}$ be the covariance of a data vector $\VEC{X}$; then, the cumulative S/N of $\VEC{X}$ is defined as
\begin{eqnarray}
	\left( {\rm \frac{S}{N} } \right)^2 = \VEC{X}^{T} \MAT{C}^{-1}_{XX} \VEC{X},
	\label{Eq:SN_X}
\end{eqnarray}
where $\MAT{C}^{-1}$ is the inverse of the covariance matrix after the covariance matrix is truncated at a maximum wavevector $k_{\rm max}$.
For example, the data vector $\VEC{X}$ is taken as the monopole power spectrum:
$\VEC{X} = \{P_0(k_1), P_0(k_2), \dots, P_0(k_{\rm max})\}$,
and usually the cumulative ${\rm S/N}$ is represented as a function of a given maximum wavenumber $k_{\rm max}$.
When we use the sample covariance matrix inferred from a set of the Patchy mocks,
the inverse covariance matrix is biased due to a finite number of realizations,
so we have to account for this effect by multiplying the inverse of the sample covariance matrix by the so-called 
Hartlap factor~\citep{Hartlap:2006kj}, $\left( N_{\rm mock}-N_{\rm bin}-2 \right)/\left( N_{\rm mock}-1 \right)$,
where $N_{\rm mock}$ and $N_{\rm bin}$ are the number of mock realizations and the number of data bins, respectively.
(see also \cite{Sellentin:2015waz} as a recent work.)

We plot the ${\rm S/N}$s of various quantities in Figure~\ref{fig:SN}: 
for the power spectrum, 
the ${\rm S/N}$s of the monopole ($P_0$), quadrupole ($P_2$) and hexadecapole ($P_4$) components are shown;
for the bispectrum, 
the ${\rm S/N}$s of three monopole components ($B_{000}$, $B_{110}$ and $B_{220}$), two quadrupole components ($B_{202}$ and $B_{112}$) 
and one hexadecapole component ($B_{404}$) are shown.
In each panel, 
we compare three predictions of the ${\rm S/N}$:
two PT calculations using the Gaussian covariance (black dashed line), which sets the maximum signal to noise available in the absence of nonlinearity, and the full model covariance with the non-Gaussian part (red solid line),
and the measurement from the Patchy mocks (blue points).
For a fair comparison of the ${\rm S/N}$s computed by the three different covariance matrices,
we use the same signal to compute the ${\rm S/N}$
by adopting the mean of the measurement of the power and bispecrum multipoles in the Patchy mocks as the signal.

Overall, the PT calculations including the non-Gaussian part are in excellent agreement with the mock results. In particular, for the ${\rm S/N}$s of $P_0$, $P_2$, $B_{000}$ and $B_{202}$, the PT calculations agree with the mock results within $10\%$ accuracy at $k_{\rm max}=0.2\hk$. For the S/Ns of $B_{110}$ and $B_{112}$, we find discrepancy between the PT calculations and the mock measurements at $k_{\rm max}=0.2\hk$ at a $\sim20\%$ level. It could be due to non-linear corrections such as gravitational clustering, RSDs and higher order biases. As expected, the non-Gaussian term always suppresses the ${\rm S/N}$ compared to the Gaussian prediction. This effect is of particular significance for the bispectrum case. For instance, for the lowest order of the monopole bispectrum components, $B_{000}$, the Gaussian covariance predicts the value of ${\rm S/N}=200$ at $k_{\rm max}=0.2\hk$, while the full covariance predicts ${\rm S/N}=50$ at the scale. Therefore, unless we correctly take account of the non-Gaussian part, we would overestimate the S/N of the bispectrum by a factor of about $4$ for the BOSS survey. The much more optimistic forecasts of the bispectrum analysis in the literature attribute likely to missing the non-Gaussian terms \cite[e.g.,][]{Sefusatti:2006pa,Sefusatti:2007ih,Gagrani:2016rfy,Tellarini:2016sgp,Yankelevich:2018uaz,Karagiannis:2018jdt} or to the complex dependence and degeneracy of cosmological parameters in the process of error propagation \cite[e.g.,][]{Sefusatti:2006pa}.

We finally conclude this section by computing the total signal-to-noise ratio estimated from the joint analysis of the power spectrum and the bispectrum.
In doing so, we should clarify what consistent signal we are extracting through this joint signal to noise analysis.
Since both the power spectrum and the bispectrum consist of the density perturbation $\delta$, in this paper we shall compute the ${\rm S/N}$ of the amplitude of the density perturbation;
in other words, we compute the detectability of the redshift-space density perturbation using the power spectrum and the bispectrum.
For this purpose, we introduce for notational convenience an amplitude parameter of the density fluctuation, $A_0$.
A formal definition of that parameter is $\delta(\VEC{k};A_0) = A_0\, \delta(\VEC{k})$, 
with the understanding that we work at the fiducial value $A_{0, \rm fid}=1$;
thus, the power spectrum and the bispectrum can be represented as 
$P(\VEC{k};A_0) = A_0^2\, P(\VEC{k})$ and $B(\VEC{k}_1,\VEC{k}_2;A_0) = A_0^3\, B(\VEC{k}_1,\VEC{k}_2)$, respectively.
The ${\rm S/N}$ of $A_0$ can be defined by the Fisher matrix:
\begin{eqnarray}
	\left( {\rm \frac{S}{N} } \right)^2_{A_0}
	= F_{\ln A_0 \ln A_0}  
	= \frac{\partial \VEC{X}^{T}}{\partial \ln A_0} \MAT{C}_{XX}^{-1}\frac{\partial \VEC{X}}{\partial \ln A_0}.
\end{eqnarray}
For instance, if we try to detect the galaxy density fluctuation only using the monopole power spectrum $P_0$,
we take the data vector $\VEC{X}$ as $\VEC{X} = \{A_0^2\, P_0(k_i)\}$ for $i=1,2,\dots$;
then, the ${\rm S/N}$ of $A_0$ is straightforwardly related to that of $P_0$ as $(\rm S/N)_{A_0}= 2\, (\rm S/N)_{P_0}$,
where the $({\rm S/N})_{P_0}$ is that plotted in the upper left panel of Figure~\ref{fig:SN}.
If the density fluctuation of galaxies was a purely Gaussian random field, $({\rm S/N})_{A_0}$ should be described by the Gaussian prediction of $({\rm S/N})_{P_0}$, where we ignore the quadrupole component because of its smallness.
When one wants to compute $({\rm S/N})_{A_0}$ through the joint analysis of $P_0$ and $B_{000}$,
the data vector is taken as $\VEC{X}_{P+B}=\{A_0^2\, P_0(k_i),\, A_0^3\, B_{000}(k_i,k_i)\}$ for $i=1,2,\dots$.
According to the above discussion, we finally define the total ${\rm S/N}$ as follows:
\begin{eqnarray}
	\left( {\rm \frac{S}{N} } \right)^2_{\rm total} &=&
	\frac{1}{4} \frac{\partial \VEC{X}_{\rm P+B}^{T}}{\partial \ln A_0} \MAT{C}_{\rm total}^{-1} \frac{\partial \VEC{X}_{\rm P+B}}{\partial \ln A_0},
\end{eqnarray}
where the pre-factor $(1/4)$ is for a direct comparison with the ${\rm S/N}$ of the power spectrum,
the total ${\rm S/N}$ reducing to $({\rm S/N})_{P_0}$ in the absence of the bispectrum.
The total covariance matrix 
$\MAT{C}_{\rm total}$ consists of the power spectrum auto-covariance $\MAT{C}_{PP}$, the bispectrum auto-covariance $\MAT{C}_{BB}$,
and the cross-covariances between the power and bispectra, $\MAT{C}_{PB}$ and $\MAT{C}_{BP}$, given by
\begin{eqnarray}
	\MAT{C}_{\rm total} = 
	\left( \begin{smallmatrix} \MAT{C}_{PP} & \MAT{C}_{PB} \\ \MAT{C}_{BP} & \MAT{C}_{BB} \end{smallmatrix}  \right).
	\label{Eq:SN_total}
\end{eqnarray}
The definition of the total ${\rm S/N}$ presented here is just one suggestion and other definitions may be used.
Of course, to correctly investigate how the bispectrum measurements improve the constraints on cosmological parameters,
we have to conduct a fisher analysis for the cosmological parameters that we want to know,
but we leave this topic for future work.

Figure~\ref{fig:SN_total} shows how the bispectrum measurements reproduce the Gaussian prediction of the ${\rm S/N}$ (black dashed line).
In particular, to see how higher multipoles of the monopole bispectrum, $B_{110}$, $B_{220}$ and $B_{330}$,
contribute to the total ${\rm S/N}$, we have added them in turn, which are plotted in different colors.
We find from the mock results that adding the lowest multipole of the monopole bispectrum $B_{000}$ increases the total ${\rm S/N}$,
but higher multipoles do not.
In the PT calculations, $B_{110}$ and $B_{220}$ contribute to the total ${\rm S/N}$ a little bit,
but it could be due to the failure of the PT calculations.
In both cases, we can conclude that the dominant contribution to the total ${\rm S/N}$ comes from $P_0$ and $B_{000}$,
which implies that the bispectrum cosmological information such as the amplitude of the density field is mainly included in the lower multipole 
or a few first multipoles of the decomposed bispectrum via Eq.~(\ref{Eq:PB_multipole}) \cite[see also][]{Gagrani:2017aa}.
Finally, we note that adding the bispectrum information does not completely reproduce the Gaussian information, and the associated total ${\rm S/N}$ is about $1.4$ times smaller than the Gaussian prediction at $k=0.2\hk$ in the Patchy mocks. We note that we have used here only the $k_1=k_2$ elements of the bispectrum multipoles to compute the ${\rm S/N}$. As shown in~\cite{Sugiyama2018}, adding the $k_1\neq k_2$ elements however increases the total ${\rm S/N}$ by $\sim 10\%$. This result would motivate to study higher order statistics, e.g., the trispectrum, for extracting the full information on the galaxy clustering~\cite[e.g.,][]{Carron:2015hha}.

\section{Discussion and conclusions}
\label{Sec:Conclusions}

In this paper, using perturbation theory, we have developed analytical models for the auto-covariance matrices of both the power spectrum and the bispectrum, as well as for the cross covariance between the power spectrum and the bispectrum, including the full non-Gaussian parts up to the 6-point function, the RSD effect, the linear bias and the shot-noise corrections. As we use the tree-level solutions of the standard perturbation theory, the PT calculations presented here are not involved with any fitting functions and parameters that should be calibrated by $N$-body simulations and depend only on the standard cosmological parameters and the linear bias parameter. The PT calculations have been validated by comparing them with the sample covariance matrices measured from the Patchy mocks~\citep{Klypin:2014kpa,Kitaura:2015uqa} corresponding to the BOSS North Galactic Cap in the redshift range of $0.4<z<0.6$.

The anisotropic signal along the LOS direction, which is induced by the RSD or AP effect, is of crucial importance in the cosmological analysis of galaxy redshift surveys. We have decomposed the power spectrum and the bispectrum into multipole components regarding the angle relevant to the LOS direction. In particular, for the bispectrum we have adopted the decomposition formalism proposed by \citet{Sugiyama2018} which is based on the tri-polar spherical harmonics. In Figures~\ref{fig:covPP}, \ref{fig:covPB} and \ref{fig:covBB}, we have focused especially on the monopole power spectrum ($P_0$), the quadrupole power spectrum ($P_2$), the leading order of the monopole bispectrum ($B_{000}$) and the leading order of the quadrupole bispectrum ($B_{202}$). For the covariance matrices associated with the monopole components, ${\rm Cov}\left[ P_0,P_0 \right]$ and ${\rm Cov}\left[ P_0,B_{000} \right]$, and with the quadrupole components, ${\rm Cov}\left[ P_2,P_2 \right]$ and ${\rm Cov}\left[ P_2,B_{202} \right]$, \Red{we find that our PT calculation reasonably explains the mock results.} Consequently, the computed signal-to-noise ratios of $P_0$, $P_2$, $B_{000}$ and $B_{202}$ from the PT calculations agree with those estimated from the Patchy mocks within $10\%$ accuracy at $k_{\rm max}=0.2\hk$ (Figure~\ref{fig:SN}). We thus believe that the analytic approach provided in this paper greatly advances our theoretical understanding of the covariance matrices for the power spectrum and the bispectrum, and will be useful for other forthcoming galaxy surveys.

One may wonder why our simple treatment (tree-level solution with the linear bias, no FOG suppression etc) works so well even for the bispectrum covariance as well as for the power spectrum covariance. We believe that our careful treatment of shot-noise terms on the covariance matrix provides a partial explanation. For the first time, we have derived the covariance matrices of the power spectrum and bispectrum measurements \textit{after} the shot-noise subtraction (Section~\ref{Sec:CovarianceModel}). 
\Red{Since we commonly measure the power spectrum and the bispectrum with the shot-noise removed, we should estimate the corresponding covariance matrix; otherwise, we will overestimate the covariance matrix from the PT calculations due to the super-sample effect associated with the shot-noise term (Section~\ref{subsec:shot-noise}). To estimate the shot noise in our PT calculation, however, we adopt the global mean number density, i.e., the mean of the total 2048 realizations. Then we compare it with the Patchy mock result for which we use the mean number density estimated differently for each realization,i.e., the local mean number density, as we would do for a real data. The effect of this discrepancy corresponds to  the SSC terms that we ignore in this paper and left for a future study.} From such treatment of the shot-noise effect on the covariance matrix, we find that the shot-noise term becomes larger than the non shot-noise term in the covariance on smaller scales than $k\sim0.15\hk$ for both cases of the power spectrum and the bispectrum. At $k=0.2\hk$, on the other hand, the shot-noise term is about 1.5 times larger than the covariance without the shot-noise term (Figure~\ref{fig:cov_Shotnoise}). This result implies that the covariance matrix of the galaxy clustering is not dominated on small scales by higher-order nonlinear corrections, e.g., loop integrals or higher order biases, but rather by the shot-noise contributions. Clearly, we caution that this finding may be the case only for a BOSS-like survey configuration with $\bar{n}\sim 3\times10^{-4}$ and $b\sim 2$ at $0.2<z<0.75$. Therefore, it will be important to revisit the impact of the shot-noise term to the covariance matrix for future surveys, e.g., for Emission line galaxies measured from PFS and DESI, which will have higher number density and lower linear bias at higher redshifts.

Despite of the overall success, we have identified three concerns with our PT calculations. First, for the cross-covariance matrices between the monopole and quadrupole components, ${\rm Cov}\left[ P_0,P_2 \right]$, ${\rm Cov}\left[ P_0, B_{202} \right]$, ${\rm Cov}\left[ P_2,B_{000} \right]$ and ${\rm Cov}\left[ B_{000},B_{202} \right]$, the PT calculations start to depart from the mock results at $k\sim0.1\hk$. This could be due to the lack of higher order corrections in perturbation theory such as the FOG effect. Second, non-linear bias effects may become dominant for galaxy samples different from the sample used in this paper, because they should be included in the non-Gaussian terms even in the tree-level solution. Third, we have not taken account of any window function effect on the covariance calculations. In particular, the finite-sized survey window would generate an additional contribution to the covariance, the so-called super-sample covariance (e.g., \citealt{Takada:2013bfn}). If one can resolve these problems, the PT approach will be established as an alternative way to estimate the covariance matrix and will allow us to investigate the parameter-dependence of the covariance matrix in the cosmological analysis. \Red{While the PT approach is relevant for spectroscopic surveys of large-scale structure, entirely different methods of covariance calculation will be needed for cosmological probes such as weak lensing and galaxy cluster number counts; e.g., \citet{Hikage:2018qbn} used the halo model approach to calculate the covariance matrix for the weak lensing shear analysis from the Hyper Suprime-Cam (HSC) survey~\citep{Aihara:2017paw}. \citet{Takada:2007fq} investigates the covariance relevant to the cluster counts also based on the halo model approach.}

The obvious next step is to present a realistic forecast of cosmological parameters from the joint analysis of the power spectrum and the bispectrum using the covariance estimates from our PT calculations. As shown in Figure~\ref{fig:SN}, the non-Gaussian part of the bispectrum covariance strongly suppresses the corresponding signal-to-noise ratio; e.g., the ${\rm S/N}$ of $B_{000}$ including the non-Gaussian part is about $4$ times smaller than the ${\rm S/N}$ computed in the Gaussian limit at $k=0.2\hk$. Furthermore, non-Gaussian effects generate high correlation between the power spectrum and the bispectrum, where the correlation coefficient reaches up to $0.7$ (Figure~\ref{fig:covPB}). Therefore, the constraining power of the bispectrum would be overestimated unless we correctly account for the non-Gaussian covariance. As a caveat, such finding is based on the signal to noise of the clustering amplitude.  Estimating the effect of the non-Gaussian covariance on the final cosmological parameters requires the propagation of this signal to noise and might lead to a different conclusion due to complex parameter dependence and degeneracy (e.g., \citealt{Sefusatti:2006pa}). Work on a realistic forecast for cosmological parameters using the result in this paper is in progress and will be reported soon.

\section*{Acknowledgements}

NSS. acknowledges financial support from Grant-in-Aid for JSPS Fellows (No. 28-1890). 
This work was supported in part by JSPS KAKENHI Grant Number JP15H05896, JP16J01890 and 19K14703,
and by World Premier  International  Research  Center  Ini-tiative (WPI Initiative), MEXT, Japan.
Numerical computations were carried out on Cray XC50 at Center for Computational Astrophysics, National Astronomical Observatory of Japan. H.-J.S.~are supported by the U.S.~Department of Energy, Office of Science, Office of High Energy Physics under DE-SC0019091. FB is a Royal Society University Research Fellow.

\section*{Data availability}
The data underlying this article are available at the SDSS data base (\url{https://www.sdss.org/dr12/}).

\bibliographystyle{mnras}
\bibliography{ms} 

\appendix
\section{Perturbation theory}
\label{Ap:PT}

In this appendix, we show analytical expressions of galaxy statistics used in this paper in perturbation theory: namely,
the power spectrum, bispectrum, trispectrum, 5th-spectrum and 6-th spectrum.

The redshift-space galaxy (halo) density fluctuation in Fourier space can be formally expressed as
\begin{eqnarray}
        \delta(\VEC{k}) = \int d^3x e^{-i\VEC{k}\cdot\VEC{x}}\left[  e^{-i\VEC{k}\cdot\hat{n} \left( \frac{\VEC{v}(\VEC{x})\cdot\hat{n}}{aH} \right)}
        \left( 1 + \delta_{\rm real}(\VEC{x}) \right)- 1\right],
\end{eqnarray}
where $\VEC{v}$ is the physical peculiar velocity of galaxies, $\hat{n}$ is the LOS direction, and $\delta_{\rm real}(\VEC{x})$ is the real-space galaxy density fluctuation. Since we only take account of the linear bias parameter to describe the galaxy density fluctuation in this paper,
$\delta_{\rm real}$ is replaced by $b_1 \delta_{\rm m}$ with $\delta_{\rm m}$ being the matter density fluctuation.
The standard perturbation theory expands the density perturbation in terms of the linear matter density perturbation, $\delta_{\rm lin}$:
\begin{eqnarray}
        \delta(\VEC{k}) = \int\frac{d^3p_1}{(2\pi)^3}\cdots\int\frac{d^3p_n}{(2\pi)^3}
        (2\pi)^3\delta_{\rm D}\left( \VEC{k} - \VEC{p}_{1n} \right)
        Z_n(\VEC{p}_1,\dots,\VEC{p}_n) \delta_{\rm lin}(\VEC{p}_1) \cdots \delta_{\rm lin}(\VEC{p}_n)
\end{eqnarray}
where $\VEC{p}_{1n} = \VEC{p}_1+\dots+\VEC{p}_n$,
and $Z_n$ means the $n$-th order kernel function describing non-linear corrections to the density fluctuation~\citep{Scoccimarro:1999ed}.
In this appendix, we omit the LOS-dependence on all quantities that we compute:
e.g., the kernel function $Z_n(\VEC{p}_1,\dots,\VEC{p}_n,\hat{n})$ with the LOS is represented just as $Z_n(\VEC{p}_1,\dots,\VEC{p}_n)$

Using the kernel function $Z_n$, the linear redshift-space power spectrum is given by
\begin{eqnarray}
        P(\VEC{k}) = \left[Z_1(\VEC{k}) \right]^2 P_{\rm lin}(k),
\end{eqnarray}
where the first order kernel function $Z_1$ corresponds to the Kaiser factor~\citep{Kaiser1987MNRAS.227....1K}, 
and the linear matter power spectrum is given by
\begin{eqnarray}
        \langle \delta_{\rm lin}(\VEC{k})\delta_{\rm lin}(\VEC{k}') \rangle
        = (2\pi)^3\delta_{\rm D}(\VEC{k}+\VEC{k}') P_{\rm lin}(k).
\end{eqnarray}
The bispectrum consists of one second order density perturbation and two linear density perturbations, given by
\begin{eqnarray}
     B(\VEC{k}_1,\VEC{k}_2,\VEC{k}_3) = 2Z_1(\VEC{k}_1)Z_1(\VEC{k}_2)Z_2(\VEC{k}_1,\VEC{k}_2) P_{\rm lin}(k_1) P_{\rm lin}(k_2) + \mbox{(2 perms.)}
\end{eqnarray}
The trispectrum has two sources 
\begin{eqnarray}
    T(\VEC{k}_1,\VEC{k}_2,\VEC{k}_3,\VEC{k}_4) = 4\, T_{2211}(\VEC{k}_1,\VEC{k}_2,\VEC{k}_3,\VEC{k}_4) 
    + 6\, T_{3111}(\VEC{k}_1,\VEC{k}_2,\VEC{k}_3,\VEC{k}_4),
\end{eqnarray}
where $T_{2211}$ consists of two second order and two linear density perturbations,
and $T_{3111}$ one third order and three linear density fluctuations, and they are given by
\begin{eqnarray}
    T_{2211}(\VEC{k}_1,\VEC{k}_2,\VEC{k}_3,\VEC{k}_4) &=& Z_1(\VEC{k}_1)Z_1(\VEC{k}_2)Z_2(-\VEC{k}_1,\VEC{k}_{14})Z_2(-\VEC{k}_2,\VEC{k}_{23})
    P_{\rm lin}(k_1) P_{\rm lin}(k_2) P_{\rm lin}(k_{14})+ \mbox{(11 perms.)} \nonumber \\
     T_{2211}(\VEC{k}_1,\VEC{k}_2,\VEC{k}_3,\VEC{k}_4) &=&Z_1(\VEC{k}_1)Z_1(\VEC{k}_2)Z_1(\VEC{k}_3) Z_3(\VEC{k}_1,\VEC{k}_2,\VEC{k}_3)
     P_{\rm lin}(k_1) P_{\rm lin}(k_2) P_{\rm lin}(k_3) + \mbox{(3 perms.)}
\end{eqnarray}
with $\VEC{k}_{14} = \VEC{k}_1+\VEC{k}_4$ and $\VEC{k}_{23}=\VEC{k}_2+\VEC{k}_3$. Similarly, the $5$-th spectrum has three sources:
\begin{eqnarray}
    P_5(\VEC{k}_1,\VEC{k}_2,\VEC{k}_3,\VEC{k}_4,\VEC{k}_5) = 
    8\, T_{22211} + 12\, T_{32111} + 24\, T_{41111}
\end{eqnarray}
where
\begin{eqnarray}
    T_{22211} &=& Z_1(\VEC{k}_1)Z_1(\VEC{k}_2) Z_2(-\VEC{k}_1,\VEC{k}_{15}) Z_2(-\VEC{k}_2,\VEC{k}_{23}) Z_2(\VEC{k}_{23},\VEC{k}_{15}) + \mbox{(39 perms.)}
    \nonumber \\
    T_{3211} &=&Z_1(\VEC{k}_1) Z_1(\VEC{k}_2) Z_1(\VEC{k}_3) Z_2(-\VEC{k}_1,\VEC{k}_{15}) Z_3(\VEC{k}_2,\VEC{k}_3,\VEC{k}_{15}) P_{\rm lin}(k_1) P_{\rm lin}(k_2) P_{\rm lin}(k_3) P_{\rm lin}(k_{15})
         + \mbox{(35 perms.)} \nonumber \\
    T_{4111} &=& Z_1(\VEC{k}_1) Z_1(\VEC{k}_2) Z_1(\VEC{k}_3)Z_1(\VEC{k}_4)Z_4(\VEC{k}_1,\VEC{k}_2,\VEC{k}_3,\VEC{k}_4)
         P_{\rm lin}(k_1) P_{\rm lin}(k_2) P_{\rm lin}(k_3) P_{\rm lin}(k_4) 
         + \mbox{(4 perms.)} \\
\end{eqnarray}
Finally, the 6th spectrum is represented as
\begin{eqnarray}
    P_6(\VEC{k}_1,\VEC{k}_2,\VEC{k}_3,\VEC{k}_4,\VEC{k}_5,\VEC{k}_6) 
    = 16\,T_{222211} + 24\, T_{322111a} + 24\, T_{322111b} + 36\, T_{331111} + 48\, T_{42111},
\end{eqnarray}
where
\begin{eqnarray}
	T_{222211} &=&  Z_1(\VEC{k}_1)Z_1(\VEC{k}_2)Z_2(-\VEC{k}_1,\VEC{k}_{13})Z_2(-\VEC{k}_2,\VEC{k}_{24})Z_2(\VEC{k}_{13},\VEC{k}_{246})Z_2(\VEC{k}_{24},\VEC{k}_{135})  \nonumber \\
        &\times& P_{\rm lin}(k_1)P_{\rm lin}(k_2)P_{\rm lin}(k_{13})P_{\rm lin}(k_{24})P_{\rm lin}(k_{135})  + \mbox{(359 perms.)} \nonumber \\
	T_{322111a} &=& Z_1(\VEC{k}_1)Z_1(\VEC{k}_2)Z_1(\VEC{k}_{3})Z_2(-\VEC{k}_3,\VEC{k}_{34})Z_2(\VEC{k}_{34},\VEC{k}_{126})Z_3(-\VEC{k}_1,-\VEC{k}_2,\VEC{k}_{126})  \nonumber \\
        &\times& P_{\rm lin}(k_1)P_{\rm lin}(k_2)P_{\rm lin}(k_{3})P_{\rm lin}(k_{34})P_{\rm lin}(k_{126}) + \mbox{(359 perms.)} \nonumber \\
	T_{322111b} &=& Z_1(\VEC{k}_1)Z_1(\VEC{k}_2)Z_1(\VEC{k}_{3})Z_2(-\VEC{k}_3,\VEC{k}_{34})Z_2(-\VEC{k}_2,\VEC{k}_{25})Z_3(\VEC{k}_1,\VEC{k}_{34},\VEC{k}_{25})  \nonumber \\
        &\times& P_{\rm lin}(k_1)P_{\rm lin}(k_2)P_{\rm lin}(k_{3})P_{\rm lin}(k_{34})P_{\rm lin}(k_{25})  + \mbox{(359 perms.)} \nonumber \\
	T_{331111} &=& Z_1(\VEC{k}_1)Z_1(\VEC{k}_2)Z_1(\VEC{k}_{3})Z_1(\VEC{k}_4)Z_3(-\VEC{k}_3,-\VEC{k}_4,\VEC{k}_{345})Z_3(-\VEC{k}_1,-\VEC{k}_2,\VEC{k}_{126}) \nonumber \\
        &\times& P_{\rm lin}(k_1)P_{\rm lin}(k_2)P_{\rm lin}(k_{3})P_{\rm lin}(k_{4})P_{\rm lin}(k_{345})  + \mbox{(89 perms.)} \nonumber \\
	T_{421111}&=& Z_1(\VEC{k}_1)Z_1(\VEC{k}_2)Z_1(\VEC{k}_{3})Z_1(\VEC{k}_4)Z_2(-\VEC{k}_4,\VEC{k}_{45})Z_4(\VEC{k}_1,\VEC{k}_2,\VEC{k}_3,\VEC{k}_{45}) \nonumber \\
        &\times& P_{\rm lin}(k_1)P_{\rm lin}(k_2)P_{\rm lin}(k_{3})P_{\rm lin}(k_{4})P_{\rm lin}(k_{45})  + \mbox{(119 perms.)}.
\end{eqnarray}

\section{Full expressions of ${\rm Cov}\left[ B,B \right]_{PPP}$, ${\rm Cov}\left[ B,B \right]_{BB}$, ${\rm Cov}\left[ B,B \right]_{PT}$ and ${\rm Cov}\left[ B,B \right]_{P_6}$}
\label{Ap:equations}

In this appendix, we summarize the full expression of the bispectrum covariance.

The $PPP$ (\ref{Eq:covPPP_perm}), $BB$ (\ref{Eq:covBB_term}) and $PT$ (\ref{Eq:covPT_term}) terms are given by
\begin{eqnarray}
	{\rm Cov}\big[ \widehat{B}(\VEC{k}_1,\VEC{k}_2,\VEC{k}_3),\widehat{B}(\VEC{k}'_1,\VEC{k}'_2,\VEC{k}'_3) \big]_{PPP}
	&=& \Bigg[\frac{(2\pi)^3\delta_{\rm D}\left( \VEC{k}_1+\VEC{k}'_1 \right)(2\pi)^3\delta_{\rm D}\left( \VEC{k}_2+\VEC{k}'_2 \right)}{V} 
		+\frac{(2\pi)^3\delta_{\rm D}\left( \VEC{k}_2+\VEC{k}'_1 \right)(2\pi)^3\delta_{\rm D}\left( \VEC{k}_1+\VEC{k}'_2 \right)}{V}  \nonumber \\
	&+&      \frac{(2\pi)^3\delta_{\rm D}\left( \VEC{k}_2+\VEC{k}'_1 \right)(2\pi)^3\delta_{\rm D}\left( \VEC{k}_3+\VEC{k}'_2 \right)}{V} 
		+\frac{(2\pi)^3\delta_{\rm D}\left( \VEC{k}_3+\VEC{k}'_1 \right)(2\pi)^3\delta_{\rm D}\left( \VEC{k}_2+\VEC{k}'_2 \right)}{V}  \nonumber \\
	&+&      \frac{(2\pi)^3\delta_{\rm D}\left( \VEC{k}_1+\VEC{k}'_1 \right)(2\pi)^3\delta_{\rm D}\left( \VEC{k}_3+\VEC{k}'_2 \right)}{V} 
		+\frac{(2\pi)^3\delta_{\rm D}\left( \VEC{k}_3+\VEC{k}'_1 \right)(2\pi)^3\delta_{\rm D}\left( \VEC{k}_1+\VEC{k}'_2 \right)}{V}  \Bigg]\nonumber \\
	&\times&
	P^{(\rm N)}(\VEC{k}_1) P^{(\rm N)}(\VEC{k}_2) P^{(\rm N)}(\VEC{k}_3),
	\label{EqAp:covBB_PPP}
\end{eqnarray}
\begin{eqnarray}
	 {\rm Cov}\big[ \widehat{B}(\VEC{k}_1,\VEC{k}_2,\VEC{k}_3),\widehat{B}(\VEC{k}'_1,\VEC{k}'_2,\VEC{k}'_3) \big]_{BB} 
	&=& \frac{(2\pi)^3\delta_{\rm D}\left( \VEC{k}_1-\VEC{k}'_1 \right)}{V} B^{(\rm N)}(\VEC{k}_1,\VEC{k}_2,\VEC{k}_3) B^{(\rm N)}(\VEC{k}'_1,\VEC{k}'_2,\VEC{k}'_3) \nonumber \\
	&+& \frac{(2\pi)^3\delta_{\rm D}\left( \VEC{k}_2-\VEC{k}'_1 \right)}{V} B^{(\rm N)}(\VEC{k}_2,\VEC{k}_1,\VEC{k}_3) B^{(\rm N)}(\VEC{k}'_1,\VEC{k}'_2,\VEC{k}'_3) \nonumber \\
	&+& \frac{(2\pi)^3\delta_{\rm D}\left( \VEC{k}_3-\VEC{k}'_1 \right)}{V} B^{(\rm N)}(\VEC{k}_3,\VEC{k}_1,\VEC{k}_2) B^{(\rm N)}(\VEC{k}'_1,\VEC{k}'_2,\VEC{k}'_3)  \nonumber \\ 
	&+& \frac{(2\pi)^3\delta_{\rm D}\left( \VEC{k}_1-\VEC{k}'_2 \right)}{V} B^{(\rm N)}(\VEC{k}_1,\VEC{k}_2,\VEC{k}_3) B^{(\rm N)}(\VEC{k}'_2,\VEC{k}'_1,\VEC{k}'_3) \nonumber \\
	&+& \frac{(2\pi)^3\delta_{\rm D}\left( \VEC{k}_2-\VEC{k}'_2 \right)}{V} B^{(\rm N)}(\VEC{k}_2,\VEC{k}_1,\VEC{k}_3) B^{(\rm N)}(\VEC{k}'_2,\VEC{k}'_1,\VEC{k}'_3)   \nonumber \\
	&+& \frac{(2\pi)^3\delta_{\rm D}\left( \VEC{k}_3-\VEC{k}'_2 \right)}{V} B^{(\rm N)}(\VEC{k}_3,\VEC{k}_1,\VEC{k}_2) B^{(\rm N)}(\VEC{k}'_2,\VEC{k}'_1,\VEC{k}'_3)\nonumber \\
	&+& \frac{(2\pi)^3\delta_{\rm D}\left( \VEC{k}_1-\VEC{k}'_3 \right)}{V} B^{(\rm N)}(\VEC{k}_1,\VEC{k}_2,\VEC{k}_3) B^{(\rm N)}(\VEC{k}'_3,\VEC{k}'_1,\VEC{k}'_2)   \nonumber \\
	&+& \frac{(2\pi)^3\delta_{\rm D}\left( \VEC{k}_2-\VEC{k}'_3 \right)}{V} B^{(\rm N)}(\VEC{k}_2,\VEC{k}_1,\VEC{k}_3) B^{(\rm N)}(\VEC{k}'_3,\VEC{k}'_1,\VEC{k}'_2) \nonumber \\
	&+& \frac{(2\pi)^3\delta_{\rm D}\left( \VEC{k}_3-\VEC{k}'_3 \right)}{V} B^{(\rm N)}(\VEC{k}_3,\VEC{k}_1,\VEC{k}_2) B^{(\rm N)}(\VEC{k}'_3,\VEC{k}'_1,\VEC{k}'_2),
	\label{EqAp:covBB_BB}
\end{eqnarray}
and
\begin{eqnarray}
	 {\rm Cov}\big[ \widehat{B}(\VEC{k}_1,\VEC{k}_2,\VEC{k}_3),\widehat{B}(\VEC{k}'_1,\VEC{k}'_2,\VEC{k}'_3) \big]_{PT} 
        &=& \frac{(2\pi)^3\delta_{\rm D}\left( \VEC{k}_1+\VEC{k}'_1 \right)}{V} P^{(\rm N)}(\VEC{k}_1) T^{(\rm N)}(\VEC{k}_2, \VEC{k}_3,\VEC{k}'_2,\VEC{k}'_3) \nonumber \\
	&+& \frac{(2\pi)^3\delta_{\rm D}\left( \VEC{k}_2+\VEC{k}'_1 \right)}{V} P^{(\rm N)}(\VEC{k}_2) T^{(\rm N)}(\VEC{k}_1, \VEC{k}_3,\VEC{k}'_2,\VEC{k}'_3) \nonumber \\
	&+& \frac{(2\pi)^3\delta_{\rm D}\left( \VEC{k}_3+\VEC{k}'_1 \right)}{V} P^{(\rm N)}(\VEC{k}_3) T^{(\rm N)}(\VEC{k}_1, \VEC{k}_2,\VEC{k}'_2,\VEC{k}'_3)  \nonumber \\ 
	&+& \frac{(2\pi)^3\delta_{\rm D}\left( \VEC{k}_1+\VEC{k}'_2 \right)}{V} P^{(\rm N)}(\VEC{k}_1) T^{(\rm N)}(\VEC{k}_2, \VEC{k}_3,\VEC{k}'_1,\VEC{k}'_3) \nonumber \\
	&+& \frac{(2\pi)^3\delta_{\rm D}\left( \VEC{k}_2+\VEC{k}'_2 \right)}{V} P^{(\rm N)}(\VEC{k}_2) T^{(\rm N)}(\VEC{k}_1, \VEC{k}_3,\VEC{k}'_1,\VEC{k}'_3)   \nonumber \\
	&+& \frac{(2\pi)^3\delta_{\rm D}\left( \VEC{k}_3+\VEC{k}'_2 \right)}{V} P^{(\rm N)}(\VEC{k}_3) T^{(\rm N)}(\VEC{k}_1, \VEC{k}_2,\VEC{k}'_1,\VEC{k}'_3)\nonumber \\
	&+& \frac{(2\pi)^3\delta_{\rm D}\left( \VEC{k}_1+\VEC{k}'_3 \right)}{V} P^{(\rm N)}(\VEC{k}_1) T^{(\rm N)}(\VEC{k}_2, \VEC{k}_3,\VEC{k}'_1,\VEC{k}'_2)   \nonumber \\
	&+& \frac{(2\pi)^3\delta_{\rm D}\left( \VEC{k}_2+\VEC{k}'_3 \right)}{V} P^{(\rm N)}(\VEC{k}_2) T^{(\rm N)}(\VEC{k}_1, \VEC{k}_3,\VEC{k}'_1,\VEC{k}'_2) \nonumber \\
	&+& \frac{(2\pi)^3\delta_{\rm D}\left( \VEC{k}_3+\VEC{k}'_3 \right)}{V} P^{(\rm N)}(\VEC{k}_3) T^{(\rm N)}(\VEC{k}_1, \VEC{k}_2,\VEC{k}'_1,\VEC{k}'_2).
	\label{EqAp:covBB_PT}
\end{eqnarray}

For the $P_6$ term, The second, third and forth lines on the RHS of Eq~(\ref{Eq:covP6_term}) are respectively given by
\begin{eqnarray}
	&& \left[ P_5(\VEC{k}_1+\VEC{k}'_1,\VEC{k}_2,\VEC{k}_3,\VEC{k}'_2,\VEC{k}'_3)+ \mbox{(8 perms.)} \right] \nonumber \\
	&=&  P_5(\VEC{k}_1+\VEC{k}'_1,\VEC{k}_2,\VEC{k}_3,\VEC{k}'_2,\VEC{k}'_3)
	   + P_5(\VEC{k}_1+\VEC{k}'_2,\VEC{k}_2,\VEC{k}_3,\VEC{k}'_1,\VEC{k}'_3)
	   + P_5(\VEC{k}_1+\VEC{k}'_3,\VEC{k}_2,\VEC{k}_3,\VEC{k}'_1,\VEC{k}'_2) \nonumber \\
	&+&  P_5(\VEC{k}_2+\VEC{k}'_1,\VEC{k}_1,\VEC{k}_3,\VEC{k}'_2,\VEC{k}'_3)
	   + P_5(\VEC{k}_2+\VEC{k}'_2,\VEC{k}_1,\VEC{k}_3,\VEC{k}'_1,\VEC{k}'_3)
	   + P_5(\VEC{k}_2+\VEC{k}'_3,\VEC{k}_1,\VEC{k}_3,\VEC{k}'_1,\VEC{k}'_2) \nonumber \\
	&+&  P_5(\VEC{k}_3+\VEC{k}'_1,\VEC{k}_1,\VEC{k}_2,\VEC{k}'_2,\VEC{k}'_3)
	   + P_5(\VEC{k}_3+\VEC{k}'_2,\VEC{k}_1,\VEC{k}_2,\VEC{k}'_1,\VEC{k}'_3)
	   + P_5(\VEC{k}_3+\VEC{k}'_3,\VEC{k}_1,\VEC{k}_2,\VEC{k}'_1,\VEC{k}'_2),
\end{eqnarray}
\begin{eqnarray}
	&& \left[ T(\VEC{k}_1+\VEC{k}'_1,\VEC{k}_2+\VEC{k}'_2,\VEC{k}_3,\VEC{k}'_3)+ \mbox{(17 perms.)} \right] \nonumber \\
	&=&  T(\VEC{k}_1+\VEC{k}'_1,\VEC{k}_2+\VEC{k}'_2,\VEC{k}_3,\VEC{k}'_3)
	   + T(\VEC{k}_1+\VEC{k}'_1,\VEC{k}_2+\VEC{k}'_3,\VEC{k}_3,\VEC{k}'_2)
	   + T(\VEC{k}_1+\VEC{k}'_2,\VEC{k}_2+\VEC{k}'_3,\VEC{k}_3,\VEC{k}'_1) \nonumber \\
	&+&  T(\VEC{k}_1+\VEC{k}'_2,\VEC{k}_2+\VEC{k}'_1,\VEC{k}_3,\VEC{k}'_3)
	   + T(\VEC{k}_1+\VEC{k}'_3,\VEC{k}_2+\VEC{k}'_1,\VEC{k}_3,\VEC{k}'_2)
	   + T(\VEC{k}_1+\VEC{k}'_3,\VEC{k}_2+\VEC{k}'_2,\VEC{k}_3,\VEC{k}'_1) \nonumber \\
	&+&  T(\VEC{k}_1+\VEC{k}'_1,\VEC{k}_3+\VEC{k}'_2,\VEC{k}_2,\VEC{k}'_3)
	   + T(\VEC{k}_1+\VEC{k}'_1,\VEC{k}_3+\VEC{k}'_3,\VEC{k}_2,\VEC{k}'_2)
	   + T(\VEC{k}_1+\VEC{k}'_2,\VEC{k}_3+\VEC{k}'_3,\VEC{k}_2,\VEC{k}'_1) \nonumber \\
	&+&  T(\VEC{k}_1+\VEC{k}'_2,\VEC{k}_3+\VEC{k}'_1,\VEC{k}_2,\VEC{k}'_3)
	   + T(\VEC{k}_1+\VEC{k}'_3,\VEC{k}_3+\VEC{k}'_1,\VEC{k}_2,\VEC{k}'_2)
	   + T(\VEC{k}_1+\VEC{k}'_3,\VEC{k}_3+\VEC{k}'_2,\VEC{k}_2,\VEC{k}'_1) \nonumber \\
	&+&  T(\VEC{k}_2+\VEC{k}'_1,\VEC{k}_3+\VEC{k}'_2,\VEC{k}_1,\VEC{k}'_3)
	   + T(\VEC{k}_2+\VEC{k}'_1,\VEC{k}_3+\VEC{k}'_3,\VEC{k}_1,\VEC{k}'_2)
	   + T(\VEC{k}_2+\VEC{k}'_2,\VEC{k}_3+\VEC{k}'_3,\VEC{k}_1,\VEC{k}'_1) \nonumber \\
	&+&  T(\VEC{k}_2+\VEC{k}'_2,\VEC{k}_3+\VEC{k}'_1,\VEC{k}_1,\VEC{k}'_3)
	   + T(\VEC{k}_2+\VEC{k}'_3,\VEC{k}_3+\VEC{k}'_1,\VEC{k}_1,\VEC{k}'_2)
	   + T(\VEC{k}_2+\VEC{k}'_3,\VEC{k}_3+\VEC{k}'_2,\VEC{k}_1,\VEC{k}'_1),
\end{eqnarray}
and
\begin{eqnarray}
	&& \left[ B(\VEC{k}_1+\VEC{k}'_1,\VEC{k}_2+\VEC{k}'_2,\VEC{k}_3+\VEC{k}'_3)+ \mbox{(5 perms.)} \right] \nonumber \\
	&=& B(\VEC{k}_1+\VEC{k}'_1,\VEC{k}_2+\VEC{k}'_2,\VEC{k}_3+\VEC{k}'_3)
	  + B(\VEC{k}_1+\VEC{k}'_1,\VEC{k}_2+\VEC{k}'_3,\VEC{k}_3+\VEC{k}'_2)
	  + B(\VEC{k}_1+\VEC{k}'_2,\VEC{k}_2+\VEC{k}'_1,\VEC{k}_3+\VEC{k}'_3) \nonumber \\
	&+& B(\VEC{k}_1+\VEC{k}'_2,\VEC{k}_2+\VEC{k}'_3,\VEC{k}_3+\VEC{k}'_1)
	  + B(\VEC{k}_1+\VEC{k}'_3,\VEC{k}_2+\VEC{k}'_1,\VEC{k}_3+\VEC{k}'_2)
	  + B(\VEC{k}_1+\VEC{k}'_3,\VEC{k}_2+\VEC{k}'_2,\VEC{k}_3+\VEC{k}'_1).
\end{eqnarray}

We can analytical calculate the delta function appearing in Eqs.~(\ref{EqAp:covBB_PPP})
when we compute the covariance of the bispectrum multipoles defined in Eq.~(\ref{Eq:cov_multipole}).
Then, we have
\begin{eqnarray}
	&&{\rm Cov}\left[ B_{\ell_1\ell_2L}(k_1,k_2), \, B_{\ell'_1\ell'_2L'}(k'_1,k'_2) \right]_{PPP} \nonumber \\
	&=& M_{\ell_1\ell_2L}^{\ell_1'\ell_2'L'} \, V
	\int \frac{d \cos\theta_{k_1}}{2}\int \frac{d \hat{k}_2}{4\pi}{\cal S}_{\ell_1\ell_2L}(\hat{k}_1,\hat{k}_2,\hat{n})
	 \nonumber \\
	&\times& 
	\Bigg\{ 
	  {\cal S}_{\ell'_1\ell'_2L'}(\hat{k}_1,\hat{k}_2,\hat{n})\frac{W(k_1,k_1')}{\widetilde{N}_{\rm mode}(k_1,k_1')}\frac{W(k_2,k_2')}{\widetilde{N}_{\rm mode}(k_2,k_2')}
	+ {\cal S}_{\ell'_1\ell'_2L'}(\hat{k}_2,\hat{k}_1,\hat{n})\frac{W(k_2,k_1')}{\widetilde{N}_{\rm mode}(k_2,k_1')}\frac{W(k_1,k_2')}{\widetilde{N}_{\rm mode}(k_1,k_2')} \nonumber \\
	&+& \hspace{0.2cm}
	  {\cal S}_{\ell'_1\ell'_2L'}(\hat{k}_1,\hat{k}_3,\hat{n})\frac{W(k_1,k_1')}{\widetilde{N}_{\rm mode}(k_1,k_1')}\frac{W(k_3,k_2')}{\widetilde{N}_{\rm mode}(k_3,k_2')}
	+ {\cal S}_{\ell'_1\ell'_2L'}(\hat{k}_3,\hat{k}_1,\hat{n})\frac{W(k_3,k_1')}{\widetilde{N}_{\rm mode}(k_3,k_1')}\frac{W(k_1,k_2')}{\widetilde{N}_{\rm mode}(k_1,k_2')} \nonumber \\
	&+& \hspace{0.2cm}
	  {\cal S}_{\ell'_1\ell'_2L'}(\hat{k}_2,\hat{k}_3,\hat{n})\frac{W(k_2,k_1')}{\widetilde{N}_{\rm mode}(k_2,k_1')}\frac{W(k_3,k_2')}{\widetilde{N}_{\rm mode}(k_3,k_2')} 
	+ {\cal S}_{\ell'_1\ell'_2L'}(\hat{k}_3,\hat{k}_2,\hat{n})\frac{W(k_3,k_1')}{\widetilde{N}_{\rm mode}(k_3,k_1')}\frac{W(k_2,k_2')}{\widetilde{N}_{\rm mode}(k_2,k_2')} 
	\Bigg\} \nonumber \\
	&\times& P^{(\rm N)}(\VEC{k}_1) P^{(\rm N)}(\VEC{k}_2) P^{(\rm N)}(\VEC{k}_3),
\end{eqnarray}
where $M_{\ell_1\ell_2L}^{\ell'_1\ell'_2L'} = (2\ell_1+1)(2\ell_2+1)(2L+1)(2\ell'_1+1)(2\ell'_2+1)(2L'+1)
\left( \begin{smallmatrix} \ell_1 & \ell_2 & L \\ 0 & 0 & 0 \end{smallmatrix}  \right)
\left( \begin{smallmatrix} \ell'_1 & \ell'_2 & L' \\ 0 & 0 & 0 \end{smallmatrix}  \right)$,
and ${\cal S}$, $\widetilde{N}_{\rm mode}$ and $W$ are defined in Eqs.~(\ref{Eq:Slll}), (\ref{Eq:new_Nmode}) and (\ref{Eq:new_delta}), respectively.
Similarly, inserting Eqs.~(\ref{EqAp:covBB_BB}) and (\ref{EqAp:covBB_PT}) in Eq.~(\ref{Eq:cov_multipole}), 
we finally derive
\begin{eqnarray}
	&&{\rm Cov}\left[ B_{\ell_1\ell_2L}(k_1,k_2), \, B_{\ell'_1\ell'_2L'}(k'_1,k'_2) \right]_{BB} \nonumber \\
	&=&
	M_{\ell_1\ell_2L}^{\ell_1'\ell_2'L'} \int \frac{d \cos \theta_{k_1}}{2}\int \frac{d\hat{k}_2}{4\pi}\int \frac{d\hat{k}'_2}{4\pi}
	{\cal S}_{\ell_1\ell_2L}(\hat{k}_1,\hat{k}_2,\hat{n})\,  {\cal S}_{\ell'_1\ell'_2L'}(\hat{k}_1,\hat{k}'_2,\hat{n})\,
	\frac{W\left( k_1,k'_1 \right)}{\widetilde{N}_{\rm mode}(k_1,k'_1)} 
	B^{(\rm N)}(\VEC{k}_1,\VEC{k}_2, \VEC{k}_3) B^{(\rm N)}(\VEC{k}_1, \VEC{k}'_2, \VEC{k}'_3) \nonumber \\
	&+& 
	M_{\ell_1\ell_2L}^{\ell_1'\ell_2'L'} \int \frac{d \cos \theta_{k_1}}{2}\int \frac{d\hat{k}_2}{4\pi}\int \frac{d\hat{k}'_2}{4\pi}
	{\cal S}_{\ell_1\ell_2L}(\hat{k}_1,\hat{k}_2,\hat{n})\,  {\cal S}_{\ell'_1\ell'_2L'}(\hat{k}_2,\hat{k}'_2,\hat{n})\,
	\frac{W\left( k_2,k'_1 \right)}{\widetilde{N}_{\rm mode}(k_2,k'_1)} 
	B^{(\rm N)}(\VEC{k}_2,\VEC{k}_1, \VEC{k}_3) B^{(\rm N)}(\VEC{k}_2, \VEC{k}'_2, \VEC{k}'_3) \nonumber \\
	&+& 
	M_{\ell_1\ell_2L}^{\ell_1'\ell_2'L'} \int \frac{d \cos \theta_{k_1}}{2}\int \frac{d\hat{k}_2}{4\pi}\int \frac{d\hat{k}'_2}{4\pi}
	{\cal S}_{\ell_1\ell_2L}(\hat{k}_1,\hat{k}_2,\hat{n})\,  {\cal S}_{\ell'_1\ell'_2L'}(\hat{k}_3,\hat{k}'_2,\hat{n})\,
	\frac{W\left( k_3,k'_1 \right)}{\widetilde{N}_{\rm mode}(k_3,k'_1)} 
	B^{(\rm N)}(\VEC{k}_3,\VEC{k}_1, \VEC{k}_2) B^{(\rm N)}(\VEC{k}_3, \VEC{k}'_2, \VEC{k}'_3) \nonumber \\
	&+&
	M_{\ell_1\ell_2L}^{\ell_1'\ell_2'L'} \int \frac{d \cos \theta_{k_1}}{2}\int \frac{d\hat{k}_2}{4\pi}\int \frac{d\hat{k}'_1}{4\pi}
	{\cal S}_{\ell_1\ell_2L}(\hat{k}_1,\hat{k}_2,\hat{n})\,  {\cal S}_{\ell'_1\ell'_2L'}(\hat{k}'_1,\hat{k}_1,\hat{n})\,
	\frac{W\left( k_1,k'_2 \right)}{\widetilde{N}_{\rm mode}(k_1,k'_2)} 
	B^{(\rm N)}(\VEC{k}_1,\VEC{k}_2, \VEC{k}_3) B^{(\rm N)}(\VEC{k}_1, \VEC{k}'_1, \VEC{k}'_3) \nonumber \\
	&+& 
	M_{\ell_1\ell_2L}^{\ell_1'\ell_2'L'} \int \frac{d \cos \theta_{k_1}}{2}\int \frac{d\hat{k}_2}{4\pi}\int \frac{d\hat{k}'_1}{4\pi}
	{\cal S}_{\ell_1\ell_2L}(\hat{k}_1,\hat{k}_2,\hat{n})\,  {\cal S}_{\ell'_1\ell'_2L'}(\hat{k}'_1,\hat{k}_2,\hat{n})\,
	\frac{W\left( k_2,k'_2 \right)}{\widetilde{N}_{\rm mode}(k_2,k'_2)} 
	B^{(\rm N)}(\VEC{k}_2,\VEC{k}_1, \VEC{k}_3) B^{(\rm N)}(\VEC{k}_2, \VEC{k}'_1, \VEC{k}'_3) \nonumber \\
	&+& 
	M_{\ell_1\ell_2L}^{\ell_1'\ell_2'L'} \int \frac{d \cos \theta_{k_1}}{2}\int \frac{d\hat{k}_2}{4\pi}\int \frac{d\hat{k}'_1}{4\pi}
	{\cal S}_{\ell_1\ell_2L}(\hat{k}_1,\hat{k}_2,\hat{n})\,  {\cal S}_{\ell'_1\ell'_2L'}(\hat{k}'_1,\hat{k}_3,\hat{n})\,
	\frac{W\left( k_3,k'_2 \right)}{\widetilde{N}_{\rm mode}(k_3,k'_2)} 
	B^{(\rm N)}(\VEC{k}_3,\VEC{k}_1, \VEC{k}_2) B^{(\rm N)}(\VEC{k}_3, \VEC{k}'_1, \VEC{k}'_3) \nonumber \\
	&+&
	M_{\ell_1\ell_2L}^{\ell_1'\ell_2'L'} \int \frac{d \cos \theta_{k_2}}{2}\int \frac{d\hat{k}_1'}{4\pi}\int \frac{d\hat{k}'_2}{4\pi}
	{\cal S}_{\ell_1\ell_2L}(\hat{k}'_3,\hat{k}_2,\hat{n})\,  {\cal S}_{\ell'_1\ell'_2L'}(\hat{k}'_1,\hat{k}'_2,\hat{n})\,
	\frac{W\left( k_1,k'_3 \right)}{\widetilde{N}_{\rm mode}(k_1,k'_3)} 
	B^{(\rm N)}(\VEC{k}'_3,\VEC{k}_2, \VEC{k}_3) B^{(\rm N)}( \VEC{k}'_3,\VEC{k}_1', \VEC{k}'_2) \nonumber \\
	&+&
	M_{\ell_1\ell_2L}^{\ell_1'\ell_2'L'} \int \frac{d \cos \theta_{k_1}}{2}\int \frac{d\hat{k}_1'}{4\pi}\int \frac{d\hat{k}'_2}{4\pi}
	{\cal S}_{\ell_1\ell_2L}(\hat{k}_1,\hat{k}'_3,\hat{n})\,  {\cal S}_{\ell'_1\ell'_2L'}(\hat{k}'_1,\hat{k}'_2,\hat{n})\,
	\frac{W\left( k_2,k'_3 \right)}{\widetilde{N}_{\rm mode}(k_2,k'_3)} 
	B^{(\rm N)}(\VEC{k}'_3,\VEC{k}_1, \VEC{k}_3) B^{(\rm N)}( \VEC{k}'_3,\VEC{k}_1', \VEC{k}'_2) \nonumber \\
	&+&
	M_{\ell_1\ell_2L}^{\ell_1'\ell_2'L'} \int \frac{d \cos \theta_{k_1}}{2}\int \frac{d\hat{k}_2}{4\pi}\int \frac{d\hat{k}'_2}{4\pi}
	{\cal S}_{\ell_1\ell_2L}(\hat{k}_1,\hat{k}_2,\hat{n})\,  {\cal S}_{\ell'_1\ell'_2L'}(\hat{k}_{\alpha},\hat{k}'_2,\hat{n})\,
	\frac{W\left( k_{\alpha}, k'_1 \right)}{\widetilde{N}_{\rm mode}(k_{\alpha},k'_1)} 
	B^{(\rm N)}(\VEC{k}_3,\VEC{k}_1, \VEC{k}_2) B^{(\rm N)}( \VEC{k}_3,\VEC{k}_{\alpha}, \VEC{k}'_2) \nonumber \\
\end{eqnarray}
where in the last line we used $\delta_{\rm D}\left( \VEC{k}_3-\VEC{k}'_3 \right) = \delta_{\rm D}\left( \VEC{k}_{\alpha} - \VEC{k}_1' \right)$ 
with $\VEC{k}_{\alpha} \equiv \VEC{k}_1+\VEC{k}_2-\VEC{k}_2'$, and 
\begin{eqnarray}
	&&{\rm Cov}\left[ B_{\ell_1\ell_2L}(k_1,k_2), \, B_{\ell'_1\ell'_2L'}(k'_1,k'_2) \right]_{PT} \nonumber \\
	&=&
	M_{\ell_1\ell_2L}^{\ell_1'\ell_2'L'} \int \frac{d \cos \theta_{k_1}}{2}\int \frac{d\hat{k}_2}{4\pi}\int \frac{d\hat{k}'_2}{4\pi}
	{\cal S}_{\ell_1\ell_2L}(\hat{k}_1,\hat{k}_2,\hat{n})\,  {\cal S}_{\ell'_1\ell'_2L'}(-\hat{k}_1,\hat{k}'_2,\hat{n})\,
	\frac{W\left( k_1,k'_1 \right)}{\widetilde{N}_{\rm mode}(k_1,k'_1)} 
	P^{(\rm N)}(\VEC{k}_1) T^{(\rm N)}(\VEC{k}_2, \VEC{k}_3, \VEC{k}'_2, \VEC{k}'_3) \nonumber \\
	&+&
	M_{\ell_1\ell_2L}^{\ell_1'\ell_2'L'} \int \frac{d \cos \theta_{k_1}}{2}\int \frac{d\hat{k}_2}{4\pi}\int \frac{d\hat{k}'_2}{4\pi}
	{\cal S}_{\ell_1\ell_2L}(\hat{k}_1,\hat{k}_2,\hat{n})\,  {\cal S}_{\ell'_1\ell'_2L'}(-\hat{k}_2,\hat{k}'_2,\hat{n})\,
	\frac{W\left( k_2,k'_1 \right)}{\widetilde{N}_{\rm mode}(k_2,k'_1)} 
	P^{(\rm N)}(\VEC{k}_2) T^{(\rm N)}(\VEC{k}_1, \VEC{k}_3, \VEC{k}'_2, \VEC{k}'_3) \nonumber \\
	&+&
	M_{\ell_1\ell_2L}^{\ell_1'\ell_2'L'} \int \frac{d \cos \theta_{k_1}}{2}\int \frac{d\hat{k}_2}{4\pi}\int \frac{d\hat{k}'_2}{4\pi}
	{\cal S}_{\ell_1\ell_2L}(\hat{k}_1,\hat{k}_2,\hat{n})\,  {\cal S}_{\ell'_1\ell'_2L'}(-\hat{k}_3,\hat{k}'_2,\hat{n})\,
	\frac{W\left( k_3,k'_1 \right)}{\widetilde{N}_{\rm mode}(k_3,k'_1)} 
	P^{(\rm N)}(\VEC{k}_3) T^{(\rm N)}(\VEC{k}_1, \VEC{k}_2, \VEC{k}'_2, \VEC{k}'_3) \nonumber \\
	&+&
	M_{\ell_1\ell_2L}^{\ell_1'\ell_2'L'} \int \frac{d \cos \theta_{k_1}}{2}\int \frac{d\hat{k}_2}{4\pi}\int \frac{d\hat{k}'_1}{4\pi}
	{\cal S}_{\ell_1\ell_2L}(\hat{k}_1,\hat{k}_2,\hat{n})\,  {\cal S}_{\ell'_1\ell'_2L'}(\hat{k}'_1,-\hat{k}_1,\hat{n})\,
	\frac{W\left( k_1,k'_2 \right)}{\widetilde{N}_{\rm mode}(k_1,k'_2)} 
	P^{(\rm N)}(\VEC{k}_1) T^{(\rm N)}(\VEC{k}_2, \VEC{k}_3, \VEC{k}'_1, \VEC{k}'_3) \nonumber \\
	&+&
	M_{\ell_1\ell_2L}^{\ell_1'\ell_2'L'} \int \frac{d \cos \theta_{k_1}}{2}\int \frac{d\hat{k}_2}{4\pi}\int \frac{d\hat{k}'_1}{4\pi}
	{\cal S}_{\ell_1\ell_2L}(\hat{k}_1,\hat{k}_2,\hat{n})\,  {\cal S}_{\ell'_1\ell'_2L'}(\hat{k}'_1,-\hat{k}_2,\hat{n})\,
	\frac{W\left( k_2,k'_2 \right)}{\widetilde{N}_{\rm mode}(k_2,k'_2)} 
	P^{(\rm N)}(\VEC{k}_2) T^{(\rm N)}(\VEC{k}_1, \VEC{k}_3, \VEC{k}'_1, \VEC{k}'_3) \nonumber \\
	&+&
	M_{\ell_1\ell_2L}^{\ell_1'\ell_2'L'} \int \frac{d \cos \theta_{k_1}}{2}\int \frac{d\hat{k}_2}{4\pi}\int \frac{d\hat{k}'_1}{4\pi}
	{\cal S}_{\ell_1\ell_2L}(\hat{k}_1,\hat{k}_2,\hat{n})\,  {\cal S}_{\ell'_1\ell'_2L'}(\hat{k}'_1,-\hat{k}_3,\hat{n})\,
	\frac{W\left( k_3,k'_2 \right)}{\widetilde{N}_{\rm mode}(k_3,k'_2)} 
	P^{(\rm N)}(\VEC{k}_3) T^{(\rm N)}(\VEC{k}_1, \VEC{k}_2, \VEC{k}'_1, \VEC{k}'_3) \nonumber \\
	&+&
	M_{\ell_1\ell_2L}^{\ell_1'\ell_2'L'} \int \frac{d \cos \theta_{k_2}}{2}\int \frac{d\hat{k}'_1}{4\pi}\int \frac{d\hat{k}'_2}{4\pi}
	{\cal S}_{\ell_1\ell_2L}(-\hat{k}'_3,\hat{k}_2,\hat{n})\,  {\cal S}_{\ell'_1\ell'_2L'}(\hat{k}'_1,\hat{k}'_2,\hat{n})\,
	\frac{W\left( k_1,k'_3 \right)}{\widetilde{N}_{\rm mode}(k_1,k'_3)} 
	P^{(\rm N)}(\VEC{k}'_3) T^{(\rm N)}(\VEC{k}_2, \VEC{k}_3, \VEC{k}'_1, \VEC{k}_2') \nonumber \\
	&+&
	M_{\ell_1\ell_2L}^{\ell_1'\ell_2'L'} \int \frac{d \cos \theta_{k_1}}{2}\int \frac{d\hat{k}'_1}{4\pi}\int \frac{d\hat{k}'_2}{4\pi}
	{\cal S}_{\ell_1\ell_2L}(\hat{k}_1, -\hat{k}'_3,\hat{n})\,  {\cal S}_{\ell'_1\ell'_2L'}(\hat{k}'_1,\hat{k}'_2,\hat{n})\,
	\frac{W\left( k_2,k'_3 \right)}{\widetilde{N}_{\rm mode}(k_2,k'_3)} 
	P^{(\rm N)}(\VEC{k}'_3) T^{(\rm N)}(\VEC{k}_1, \VEC{k}_3, \VEC{k}'_1, \VEC{k}_2') \nonumber \\
	&+&
	M_{\ell_1\ell_2L}^{\ell_1'\ell_2'L'} \int \frac{d \cos \theta_{k_1}}{2}\int \frac{d\hat{k}_2}{4\pi}\int \frac{d\hat{k}'_2}{4\pi}
	{\cal S}_{\ell_1\ell_2L}(\hat{k}_1, \hat{k}_2,\hat{n})\,  {\cal S}_{\ell'_1\ell'_2L'}(-\hat{k}_{\beta},\hat{k}'_2,\hat{n})\,
	\frac{W\left( k_{\beta},k'_1 \right)}{\widetilde{N}_{\rm mode}(k_{\beta},k'_1)} 
	P^{(\rm N)}(\VEC{k}_3) T^{(\rm N)}(\VEC{k}_1, \VEC{k}_2, -\VEC{k}_{\beta}, \VEC{k}_2'), \nonumber \\
\end{eqnarray}
where $\VEC{k}_{\beta} = \VEC{k}_1+\VEC{k}_2+\VEC{k}'_2$.

\section{Power spectrum and bispectrum}
\label{Ap:powerspectrum}

\begin{figure}
	\includegraphics[width=\columnwidth]{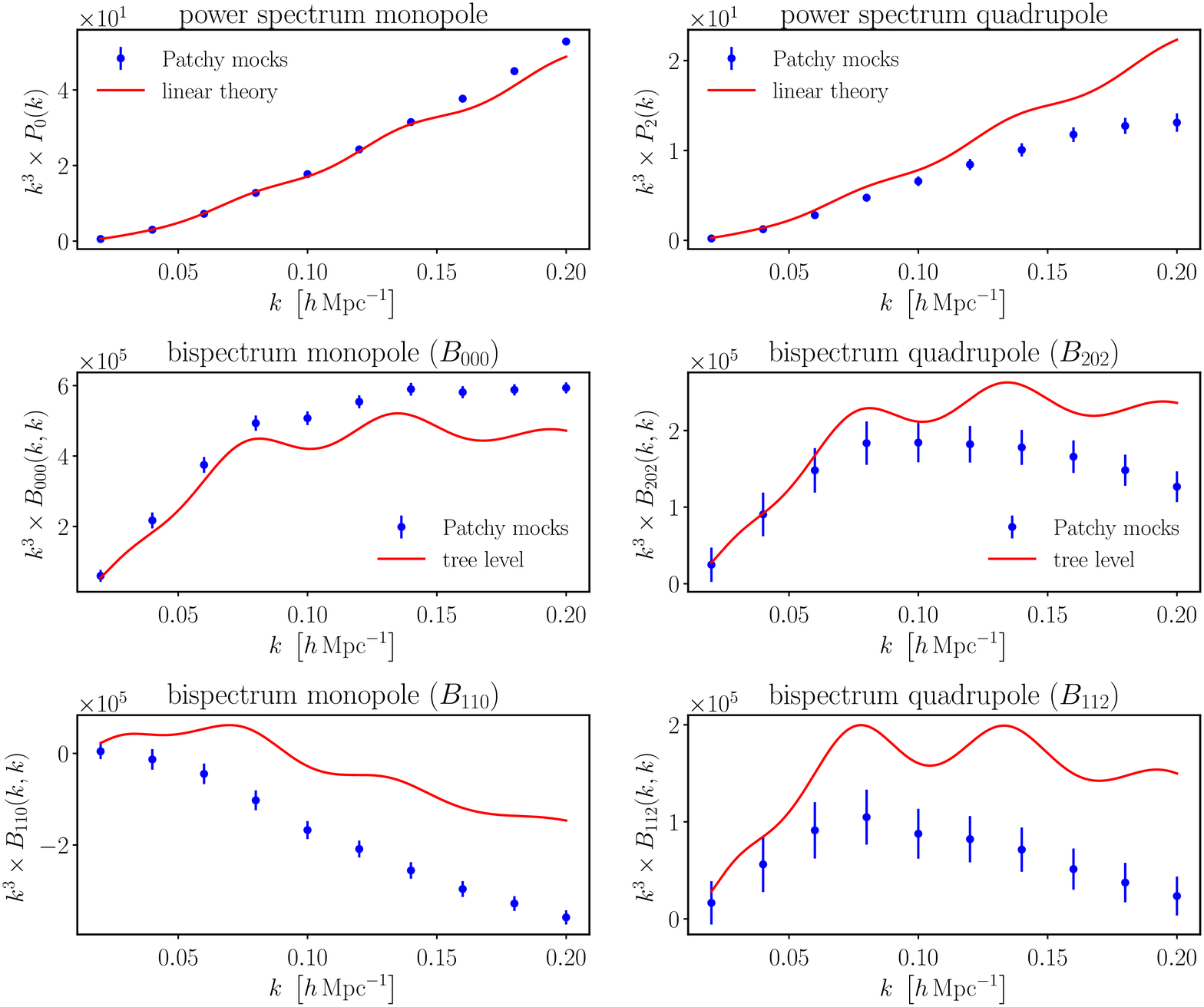}
	\caption{Comparison of the power and bispectrum multipoles predicted using perturbation theory (red lines) with the corresponding measurements from the Patchy mocks (blue points).
	}
	\label{fig:pkbk}
\end{figure}

In this appendix, we plot the power and bispectrum multipoles computed using perturbation theory (PT) at the leading order and compare them with the corresponding measurements from the Patchy mocks.

Figure~\ref{fig:pkbk} shows $P_0$ (upper left), $P_2$ (upper right), $B_{000}$ (middle left), $B_{202}$ (middle right), $B_{110}$ (lower left) and $B_{112}$ (lower right). For the monopole components of both the power and bispectra, $P_0$, $B_{000}$ and $B_{110}$,  the absolute values of their amplitudes computed by PT tend to be smaller than those of the Patchy mock measurements probably because of a lack of non-linear gravitational effects. On the other hand, for the quadrupole components, $P_2$, $B_{202}$ and $B_{112}$, the Patchy mock results becomes smaller than the PT calculations on small scales probably due to non-linear velocity effects such as the Finger-of-God effect. For $B_{110}$ and $B_{112}$, the PT solution starts to depart from the mock measurements even at larger scales $k\sim0.05\hk$ compared to the scale for $B_{000}$ and $B_{202}$, $\sim 0.1\hk$. This may be because of lack of non-linear bias effects. Investigating the impact on the non-linear bias effect on the bispectrum signal and the corresponding covariance is left for future work.

\bsp	
\label{lastpage}

\end{document}